\documentclass[useAMS,usenatbib]{mn2e}
\usepackage{graphicx}
\usepackage{aas_macros}
\usepackage{array}
\title[$N$-body simulation insights into the X-shaped bulge of the
Milky Way]{$N$-body insights into the Milky Way's X-shape}

\author[Esko Gardner et al.]{Esko
  Gardner,$^{1}$\thanks{E-mail:eskog@iki.fi} Victor P. Debattista,$^2$
  Annie C. Robin,$^1$ Sergio V\'asquez$^{3,4,5}$ and \newauthor Manuela
    Zoccali$^{3,5}$\\
  $^{1}$ Institut Utinam, CNRS UMR 6213, OSU THETA, Universit\'e de
  Franche-Comt\'e, 41bis avenue de l'Observatoire, 25000 Besan\c{c}on,
  France\\ $^2$ Jeremiah Horrocks Institute, University of Central
  Lancashire, Preston, PR1 2HE, UK\\ $^3$ Instituto de
  Astrof\'{i}sica, Facultad de F\'{i}sica, Pontificia Universidad Cat\'olica de Chile, Av.
  Vicu\~{n}a Mackenna 4860, Santiago, Chile\\ $^4$ European Southern
  Observatory, Alonso de Cordova 3107, Santiago, Chile\\ $^5$ The
  Milky Way Millennium Nucleus, Av. Vicu\~{n}a Mackenna 4860, 782-0436
  Macul, Santiago, Chile }
  \begin{document}
  \date{Draft - 2013 November}
  \pagerange{\pageref{firstpage}--\pageref{lastpage}} \pubyear{2013}
  \maketitle
  \label{firstpage}
  \begin{abstract}

    Using simulations of box/peanut- (B/P-) shaped bulges, we explore
    the nature of the X-shape of the Milky Way's bulge.  An X-shape
    can be associated with a B/P-shaped bulge driven by a bar.  By
    comparing in detail the simulations and the observations we show
    that the principal kinematic imprint of the X-shape is a minimum
    in the difference between the near and far side mean line-of-sight
    velocity along the minor axis. This minimum occurs at around $|b|
    = 4 \degr$, which is close to the lower limit at which the X-shape
    can be detected.  No coherent signature of an X-shape can be found
    in Galactocentric azimuthal velocities, vertical velocities, or
    any of the dispersions.  After scaling our simulations, we find
    that a best fit to the Bulge Radial Velocity Assay data leads to a
    bar angle of $15\degr$.  We also explore a purely geometric method
    for determining the distance to the Galactic Centre by tracing the
    arms of the X-shape.  We find that we are able to determine this
    ill-known distance to an accuracy of about 5\% with sufficiently
    accurate distance measurements for the red clump stars in the
    arms.
  \end{abstract}

\section{Introduction} 

About one-quarter of the stars in the local Universe are hosted by the
bulges of disc galaxies \citep{Persic1992, Fukugita1998}.  Bulges can
be either `classical' or `pseudo' bulges, with mixed types also
possible \citep[e.g.][]{Erwin2003, Debattista2005, Athanassoula2005,
  Nowak2010}. Classical bulges form via merging of sub-galactic
clumps, satellites and clusters \citep{Eggen1962, Tremaine1975,
  Searle1978, Kauffmann1993, Baugh1996, vandenBosch1998, Hopkins2010}.
On the other hand, pseudo bulges form via secular evolution of the
disc \citep{Combes1981, Combes1990, Raha1991, Norman1996,
  Courteau1996, Bureau1999, Debattista2004, Athanassoula2005,
  Drory2007, Kormendy2004}.  Understanding the mix of bulges present
in the Universe therefore represents an important step towards
understanding galaxy formation.

When viewed edge-on, roughly $45\%$ of galaxies host box/peanut-
(B/P-) shaped bulges \citep{Burbidge1959, Shaw1987, Lutticke2000a,
Laurikainen2011}, which are associated with bars \citep{Kuijken1995,
Bureau1999, Chung2004}. B/P-shaped bulges can also be recognized
photometrically at moderate inclinations \citep{Quillen1997,
Bettoni1994, Erwin2013} and, kinematically, even in face-on galaxies
\citep{Debattista2005, MendezAbreu2008}.  B/P-shapes are supported by
'banana' and 'anti-banana' orbits, which have been extensively studied
\citep{Pfenniger1991, Combes1990, Pfenniger1984, Pfenniger1985,
Patsis2002a}.  They arise from vertically unstable x1 orbits, even
when the non-axisymmetry driving them is weak
\citep{Patsis2002b}.

The origin of the Milky Way's bulge remains an issue of disagreement.
On the one hand the Milky Way's bulge is part of a bar
\citep{Dwek1995}, and is rapidly rotating (\citealt{Shen2010}, but see
also \citealt{Saha2012}). On the other hand, the bulge consists
largely of old \citep{Ortolani1995, Kuijken2002, Valenti2003,
Zoccali2003, Clarkson2008, Clarkson2011}, $\alpha$-enhanced stars
\citep{McWilliam97}.  It is now well-established that the Milky Way
hosts a bar \citep{Dwek1995, Binney1997, Bissantz2002, Lopez2005,
Vanhollebeke2009}. The first signs of a near/far asymmetry in the
bulge of the Milky Way were found by \cite{Rangwala2009}, who noted
differences in velocities in Baade's window and in two other fields
($l = \pm 5\degr$, $b = -3.5\degr$).  Soon thereafter, both
\cite{Nataf2010} and \cite{McWilliam2010} independently identified a
split in the red clump (RC) towards the Galactic Centre.
\cite{Nataf2010} used 267 fields taken from the OGLE-III survey
\citep{Udalski2008} at $-10\degr < l < 10\degr$ and $2\degr < |b| <
7\degr$. They found two distinct RCs towards many of these
lines-of-sight.  They also noticed that both RCs are equally
populated, and that these trends are found in the Northern and
Southern hemispheres. \cite{McWilliam2010} used four independent
photometric data sets, Two Micron All Sky Survey (2MASS) \citep{2MASS}, WFI and SOFI photometry
\citep{Zoccali2003} and OGLE \citep{Udalski2002} maps, to show that
the double RC is real and is present at $b \ga 4\degr$ in all the data
sets. Their analysis of 2MASS observations also suggested a
three-dimensional X-shape. \cite{DePropris2011}, using data from the
Galaxy and Mass Assembly (GAMA) Survey \citep{GAMA} for a field at
($l$,$b$)$ = (0\degr,-8\degr)$, found no discernible difference
between the near and the far sides in both radial velocity and radial
velocity dispersion. \cite{Saito2011} used 2MASS \citep{2MASS} data at
$ |l| \leq 8.5\degr$ and $3.5\degr \leq |b| \leq 8.5\degr$ to
construct a 3D density map of the X-shaped bulge. They found that the
X-shape is oriented at an angle of 20\degr\ to the line-of-sight. They
also found that the arms merge at $ |b| \leq 4\degr$, implying that
the X-shape is absent in Baade's Window.  \cite{Ness2012}, using the
ARGOS survey in three fields ($l$ = 0\degr\ $b = -5\degr, -7.5\degr,
-10\degr$), also found an X-shape. They found a difference in the
line-of-sight velocities of the two clumps which can be reproduced
qualitatively by an $N$-body model of a buckled bar from
\cite{Athanassoula2003}. Another comparison of a numerical model to
the Milky Way was provided by \cite{Shen2012}. When viewed like the
Milky Way, their model contains an X-shaped structure similar to the
one in the Galaxy which they show arises from the buckled
bar\footnote{Henceforth, whenever we refer to a B/P-shape we mean that
part of a bulge that has acquired a box or peanut shape.  We reserve
the term X-shape for the double-peaked density distribution in the
Milky Way resulting from a B/P-shape viewed from the Sun.}.  At higher
latitudes, $(l,b) = (0\degr,-10\degr)$, \cite{Uttenthaler2012}
measured radial velocities for the near and far RC and found that the
two distributions are indistinguishable from each other.  Recently,
\cite{Vasquez2013} obtained full space velocities for RC stars in the
near and far arms of the X-shape at $(l,b) = (0\degr,-6\degr)$.  They
find a difference in the mean line-of-sight velocities but none in the
mean vertical velocities of these stars. They compared their
observations to an $N$-body model from \citet{Debattista2005} (which we
also use in this paper), finding qualitative agreement between them.
  
This paper uses $N$-body simulations of barred galaxies to explore the
effect of B/P-bulge shapes of various strengths and at different
orientations on the observed kinematics of stars.  We also study
whether it is possible to measure the distance to the Galactic Centre
using the X-shape.


\section{Simulations}

We use the sample of high force and mass resolution barred galaxy
simulations described in \cite{Debattista2005}, who used these
simulations to explore the kinematic signature of face-on B/P bulges.
In units where $R_{\rm d} = M_{\rm d} = G = 1$ (where $R_{\rm d}$ and
$M_{\rm d}$ are the disc's exponential scale-length and mass,
respectively, and $G$ is the gravitational constant), which gives a
unit of time $(R_{\rm d}^3/GM_{\rm d})^{1/2}$, the values for the
disk$+$halo parameters are set such that the rotation curves were
always approximately flat to large radii. \cite{Debattista2005}
provides full details of the parameters of the different models.  In
scaling from simulation units to size in kiloparsecs ($L$) and
velocities in km s$^{-1}$ ($V$), the mass of the disc becomes
\begin{equation}
M = \frac{L V^2}{4.3 \times 10^{-6}
  \mathrm{~kpc~M_\odot^{-1}~(km~ s^{-1})^2}}
\end{equation}
For example, \citet{Debattista2006} suggested a scaling with $L = 2.5$
kpc and $V = 200$ km s$^{-1}$, so that the stellar mass $M = 2.3
\times 10^{10} M_\odot$.

\begin{figure}
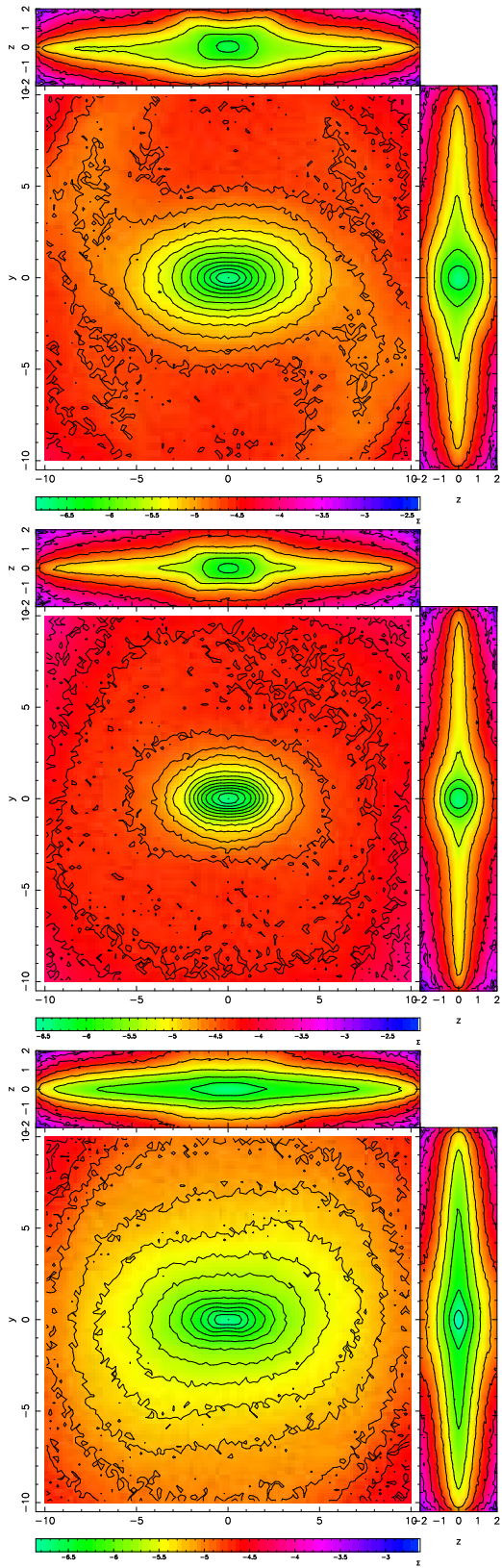

\begin{center}
    \includegraphics[width=0.85\columnwidth,angle=-90]{fig1.1.ps} \\
    \includegraphics[width=0.85\columnwidth,angle=-90]{fig1.2.ps}\\
    \includegraphics[width=0.85\columnwidth,angle=-90]{fig1.3.ps}    
\end{center}
\caption{Three orthogonal projections of the density of models
  R1 (top), B3 (middle) and R5 (bottom) as seen from outside the
  system. In each case the bar has been rotated on to the $x$-axis.}
\label{externalview}
\end{figure}

These simulations span a range of bulge, disc and halo properties
resulting in B/P shapes of varying strength. The main simulations we
consider from that paper are R1, B3 and R5. Comparison with models R2,
R6, and B2 from the same paper are presented in the Appendix. Models
in the R series had no initial bulges, while models B2 and B3 hosted
classical bulges from the start.  Models R1 and R2 both form strong
B/P shapes via the usual buckling instability. The B/P shape in model
R2 extends to a larger fraction of the bar size than in model R1.
Model R5 instead forms a quite weak B/P shape, while model R6 never
formed one.  Both model B2 and B3 formed B/P shapes. In the former
case, this is somewhat masked by the presence of a classical bulge,
but in model B3 the B/P shape is readily apparent despite the presence
of a classical bulge. Further details, including face-on and edge-on
views, of all the models can be found in \cite{Debattista2005}.  An
animation of the formation and evolution of the B/P shape in model R1
is presented in \cite{Debattista2006} (where this model is referred to
as model L2). Images of the three models considered here, R1, B3 and
R5, are shown in Fig. \ref{externalview}, for reference. These
images adopt our preferred scaling, which is described below.  It is
worth noting in Fig. \ref{externalview} that model R1 is not
symmetric about the mid-plane, having a more prominent B/P-shape at $z
> 0$ than at $z < 0$.  The main part of the paper will concentrate
in-depth on models R1 and B3 which host the clearest X-shapes. Model
R5 is an example of a weak B/P shape.

We generally compute velocities in our models as observer-centred
without correcting for the peculiar motion of the observer relative to
the local standard of rest, unless otherwise noted.  However, we also
consider azimuthal velocities (V$_\phi$) which are always in the
galactocentric reference frame.


\section{Scaling the simulations}
\label{chi2section}

The size scaling is chosen to match the size of the bar in the Milky
Way (3.43 kpc \citealt{Robin2012}). Fig. \ref{externalview} shows
three models scaled to this size with the bar rotated into the
$x$-axis.  Additionally, we use the bar as a measure of the near and
far sides of the bulge, choosing anything beyond the centre of the
galaxy, up to half a bar length, as the far side, and from the centre
of the galaxy to half a bar length towards the observer as the near
side. This is done to better refine our study to just the regions
where the X-shape is expected to lie. The scalings differ from model
to model, with $L = 2.4, 3.0$ and $3.5$ kpc in models R1, B3 and R5,
respectively.

\begin{figure}
\includegraphics[angle=-90,width=0.96\columnwidth]{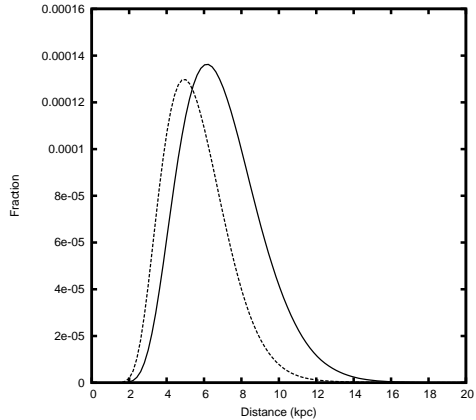}
\caption{Examples of selection functions for BRAVA stars.  We show
  only two lines-of-sight, with the nearest, ($l,b$) =
  ($-9\degr,-4\degr$), (dashed line) and the furthest, ($l,b$) =
  ($-6\degr,-8\degr$), (solid line) peak. The selection functions for
  the other fields have similar profiles and fall between these two.}
\label{selfunc} 
\end{figure}

In order to be able to compare our models to the Milky Way, we will
also have to choose what angle the model will be viewed from, as well
as the multiplicative scaling factor from simulation velocity units to
km s$^{-1}$. We used BRAVA \citep[Bulge Radial Velocity
Assay,][]{BRAVA} data of radial velocities from the fields at
$-10\degr \leq l \leq 10\degr$ at $b = -$6\degr, $-$8\degr\ and
$-10\degr \leq l \leq 22\degr$ at $b = -$4\degr.  We match by
minimizing $\chi^2$ between the means and standard deviations of the
line-of-sight velocities in the BRAVA data and in our model data. We
correct for the solar motion using the solar velocities of
\cite{Schonrich2012} and assume a Sun-Galactic Centre distance of
  8 kpc.  We convolve our models with the selection function of
M-giant stars along the various lines of sight of the BRAVA data,
including the 3D dust extinction maps from \cite{Marshall2006}.  In
order to compute the selection functions, we generate a large sample
of stars in each direction located in a solid angle of $2.1\degr
\times 2.1\degr$ around the field centre and with apparent $K$-band
magnitude between 8.2 and 9.25. Then we compute the probability for a
star to be selected by the survey as a function of distance in each
direction by evaluating how many stars have been selected in the
observed sample among all the stars present at a given distance
regardless of their magnitude.  The two selection functions with peaks
nearest ($D \simeq 5$ kpc) and furthest ($D \simeq 6$ kpc) from the
Sun are shown in Fig. \ref{selfunc}.  The full-width at half-maximum
of the selection functions are generally in the range $4-5$ kpc.

\subsection{Choosing the angle of the bar}
\begin{figure}
\begin{center}
    \includegraphics[width=0.8\columnwidth]{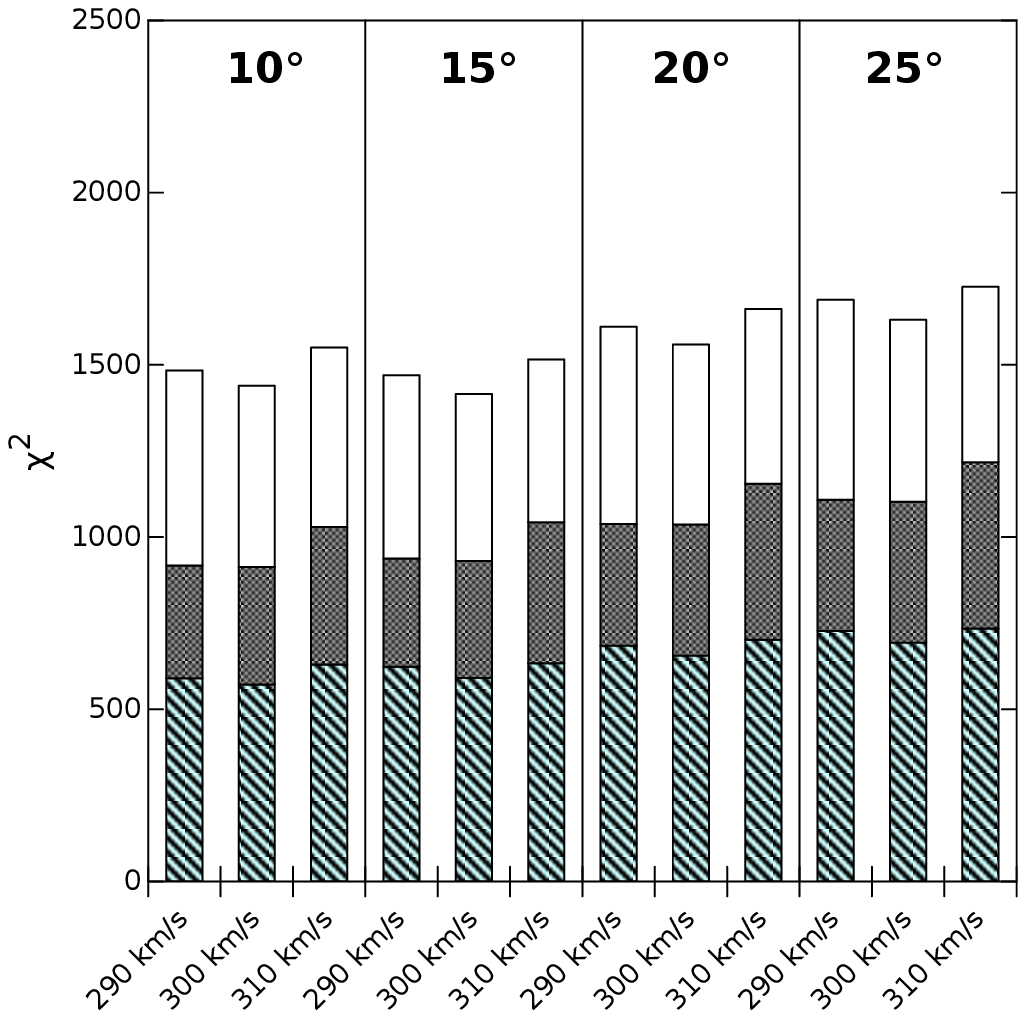} \\
    \includegraphics[width=0.8\columnwidth]{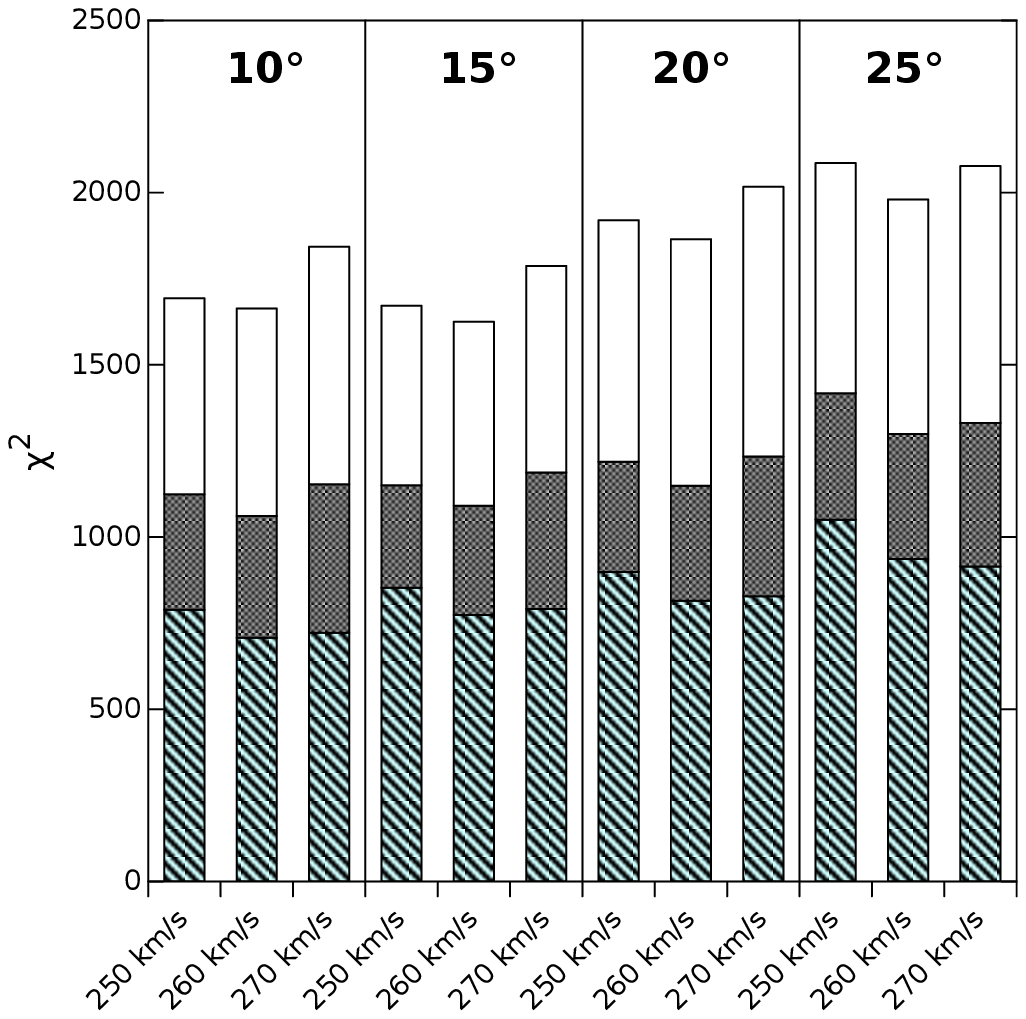}\\
    \includegraphics[width=0.8\columnwidth]{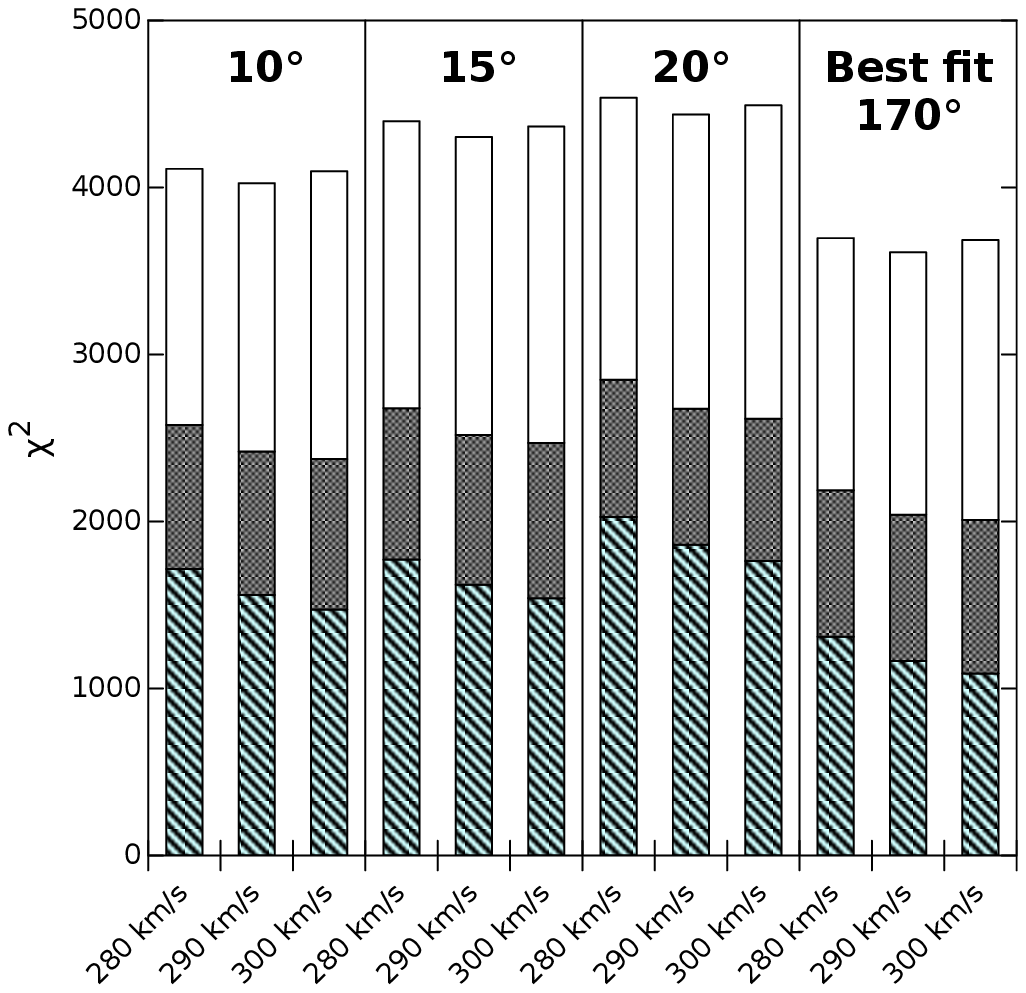}
\end{center}    
\caption{$\chi^2$ of fits to the BRAVA data for models R1 (top), B3
(middle) and R5 (bottom) varying the velocity scaling, $V$, and the
angle of the bar in the simulation. The shaded areas, from bottom to
top, show the individual $\chi^2$-values for the BRAVA fields at $b$=
$-$4\degr, $-$6\degr\ and $-$8\degr.  The angles, 10\degr, 15\degr,
20\degr\ and 25\degr\, are separated by vertical lines. For model R5 we
show the $\chi^2$-values for 10\degr, 15\degr, 20\degr\ and the best
fit at 170\degr.  Note the difference in vertical scale in the bottom
panel compared with the others.}
\label{scalingchi2} 
\end{figure}

The bar angle to the line joining the Sun to the galactic centre was
chosen by minimizing the $\chi^2$ for angles between 10\degr and 50\degr\
in steps of 1\degr. Both models R1 and B3 resulted in a $\chi^2$
minimum at a bar angle of 15\degr.  Fig. \ref{scalingchi2} shows how
the change in bar angle affects the value of $\chi^2$.  We use
  $\Delta\chi^2 = \chi^2 - \min(\chi^2)$, where $\min(\chi^2)$ is the
  minimum value of $\chi^2$, to estimate the probable $1 \sigma$
  confidence interval on the angle as $\Delta\chi^2 <
  \sqrt{2N_{obs}}$, where $N_{obs}$ is the number of observables used
  in the comparison \citep{vdBosch2009}.  We obtain $\pm 5\degr$ for
  model R1 and $\pm 10\degr$ for model B3.  Model R5 instead produces
a best fit at 170\degr. However, the $\chi^2$ of the best fit is more
than twice that of R1 or B3, and is almost triple for an angle of
15\degr.  For consistency, we will always show model R5 at a bar angle
of 15\degr.  Fig. \ref{scalingchi2} shows a breakdown of the
$\chi^2$ by fields at different $b$.  For both model R1 and B3, the
largest contribution to $\chi^2$ comes from the fields closest to the
mid-plane, $b = -4\degr$, while the $b=-6\degr$ fields contribute the
least.  This is not the case for model R5 however, which has a
contribution from the $b = -8\degr$ fields comparable, or larger than,
the ones at $b=-4\degr$.

\begin{figure}
\includegraphics[angle=-90,width=0.8\columnwidth]{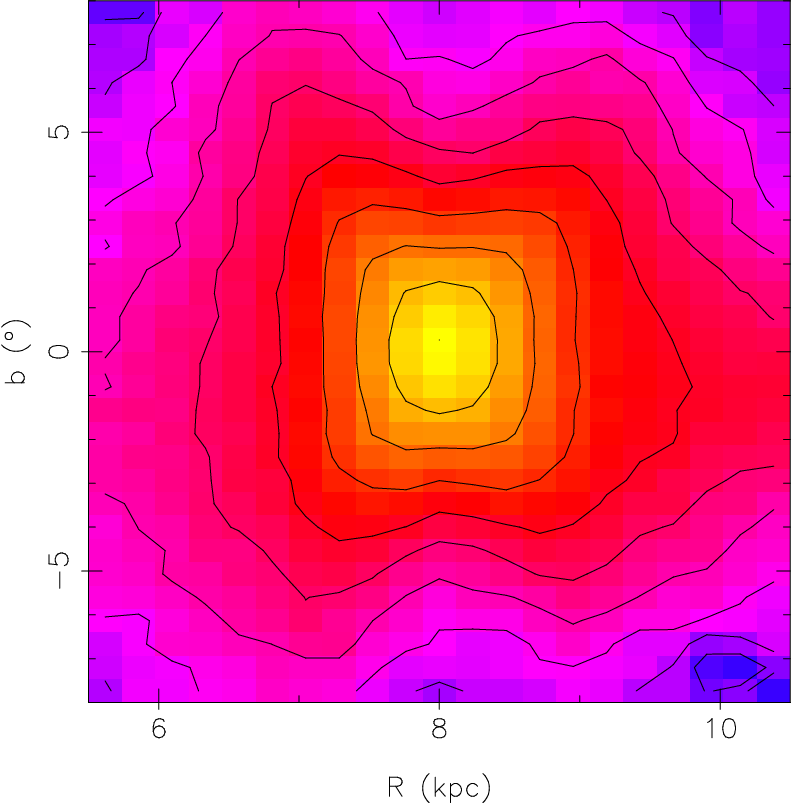}\\
\includegraphics[angle=-90,width=0.8\columnwidth]{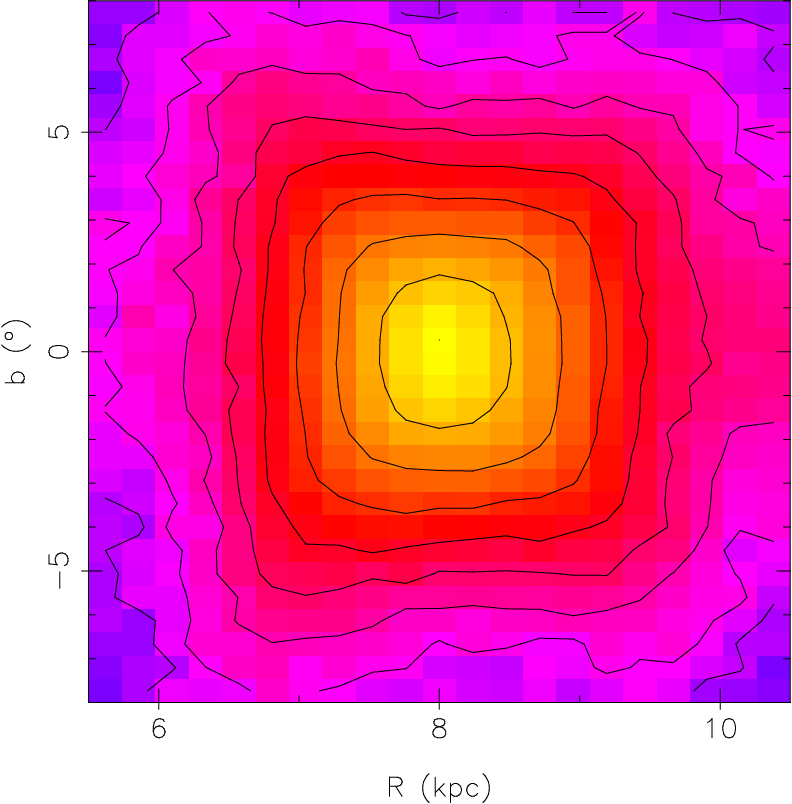}\\
\includegraphics[angle=-90,width=0.8\columnwidth]{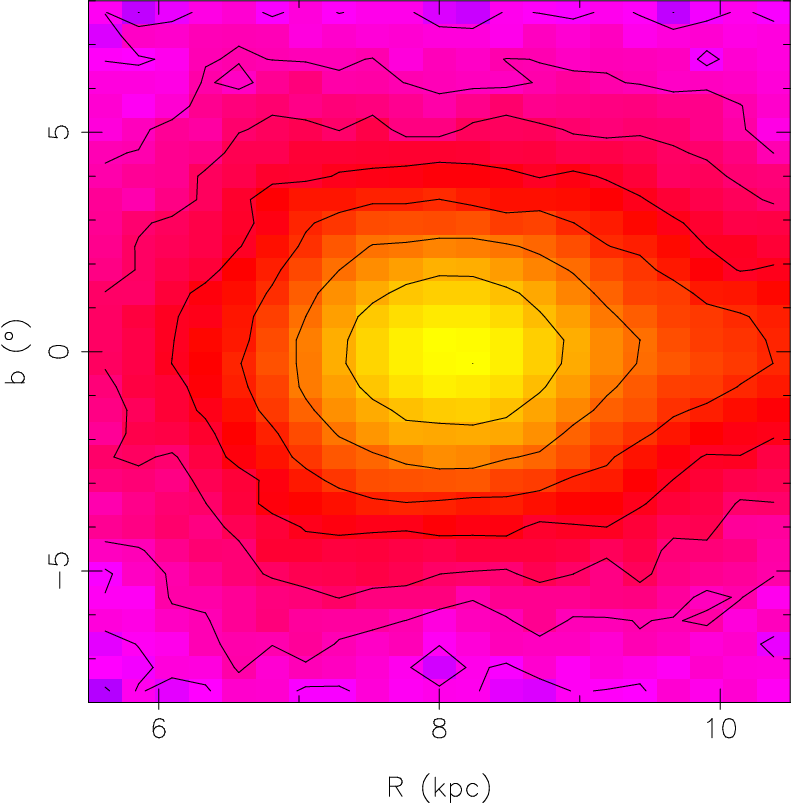}
\caption{Density of models R1 (top), B3 (middle) and R5 (bottom) in
the $l=0\degr$ plane. The plot shows the density as a function of
latitude and distance from the observer. This mimics the projection in
fig. 4 of \protect\cite{Saito2011}.  The bar angle is set to
$15\degr$ in each case.}
\label{shapes1} 
\end{figure}

The models in Fig. \ref{shapes1} are shown in the $(R,b)$ plane at
$l = 0\degr$, as in \cite{Saito2011}. The mock-observations have not
been corrected for the volume effect, instead including all particles
along a given line-of-sight, since the observations we compare with
are themselves not corrected for the volume effect.  The X-shape in
model R1 exhibits an asymmetry across $b = 0\degr$, arising from its
intrinsic asymmetry across the mid-plane.  Its arms can be traced down
to about $3\degr$, becoming boxy at lower $b$.  The X-shape in the
bulge of model B3 is much less readily apparent, though it clearly has
a boxy shape.  Model R5 has a weakly boxy shape in Fig.
\ref{externalview}, but no evidence of an X-shape in Fig.
\ref{shapes1}.

\subsection{Choosing the velocity scale}

After fitting the bar angle, we choose the velocity scale, $V$.
Although choosing the velocity scale and bar angle are intricately
linked, we verified that there was no difference which was chosen
first. The best velocity scale for R1 was $V = 299$ km s$^{-1}$, which
differs from the one we used in \cite{Vasquez2013}, 250 km s$^{-1}$,
for which we fitted only a single BRAVA field. For B3, the $\chi^2$
minimum is at $V = 257$ km s$^{-1}$. For R5, the best velocity scale
for a bar angle of 15\degr\ is $V = 291$ km s$^{-1}$.


\section{Comparison of simulations to observations}

\begin{figure*}
      \begin{tabular}{@{}l@{}l@{}l}
    \includegraphics[width=0.7\columnwidth]{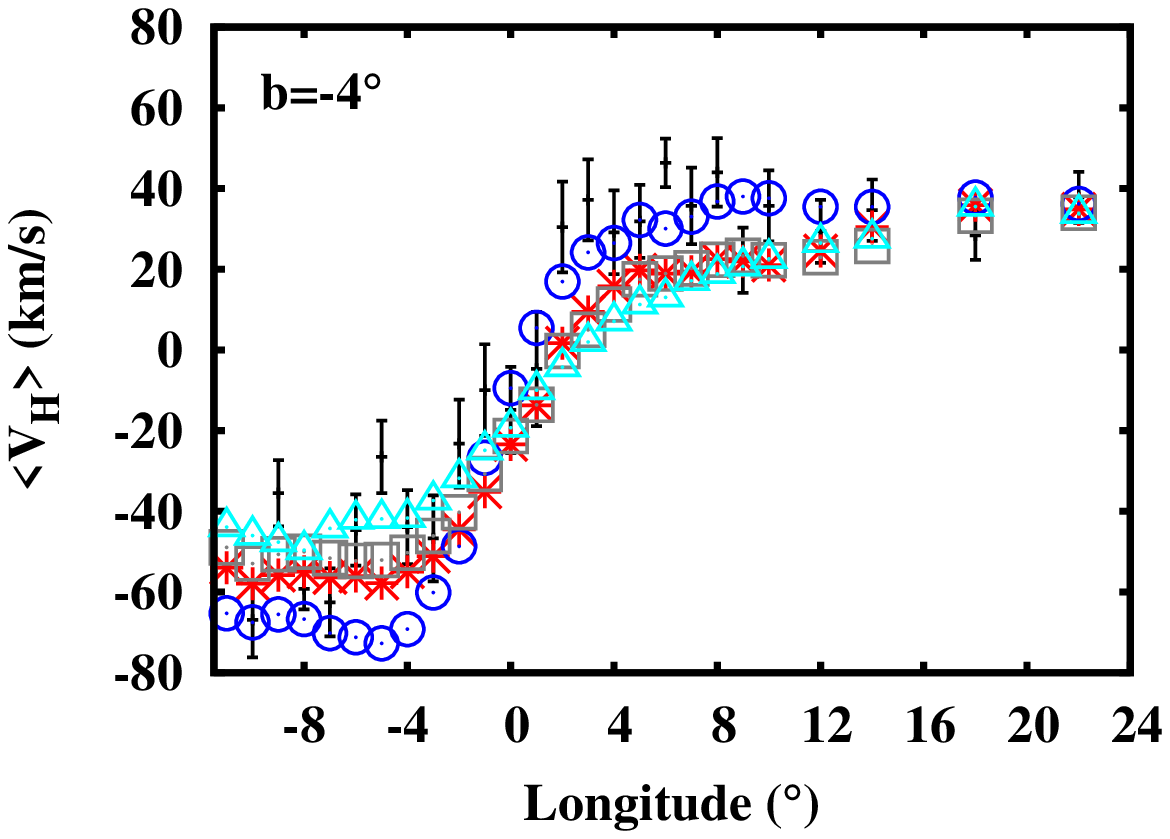} &
    \includegraphics[width=0.7\columnwidth]{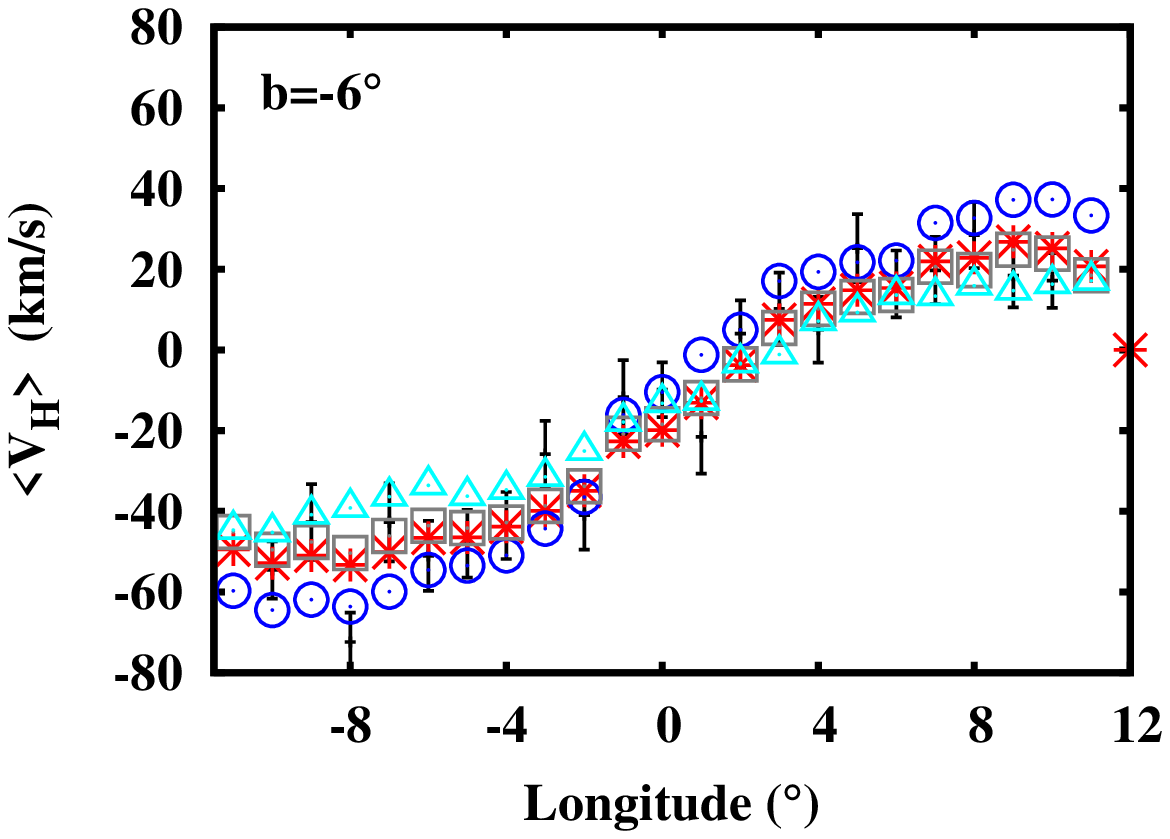} &
    \includegraphics[width=0.7\columnwidth]{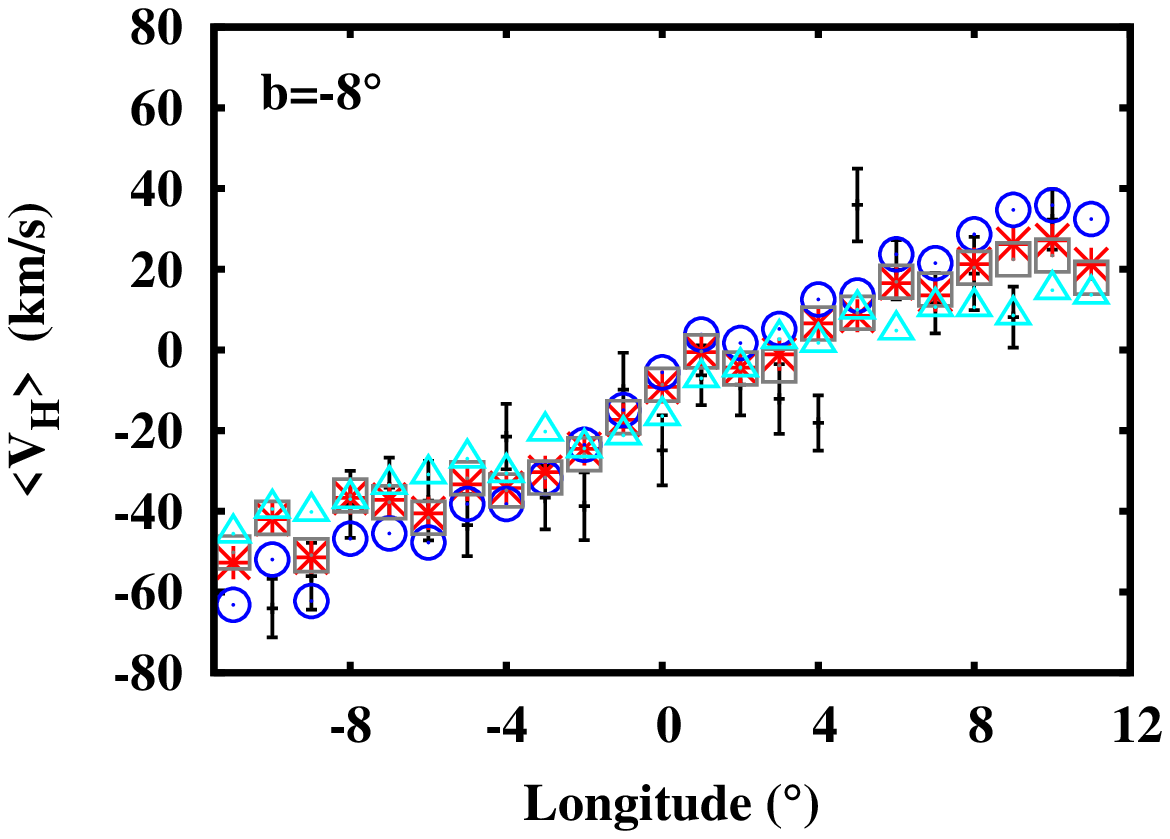}\\
    \includegraphics[width=0.7\columnwidth]{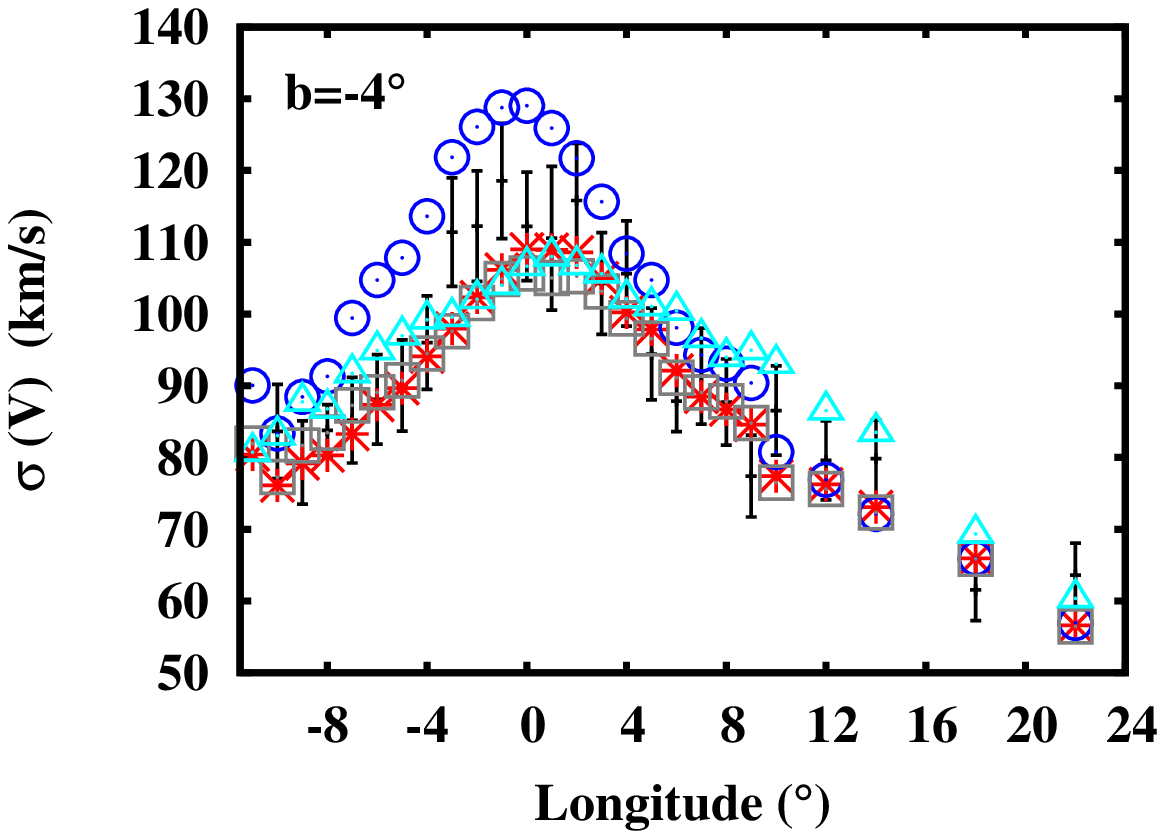} &
    \includegraphics[width=0.7\columnwidth]{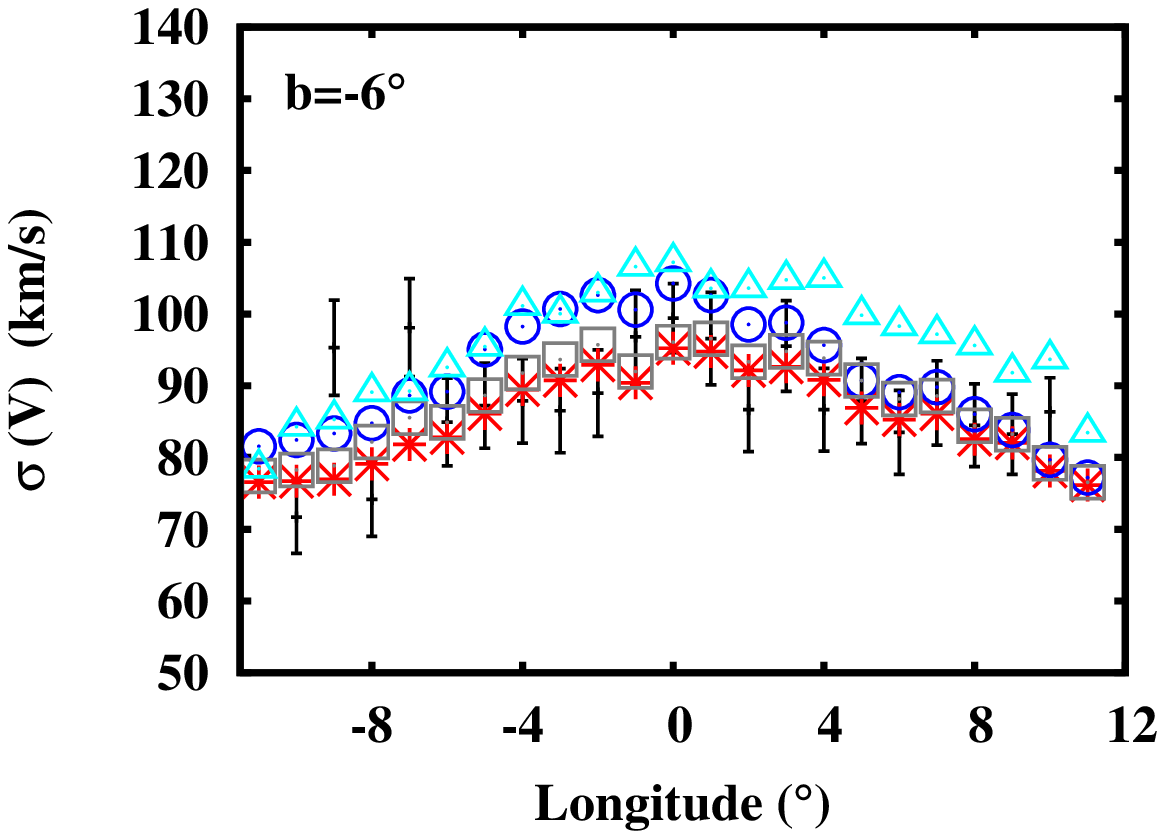} &
    \includegraphics[width=0.7\columnwidth]{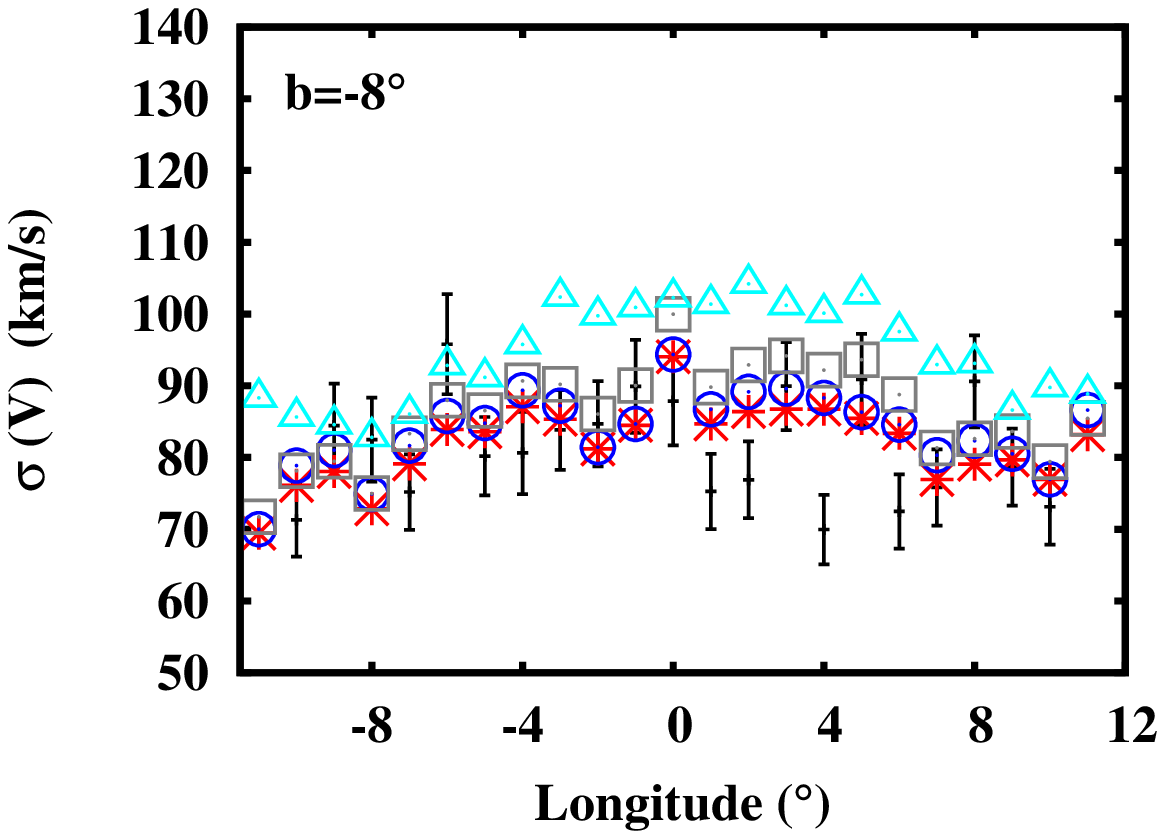}
\end{tabular}
\caption{Mean heliocentric velocities and velocity dispersions for
BRAVA fields along strips of $b = -4\degr$, $-6\degr$ and
$-8\degr$. Data: points with error bars; models : R1: grey squares,
B3: red stars, R5: cyan triangles, Fux: blue circles. The model data
have been obtained using the BGM, sampling velocities from the $N$-body
models.}
\label{modelcomparison}
\end{figure*}

In comparing the kinematics predicted by $N$-body simulations to real
observational data it should be remembered that observations are
generally not based on a volume-limited sample but rather on a
magnitude-limited sample. Sometimes, additional selections are
included as well (e.g. reddened or de-reddened colours, proper
motion-selected, temperature-selected, gravity-selected, etc.). 
  In the previous section, when comparing the $N$-body simulations to
  BRAVA data, we estimated the selection function in distance from the
  luminosity function of M giants and a 3-D extinction model. We
  assumed that the observational data trace a distance interval
without additional selection biases. This was suitable for
  computing the scaling of the $N$-body simulations. In the present
  section we investigate whether the kinematics of the particles in
  the $N$-body simulations are comparable to real stellar kinematics,
  once all selection biases, such as the varying fraction of M-giants
  with distance, are taken into account. The population synthesis
  approach is well-suited for this purpose.

\subsection{Population synthesis model comparison with BRAVA data}

We simulate the BRAVA sample with the M-giant distribution in the
Besan\c{c}on Galaxy Model (BGM) and compare the heliocentric radial
velocity distributions in different fields.  We apply the kinematics
of the $N$-body simulations to the stars in the bar component of
the BGM. The kinematics are computed using the mean and dispersion of
the particle velocities on the three axes in boxes 250 pc wide in
Cartesian $(x,y,z)$ coordinates covering the whole bar.  Then, for
each M-giant star drawn from the BGM bar population, we randomly apply
velocities from the Gaussian distribution with the mean and dispersion
of the position of the star, which, to first order, should reproduce
the velocity field of the bar in the $N$-body simulation.  Notice
  that this approximation may not be valid in rare regions where the
  distribution is very skewed.  Moreover, in some locations where the
number of particles in the $N$-body simulation is small, the
velocity is taken to be undefined since otherwise the assumed
kinematics are noisy. This can be remedied by applying some form of
smoothing to the model velocities.  In the regions dominated by the
bar, the number of particles in the simulation is sufficient for a
reliable value of the mean and the dispersion of the velocities so we
have not had to smooth the velocity field in this experiment.

To generate the stellar populations we use the version of the BGM
described in \cite{Robin2012}. This includes five components: a thin
disc, a thick disc, a stellar halo, a bar and a bulge. For simulating
the BRAVA sample, we select M giants in the apparent magnitude range
$8.2 < K_s < 9.25$, as explained in \cite{BRAVA}, assuming the
distribution of the extinction along the line-of-sight from the
\cite{Marshall2006} 3D maps.

The statistics (mean and standard deviation) of the heliocentric
radial velocity of BRAVA fields along latitudes $-$4\degr, $-$6\degr
and $-$8\degr\ (from \citealt{BRAVA}) are compared with the output of
the BGM, as described above, in Fig. \ref{modelcomparison}. We also
compare with the kinematics drawn from the earlier model of
\cite{Fux1999}. BRAVA data are shown with error bars (given by
Poisson uncertainties). The model simulations cover slightly larger
angular areas in order to minimize Poisson noise.

Globally, two of our $N$-body models, R1 and B3, perform well, indeed
better than the Fux model, while R5 fails to properly fit the
observations. At $b = -4\degr$ models R1, B3 and R5 appear slightly
shifted in velocity at $2\degr \leq l \leq 11\degr$. In this region,
the Fux model performs better. But at negative longitudes, R1, B3 and
R5 are significantly better than the Fux model, which encounters
problems, as already noted by \cite{Howard2009}. R1, B3 and R5 are
slightly different at negative longitudes, but the difference (about 5
$-$ 10 km s$^{-1}$) is small compared with the uncertainty in the
BRAVA data. The velocity dispersions from R1, B3 and R5 are good
approximations for almost all longitudes, given the uncertainties.
They differ from each other only in the central region and only by a
few km s$^{-1}$. At $b = -$8\degr\ the models do not differ much
within the uncertainties of the data, except for model R5, where the
velocity dispersions are systematically too high. At $b = -$6\degr,
the differences between the Fux model, R1, and B3 are more noticeable.
The Fux model rotates too fast and has too large velocity dispersion
at negative longitudes.  Model R5 fits the BRAVA data more poorly,
because of slower rotation and an even higher velocity dispersion than
the Fux model.
       		  
This test confirms that any selection biases are likely to be small.
Overall models R1 and B3 are good approximations to the observed
kinematics over a large range of longitudes and latitudes in the bulge
region. Model R1 gives a slightly better fit.  However, the BRAVA data
are unable to distinguish between models R1 and B3, because the sample
in each direction is small and M giants cannot distinguish the near
and far sides. Hence, further tests using RC giants, especially in
regions where the clump is double, will be performed in the near
future.

\subsection{Model comparison to other observations}

We next compare our two main B/P-shaped bulge simulations (R1 and B3)
with kinematic data from the literature.  The results of this
comparison are compiled in Table \ref{obsvssims}.
  
\begin{table*}
\caption{Comparison of observational values for the differences
between mean velocities and velocity dispersions of the near and far
sides of the bulge and the same quantities in our models. Velocities
are in km s$^{-1}$.  For the models, we consider $b<0\degr$ and
$b>0\degr$ measurements separately, the latter indicated by $(+b)$ at
the top of each column.}

\label{obsvssims}
\begin{tabular}{lll|ll|cccc|cccc}
    \hline
    Reference&$l$&$b$& $|\Delta \overline{V}|$ & $\Delta \sigma_V$
    &R1 & & (+$b$) & (+$b$)  & B3  &  & (+$b$) & (+$b$) \\
    &   &   &  &  &$|\Delta \overline{V}|$  & $\Delta \sigma_V$ &
    $|\Delta \overline{V}|$ & $\Delta \sigma_V$ & $|\Delta \overline{V}|$
    & $\Delta \sigma_V$  & $|\Delta \overline{V}|$ &$\Delta \sigma_V$ \\
    \hline
    \cite{Rangwala2009} & 5.5\degr & $-$3.5\degr & 40 $\pm$ 11 &
    & 14 & &22 & & 26  & & 24 & \\
    & $-5\degr$ & $-3.5\degr$ & 32 $\pm$ 12& & 16 & & 28 & & 27 & & 26\\

    \cite{DePropris2011} & $0\degr$ & $-8\degr$ &   10 $\pm$ 14&
    4 $\pm$ 14 & 9  & 7 & 8 & 9 & 22  & 6 & 15  & 4 \\

    \cite{Ness2012} & $0\degr$ & $-5\degr$  & 30 $\pm$ 12 &
    & 10 & & 14 & & 28 & & 28 &\\
    & $0\degr$ & $-7.5\degr$ and $-10\degr$ & 7 $\pm$ 9&
    & 5 &  &13 & & 17 & & 17 &\\
    
    \cite{Vasquez2013} & $0\degr$ & $-6\degr$ & 21 $\pm$
    14 & 10  $\pm$ 10 & 11 & 6 & 12 & 6 & 28 & 5 & 23 & 1 \\
    \hline
\end{tabular}

\end{table*}

Compared to the line-of-sight velocities in \cite{Rangwala2009}, we
find similar differences, although the one at negative $l$ does not
fit as well as the field at positive $l$; however the field at
positive instead of negative $b$ fits the data much better.  Compared
to the data of \cite{DePropris2011}, we find that R1 fits their data
better, showing very little difference in the line-of-sight velocities
and line-of-sight velocity dispersions of the near and far sides. Thus
this is not an ideal location to search for the kinematic signature of
the X-shaped bulge.  For the \cite{Ness2012} data we find a better fit
to the lower latitude field of B3, but a worse fit for the combined
two higher latitude fields, where the difference in observations is
small.

We use a different value for the local standard of rest than we did in
\cite{Vasquez2013}, adopting ($U, V, W$)$_{\odot}$ = (14, 12, 6) km
s$^{-1}$ and V$_c$ = 238 km s$^{-1}$ from \cite{Schonrich2012} to
derive heliocentric velocities. For the observations in
$(l,b)=(0\degr,-6\degr)$, the differences between the velocities of
the near and far sides for $U, V$ and $W$ are, respectively, 5 $\pm$
14, 23 $\pm$ 19 and 13 $\pm$ 16 km s$^{-1}$ for the mean and 6 $\pm$
10, 18 $\pm$ 13 and 11 $\pm$ 12 km s$^{-1}$ for the dispersions. For
R1, the differences of means are 2, 1 and 1 km s$^{-1}$, while the
differences in dispersions are 9, 5 and 3 km s$^{-1}$. For B3 the
differences are 3, 2 and 3 km s$^{-1}$ for the means and 7, 4 and 3 km
s$^{-1}$ for the dispersions. The differences are substantially lower,
but well within the errors as well as consistent with the previous
estimates for differences of R1 in \cite{Vasquez2013}.


\section{The kinematic imprint of the X-shape}

\subsection{Mapping near-/far side kinematic differences across the
  bulge}

We next examine how the kinematics of the near and far sides of the
bulge differ across the whole bulge region. We consider the square
region defined by $|l|, |b| < 10\degr$ aiming to uncover the kinematic
signature of the B/P-shaped bulge.  Given an axisymmetric system, the
kinematic differences between the near and far sides of the bulge
should be featureless. We calculate maps of the difference
between the near and far side mean and dispersion of line-of-sight,
galactocentric azimuthal and vertical velocities.

\begin{figure}
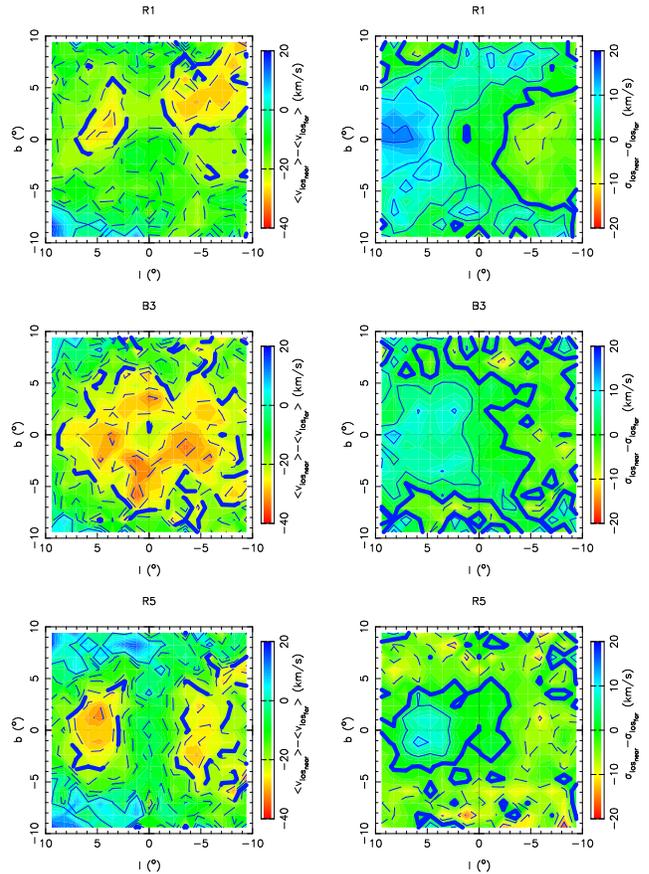

\begin{tabular}{ll}
\includegraphics[width=0.47\columnwidth]{fig6.1.ps}    &
\includegraphics[width=0.47\columnwidth]{fig6.2.ps}
\end{tabular}
\caption{Differences between near and far side line-of-sight mean
velocity (left panels) and near and far side line-of-sight velocity
dispersion (right panels). Contours are spaced by 5 km s$^{-1}$ and
the bold contour marks 0 km s$^{-1}$. Dashed contours show negative
values, while solid contours show positive values.}
\label{scaleR1B3vlos} 
\end{figure}

Fig. \ref{scaleR1B3vlos} shows the differences for the line-of-sight
velocities and dispersions.  It is immediately apparent that there is
a qualitative difference between models R1 and B3, both with X-shapes,
and model R5 without.  In model R5, the contours of the difference in
mean velocities are more or less lines of constant $l$.  In models R1
and B3 instead contours corresponding to the largest differences in
mean velocity cross the $l=0\degr$ plane at $|b| \simeq
3\degr-4\degr$.  In model R1 this leads to the striking result that
the velocity difference is larger at $b > 0\degr$, where we showed
above that the B/P-shape is also stronger, than at $b < 0\degr$. The
inescapable conclusion is that this difference is a kinematic imprint
of the X-shape.  The difference between the velocities is generally
negative (in our frame a positive velocity corresponds to a motion
away from the Sun). The line-of-sight velocity dispersion has positive
differences at positive $l$. In all three models the differences in
both the mean and dispersion of line-of-sight velocities have a peak
near ($l,b$)= ($5\degr,0\degr$).

\begin{figure}
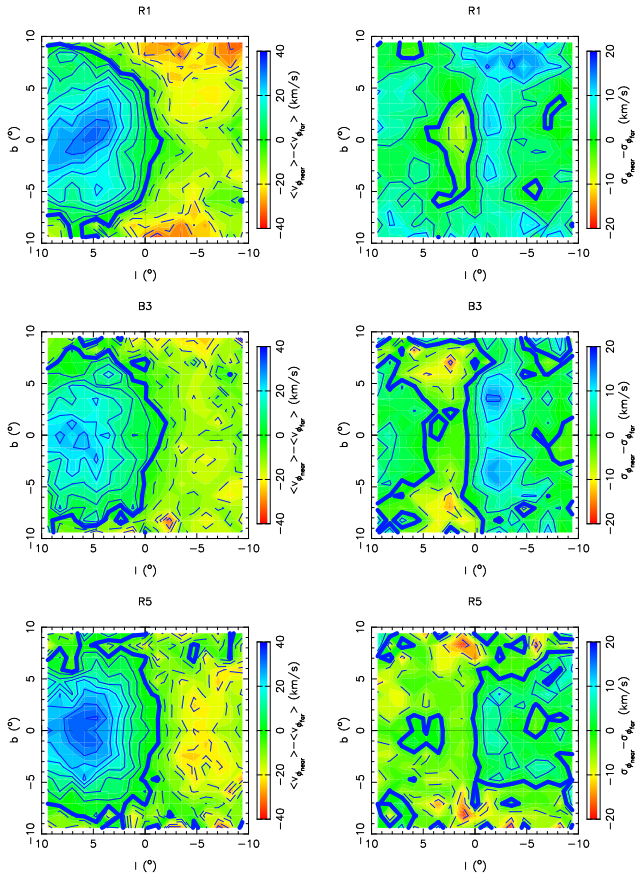

\begin{tabular}{ll}
\includegraphics[width=0.47\columnwidth]{fig7.1.ps}    &
\includegraphics[width=0.47\columnwidth]{fig7.2.ps}
\end{tabular}
\caption{Differences between near and far side mean galactocentric
  azimuthal velocity (left panels) and near and far side
  galactocentric azimuthal velocity dispersion (right panels).
  Contours are spaced by 5 km s$^{-1}$ and the bold contour marks 0 km
  s$^{-1}$. Dashed contours show negative values, while solid contours
  show positive values.}
\label{scaleR1B3vphi}
\end{figure}

\begin{figure}
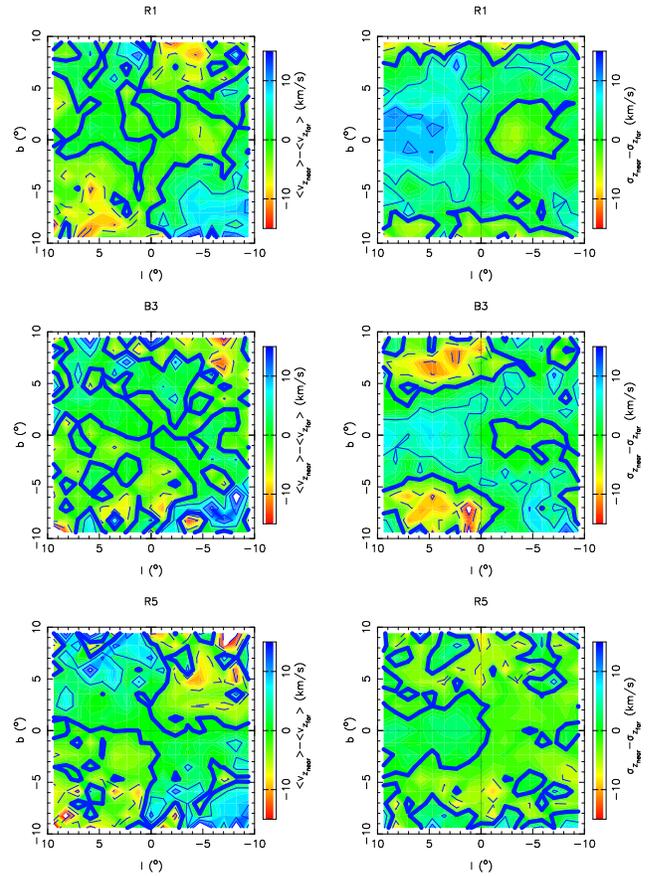

\begin{tabular}{ll}
\includegraphics[width=0.47\columnwidth]{fig8.1.ps}    &
\includegraphics[width=0.47\columnwidth]{fig8.2.ps}
\end{tabular}
\caption{Differences between near and far side mean vertical velocity
(left panels) and near and far side vertical velocity dispersion
(right panels). Contours are spaced by 5 km s$^{-1}$ and the bold
contour marks 0 km s$^{-1}$. Dashed contours show negative
values, while solid contours show positive values.}
\label{scaleR1B3vz}
\end{figure}

For galactocentric azimuthal velocities (Fig. \ref{scaleR1B3vphi}),
there is a difference between the near and far sides evident in the
average velocity (left panels) and in the velocity dispersion (right
panels), for all three models. The largest differences in the average
velocity are close to where the near side of the bar is located,
around ($l,b$) = ($6\degr,\pm1\degr$). The largest differences in the
azimuthal velocity dispersion can be found around ($l,b$) =
($-2\degr,9\degr$) for R1 and ($l,b$) = ($3\degr,\pm8\degr$) and
($l,b$) = ($-2\degr,\pm4\degr$) for B3. The differences are small for
R5, centred at ($l,b$) = ($-2\degr,0\degr$).  However, {\it
qualitatively} there is no difference between models R1 and B3 with an
X-shape, and model R5 without, in either the difference of mean
velocities or the difference of dispersions.

Vertical velocities (Fig. \ref{scaleR1B3vz}) again show no coherent
difference between the near and far sides. There are small (5-10 km
s$^{-1}$) positive differences in vertical velocity dispersion near
$b$ = 0\degr\ at positive $l$. There are a few anomalous regions in
vertical velocity dispersion for B3, located at $8\degr < l < 0\degr,
b = \pm8\degr$, which appear also in the azimuthal velocity
dispersion.
 
Overall, the result of this analysis is that the strongest imprint of
an X-shape is in the near/far side difference in mean line-of-sight
velocities.  The galactocentric tangential mean velocities and
vertical velocities do not betray the presence of an X-shape, nor do
any of the velocity dispersions to any significant extent.

\begin{figure}
    \begin{center}
    \includegraphics[width=.95\columnwidth]{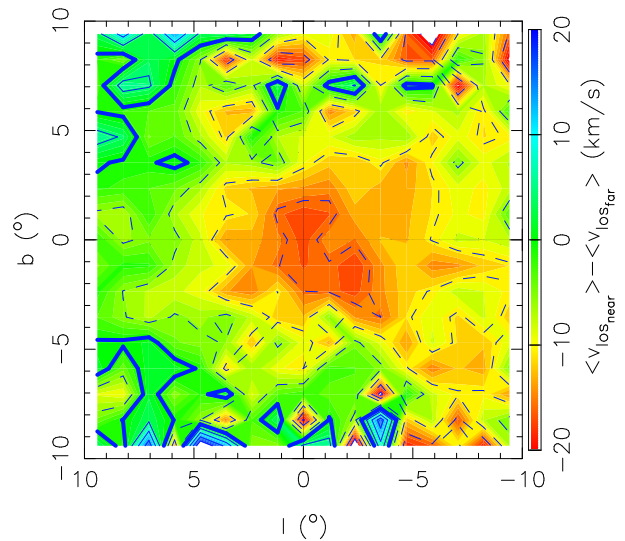}
\end{center}
\caption{Near/far side differences in mean $V_{los}$ in model B3 which
  has been rendered axisymmetric by azimuthally shuffling all
  particles.  }
\label{fig:B3shuffled}
\end{figure}

Finally, we use model B3, which had one of the strongest near/far-side
asymmetries in $V_{los}$, to demonstrate that the asymmetry is a
signature of the X-shape by removing it by axisymmetrising the model.
The resulting map of $\Delta V_{los}$, which is shown in Fig.
\ref{fig:B3shuffled}, has a minimum at $(l,b) = (0\degr,0\degr)$ and
decreases more or less monotonically with $|b|$, in stark contrast
with B3 when it is not axisymmetrised.  This adds further weight to
our interpretation of a minimum in $\Delta V_{los}$ off $b=0\degr$ as
being the kinematic signature of an X-shape.

\subsection{Line-of-sight velocities at $l = 0\degr$} 

\begin{figure}
    \begin{center}
    \includegraphics[width=.95\columnwidth]{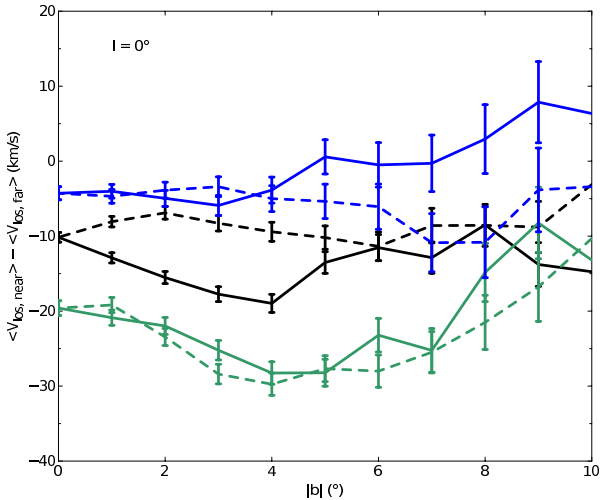}
\end{center}
\caption{Near/far side differences in mean $V_{los}$ in the $l =
0\degr$ plane for models R1 (black lines), B3 (green lines) and R5
(blue lines).  Solid (dashed) lines show $b>0\degr$ ($b<0\degr$).
}
\label{foldeddVlos}
\end{figure}

Fig. \ref{foldeddVlos} plots the near$-$far side difference in mean
$V_{los}$.  Errors in $\Delta V_{los}$ are calculated by taking
  the combination in quadrature of the errors of the means for each
  field on the near and far side.  In all models this difference is
negative throughout, presumably in part because the volumes probed are
different.  Additionally, in model B3 and in $b>0\degr$ of model R1 a
very clear minimum across $|b|$ at $l = 0\degr$ exists which is
caused by the X-shape.  No comparable minimum is present at $b <
0\degr$ in model R1 (dashed black line).  Nor is there a similar
minimum in model R5.

\begin{figure}
    \includegraphics[width=\columnwidth]{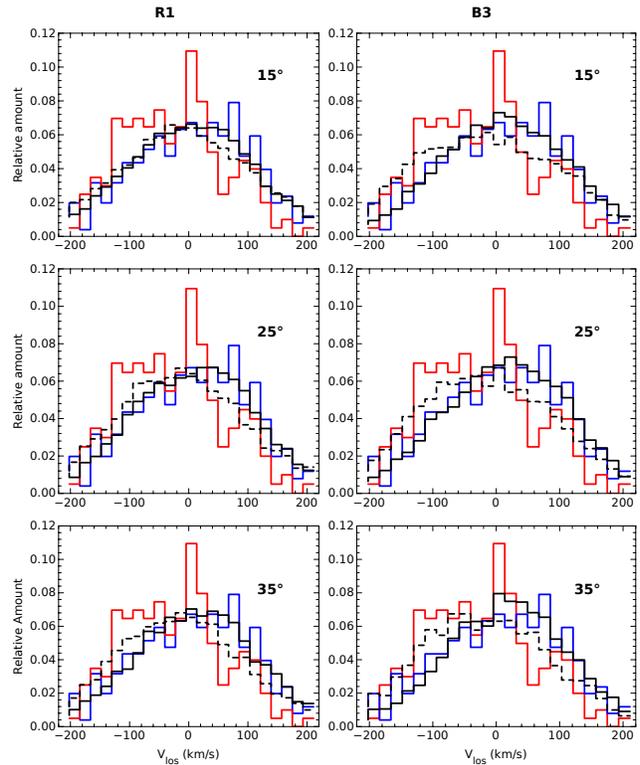}
\caption{Distributions of line-of-sight velocities for the near
(dashed lines) and far sides (solid lines) of the scaled models at bar
angles, 15\degr\ (top), 25\degr\ (middle) and 35\degr\ (bottom).
Distributions are for the field at $(l,b) = (0\degr,-6\degr)$.
The solid red and blue lines correspond to, respectively, the
observed velocities from the bright RC and the faint RC
and are identical in all six panels. The left column is for model R1
while the right one is for model B3. All distributions have been
normalized to unit area.} \label{vr}
\end{figure}

We now examine V$_{los}$ in the field $(l,b) = (0\degr,-6\degr)$ of
\cite{Vasquez2013}, and consider how bar angles of 15\degr, 25\degr\
and 35\degr\ affect the distribution.  The sub-panels of Fig.
\ref{vr} each have separate distributions, for the near (dashed lines)
and far sides (full lines). In this figure, we also show the
distribution of the observed bright (red line) and faint (blue line)
red clump stars from \cite{Vasquez2013}. The means of the
distributions of $V_{los}$ differ significantly, with the far side
having slightly more positive velocities (in our convention positive
$V_{los}$ velocities correspond to stars moving away from the
observer), while the near side stays at zero, or has negative
velocities. This is seen in the observations as well as the models. As
the bar angle increases, the negative velocity of the near side
becomes more prominent. The differences between the two means of the
radial velocity distribution are 11.6, 25.9 and 34.1 km s$^{-1}$ for
angles of 15\degr, 25\degr\ and 35\degr, respectively, for R1, and
28.4, 33.5 and 35.2 km s$^{-1}$ for the same angles in B3. Velocity
dispersions differ by less than 10 km s$^{-1}$, for both models, with
the difference becoming smaller at larger bar angles. Thus the
near/far sides asymmetry provides a sensitive probe of the bar angle.
Despite the qualitative agreement between the observational data and
the models, it is difficult to interpret quantitatively. There are two
reasons for this: (i) the significant Poisson noise in the data and
(ii) the mismatch in the definition of the near/far sides between the
models and the bright/faint RCs in \cite{Vasquez2013}.
Therefore, the bar angle cannot be derived reliably from this
comparison. Conversely, the kinematic imprint of an X-shape is not
masked by any plausible value of the bar angle. Rather surprisingly,
while the density maps at $l=0\degr$ of R1 and B3 in Fig.
\ref{shapes1} are relatively different, the two models have
qualitatively similar distributions of $V_{los}$.

\subsection{Bulge versus disc near/far side kinematic differences}

\begin{figure}
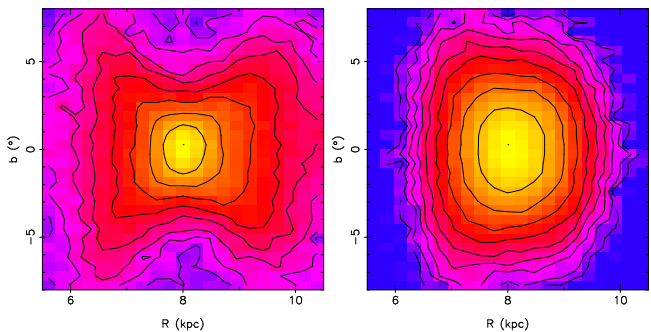

\centerline{
\includegraphics[angle=-90,width=0.5\hsize]{fig12.1.eps}
\includegraphics[angle=-90,width=0.5\hsize]{fig12.2.eps}
}
\caption{Density of the disc (left) and classical bulge (right) at
  $l=0\degr$ in model B3.}
\label{fig:denssB3bulgevsdisc} 
\end{figure}

\begin{figure}
\includegraphics[width=\hsize]{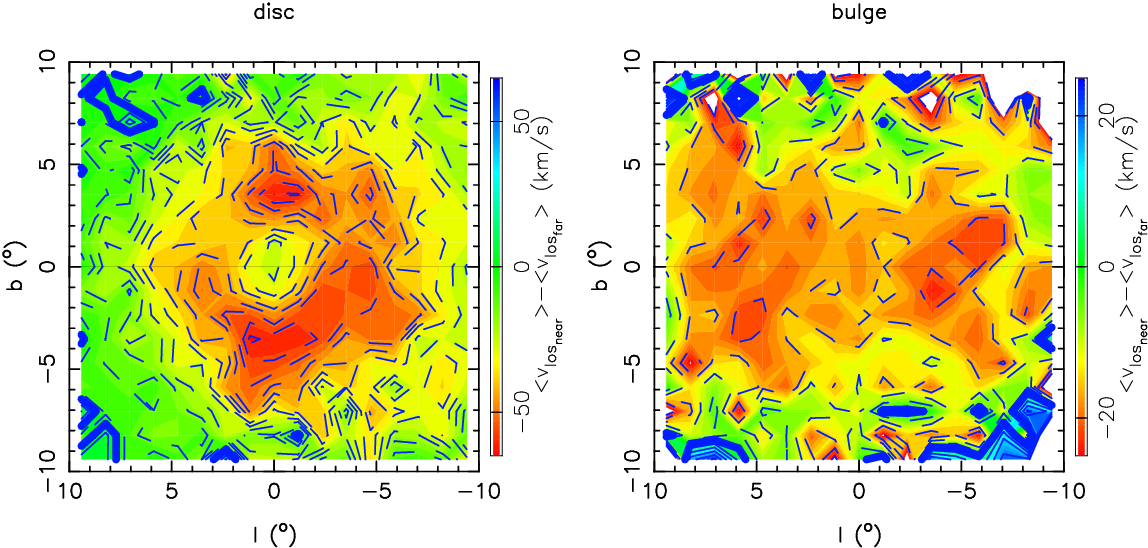}
\includegraphics[width=\hsize]{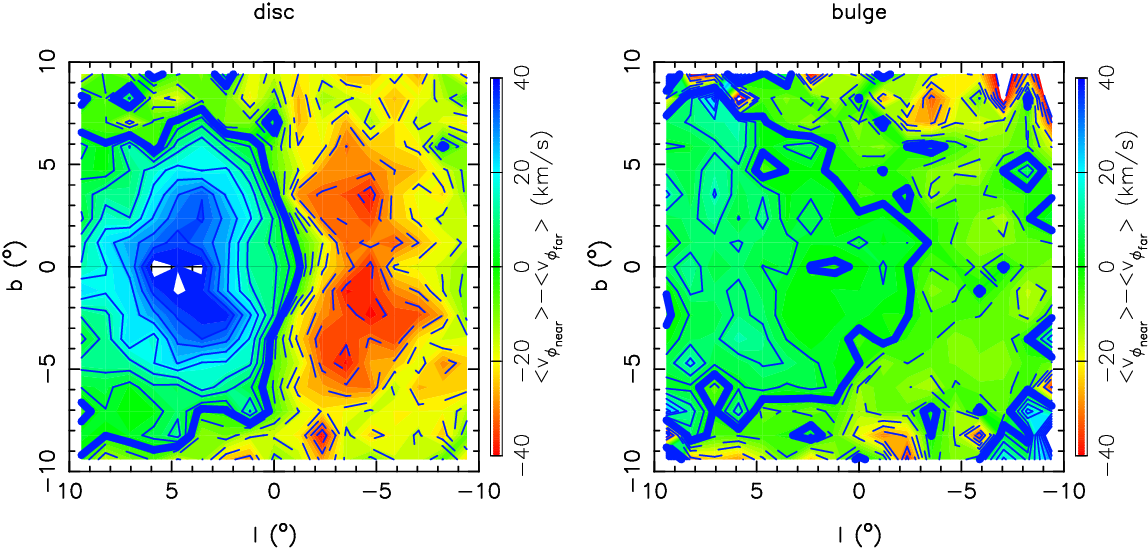}
\includegraphics[width=\hsize]{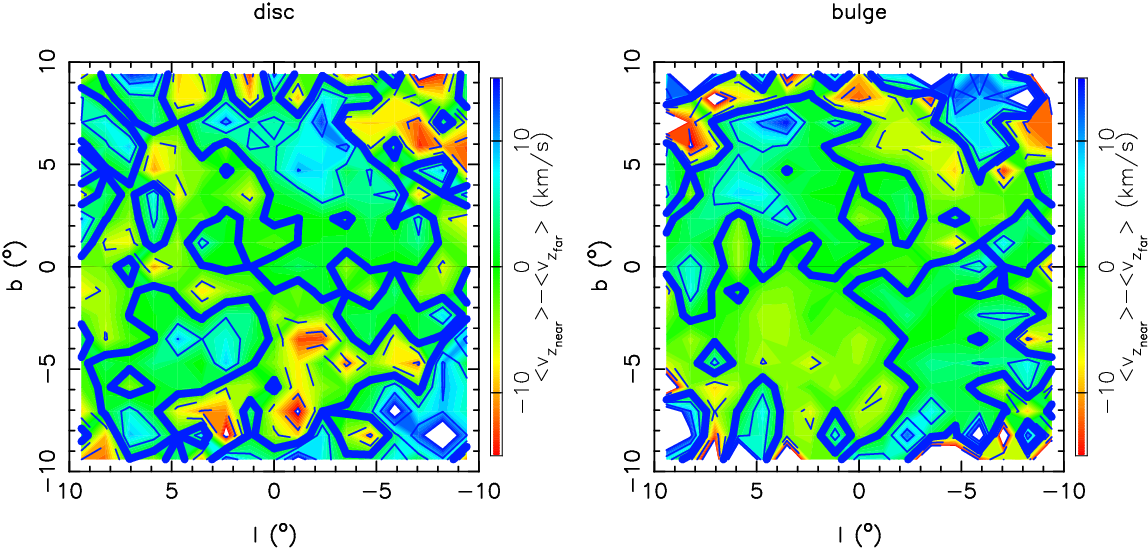}
\caption{ Differences between near and far side velocities for the
  disc (left) and classical bulge (right) components of model B3.  The
  top, middle and bottom rows show differences in the line-of-sight,
  tangential ($\phi$) and vertical ($z$) directions.  Contours are
  spaced by 5 km s$^{-1}$ and the bold contour marks 0 km s$^{-1}$.
  Dashed contours show negative values, while solid contours show
  positive values.  }
\label{fig:velsB3bulgevsdisc} 
\end{figure}

We now consider the different behaviours of the disc and classical
bulge in model B3.  Fig. \ref{fig:denssB3bulgevsdisc} presents the
density distribution at $l=0\degr$ separately for the two components.
A very prominent X-shape is present in the disc component, whereas the
bulge has a boxy shape.  Fig.  \ref{fig:velsB3bulgevsdisc} compares
the disc (left panels) and bulge (right panels) differences in mean
velocities; the most striking result is that the kinematic signature
of the X-shape, the minimum in the line-of-sight velocity along
$l=0\degr$, appears only in the disc not the bulge.  The remaining
kinematics, including the near/far side differences in velocity
dispersions (Fig.  \ref{fig:dispsB3bulgevsdisc}), merely show the
effect of a higher rotation and lower dispersion in the disc relative
to the bulge.

\begin{figure}
\includegraphics[width=\hsize]{fig14.1.ps}
\includegraphics[width=\hsize]{fig14.2.ps}
\includegraphics[width=\hsize]{fig14.3.ps}

\caption{Differences between near and far side velocity dispersions
  for the disc (left) and classical bulge (right) components of model
  B3.  The top, middle and bottom rows show differences in the
  line-of-sight, tangential ($\phi$) and vertical ($z$) directions.
  Contours are spaced by 5 km s$^{-1}$ and the bold contour marks 0 km
  s$^{-1}$. Dashed contours show negative values, while solid contours
  show positive values.}
\label{fig:dispsB3bulgevsdisc} 
\end{figure}

\subsection{Gas $+$ star formation simulation}

\begin{figure}
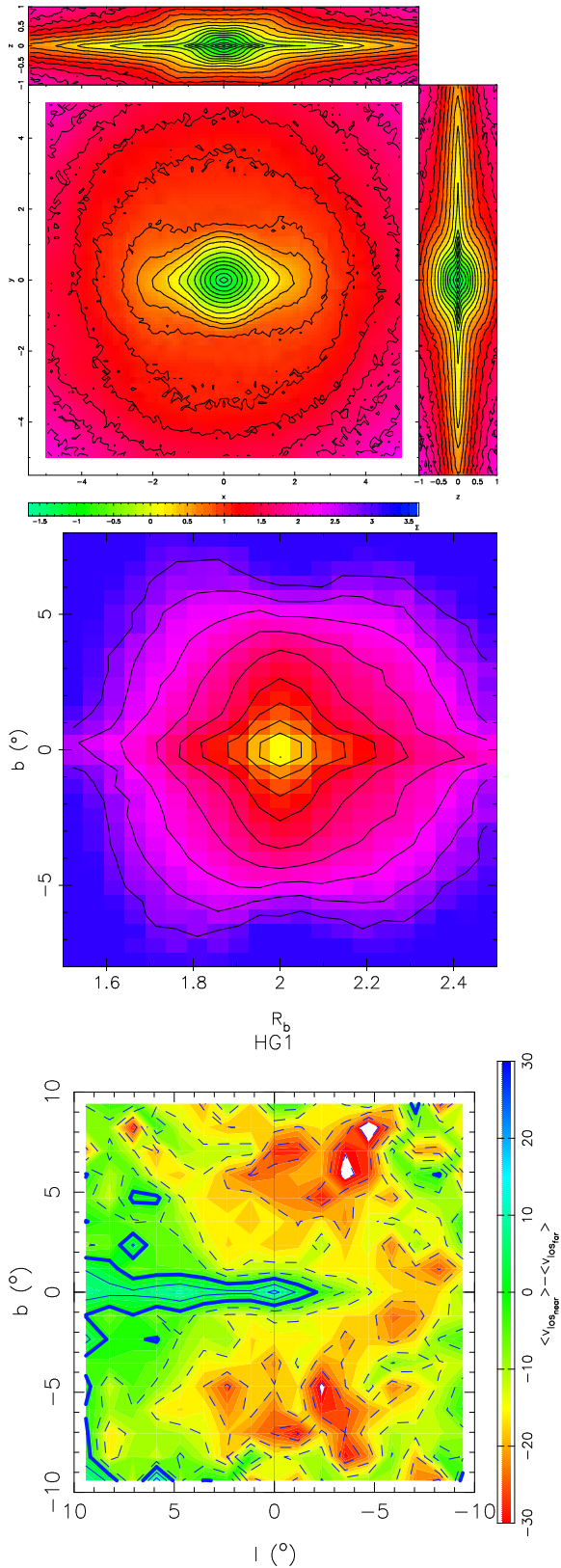

\includegraphics[angle=-90,width=0.8\hsize]{fig15.1.ps}
\includegraphics[angle=-90,width=0.8\hsize]{fig15.2.ps}
\includegraphics[width=0.9\hsize]{fig15.3.ps}
\caption{ Model HG1 with gas and star formation.  Top panel: Three
  orthogonal projections of the model as seen from outside the system
  with the bar rotated into the $x$-axis.  Middle panel: Density in
  the $l=0\degr$ plane.  The region $|b|<1\degr$ is dominated by the
  disc.  Bottom panel: Difference between near and far side mean
  line-of-sight velocity.}
\label{fig:hg1}
\end{figure}

We now consider simulation HG1 in which a disc forms from a cooling
gas corona embedded in a live dark matter halo.  This simulation has
not been published before but uses the approach described in
\citet{Roskar2008} with higher resolution.  By the end of the
simulation, at 10 Gyr, the disc consists of $\sim 1.1 \times 10^7$
particles.  A bar obviously forms with an X-shape.  The top panel of
Fig. \ref{fig:hg1} shows the density distribution; the bar has a
radius of about 3 kpc.  A boxy bulge is also visible in the
edge-on projection.  For the purposes of this paper we do not scale
this bar to the Milky Way, merely using it to demonstrate that the
kinematic signature of an X-shaped bulge is also present in this more
realistic simulation, rather than being a feature of bars forming
within frozen dark matter haloes.  The middle panel of Fig.
\ref{fig:hg1} demonstrates that the bulge has an X-shape when viewed
in the $l = 0$ plane.  This model, unlike the other models, also has a
conspicuous disc at $|b| < 1\degr$, outside the region where the Milky
Way's X-shape is found.  Finally, the bottom panel of Fig.
\ref{fig:hg1} presents a map of the difference between the near and
far side mean line-of-sight velocity.  At $|b| < 1\degr$ the
kinematics are strongly contaminated by the inner disc, so we ignore
this region in our analysis; We find that the signature of the X-shape
is still present as a minimum in the velocity difference at $|b|
\simeq 7\degr$.  Thus we are confident that a minimum in near-far side
difference in mean line-of-sight velocities at $l=0\degr$ constitutes
a signature of an X-shaped bulge.


\section{Measuring the distance to the Galactic Centre using the X-shape}

\begin{figure}
    \begin{center}
    \includegraphics[width=.95\columnwidth]{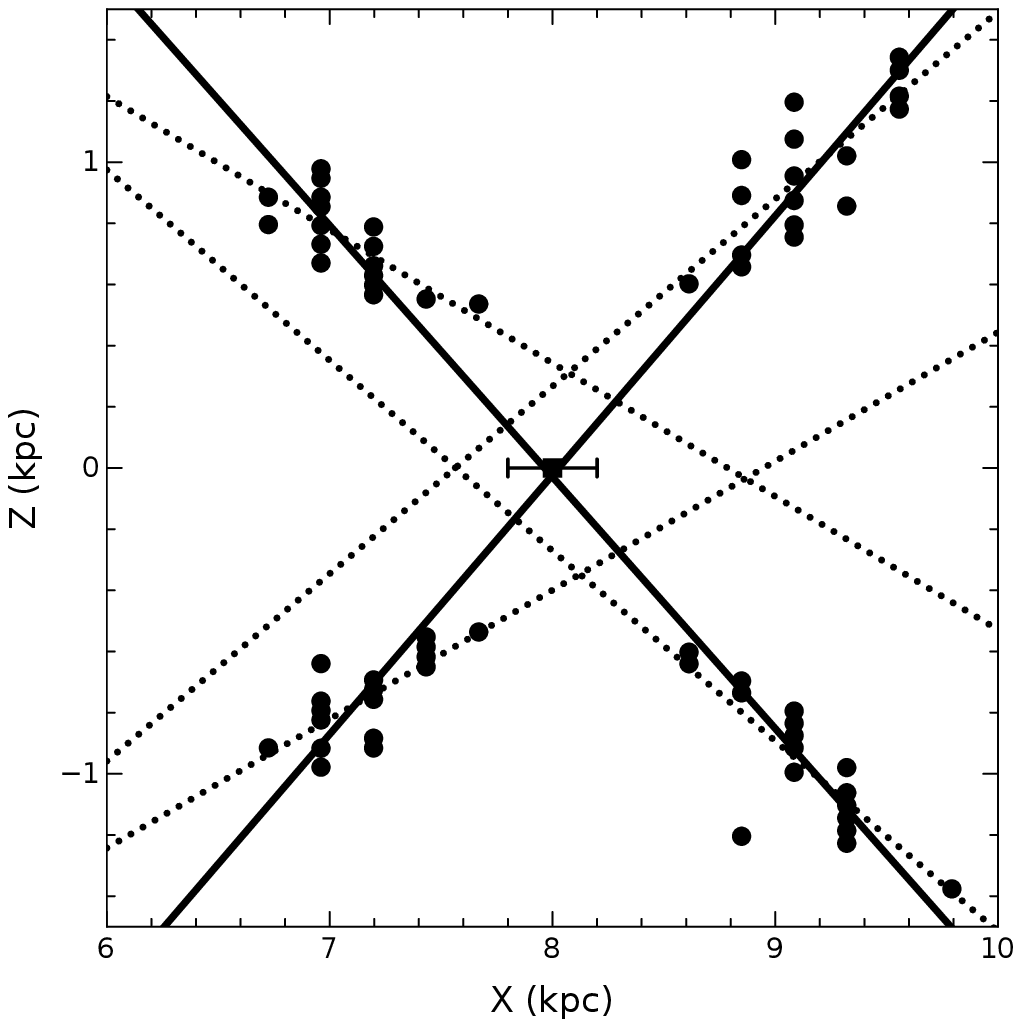} \\
    \includegraphics[width=.95\columnwidth]{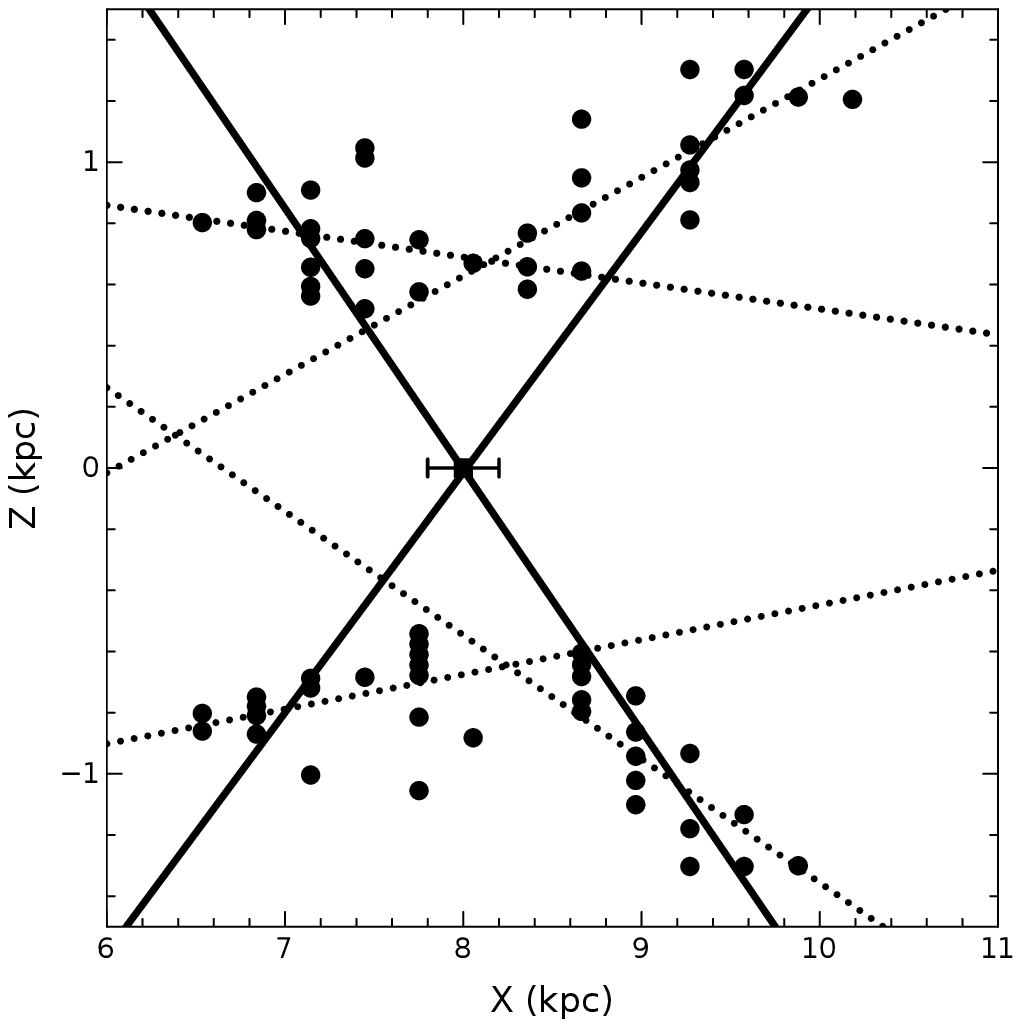} 
\end{center}
\caption{Positions of the density peaks along the line-of-sight at $l
= 0\degr$, $4\degr < |b| \le 8\degr$ for R1 (top) and B3 (bottom). The
measurements are every 0.25 deg and each point covers an area of 0.25 deg$^2$.
The dotted lines represent linear fits to each arm of the X-shape. The
solid lines show linear fits joining arms diagonal from each other.
The square denotes the position of the centre, while the error bar
indicates an error of 2.5\%. } \label{findthecenter}
\end{figure}

\cite{McWilliam2010} present a distribution of RC stars (their
fig. 8) and measure the distance to the centre of the Galaxy using
the X-shape in the RC stars traced in the $l=0\degr$ plane. We explore
the potential for using the X-shape for determining the distance to
the Galactic Centre by measuring the ridge of highest density along
the line-of-sight on the near and far sides of the bulge, excluding
any central peaks. Latitudes $|b| < 4\degr$ fail to find a peak off
the centre, therefore we include only the lines-of-sight at $|b| \ge
4\degr$ sampling every 0.25\degr\ up to $|b|$ = 8\degr, for a
total of 17 measurements per arm. Each sample covers $0.5\degr \times
0.5\degr$. Then we fit a line to each of the four arms of the X-shape.
Fig. \ref{findthecenter} shows the measured positions of the arms in
our models as traced by the peak density, while the dotted lines show
the linear fits to each arm. By using the arms alone, it is possible
to measure the distance to the galactic centre to within 2.5\%,
although the lower two arms of B3 give a result accurate to only 4\%.
The increased error in B3 arises because the fit to the two arms
closer to the observer have a much shallower slope, so that the
uncertainty in the slope translates into a relatively large variation
in the intersection with the linear fit to the far side arm.  The
solid lines in Fig. \ref{findthecenter} show the linear fits for
arms at opposite corners of the X-shape (e.g.  top-left and
bottom-right). The fit to the full X-shape almost exactly finds the
centre, to within better than a tenth of a per cent for both models. 

Either method for measuring the distance to the galactic centre works
well, although the full shape seems to constrain the system better.
We note that the asymmetry across the mid-plane of model R1 does not
translate into a problem for the galactic centre distance
determination.  In the Milky Way observations linking positive and
negative latitudes may be subject to greater uncertainty than those at
only positive or only negative latitudes.

This implies that given a good enough distance determination of the
arms of the X-shape it should be possible to measure the position of
the Galactic Centre to within a few hundred parsec.  In order to
understand how random errors would affect the distance determination,
we mapped each arm with five points ($b$ $= \pm 4\degr, 5\degr,
6\degr, 7\degr, 8$\degr), and fitted a linear relation for the
X-shape.  We then measure the galactic centre as before using the
method of fitting the diagonal arms, determining the accuracy of the
distance measurement after adding a random error uniformly distributed
between $-10$ and $10$ per cent or $-20$ and $20$ per cent to each measurement
of a peak along an arm.  We repeat this procedure one million times.
We find that linear fits are able to measure the distance to within
5\% or better at least 96.5\% per cent of the time, as seen in Table
\ref{distanceerrors}. The 20\% error limit, for \textit{Gaia},
requires an accuracy of 25 $\mu as$ at a distance of 8 kpc for an
individual star.  This level of accuracy is obtainable for stars with
magnitudes of $G < 16$ \citep{Lindegren2012}. From \cite{Robin2005}
and \cite{Reyle2008}, using three different extinction models
\citep{Schlegel1998,Schultheis1999, Marshall2006} , it is clear that
RC stars at $|b| > 4\degr$ will be reached by {\it Gaia} at G
$<$ 16, even with extinction, which is anyway low at these latitudes.
Since the method relies on locating the density peak, which relies on
distance determinations of many stars, the uncertainty on individual
peaks with \textit{Gaia} should be much less than $20\%$.  Thus this
purely geometric method, which focuses on a region of the Milky Way
which is much less extincted, promises to significantly improve our
measurement of the distance to the Galactic Centre.
 
\begin{table} 
\caption{Error distribution in the measurement of distance using the
X-shape with lines joining diagonal arms.  Each column contains the
fraction of distance measurements that are within the given error, out
of the total one million Monte Carlo experiments.}
\label{distanceerrors}
    \begin{tabular}{lllll}
    \hline
    Model & Assumed  & $\leq $2.5\% & $\leq 5\%$  & $\leq 7.5\%$  \\
	  & error &\\
    \hline
    R1    & 10\%& 0.904 & 0.999 & 0.999  \\
    R1    & 20\%& 0.825 & 0.965 & 0.996  \\ 
    B3    & 10\%& 0.939 & 0.999 & 0.999  \\ 
    B3    & 20\%& 0.846 & 0.970 & 0.996  \\
    \hline
\end{tabular}
\end{table}


\section{Discussion and conclusions}

Our analysis shows that the primary kinematic imprint of an X-shape is
in the difference between the mean line-of-sight velocities,
$V_{los}$, of the near and far sides of the bulge.  In the $l=0\degr$
plane, the presence of an X-shape leads to a minimum in this
difference at $|b| \sim 4 \degr$.  At this latitude the arms of the
X-shape can be hard to distinguish but the difference is unmistakably
due to the X-shape.  It bears repeating that the kinematic evidence of
the X-shape is not just a difference between near and far side mean
$V_{los}$, but rather a minimum in this difference.  Differences
between mean Galactocentric azimuthal velocities, while non-zero, do
not show a qualitatively different behaviour between models with and
without an X-shape.  The vertical velocities are even more unable to
distinguish whether an X-shape is present or not.  The velocity
dispersions in all these quantities also show no qualitative
difference between systems with and without an X-shape.  Therefore we
recommend better measurements of RC velocities across $3\degr
\la |b| \la 7\degr$ at $l=0\degr$.

Our comparison with simulations also highlights the relatively poor
discriminating power of the available kinematics in the bulge region.
Both model R1, with only a bar induced B/P-shaped bulge, and model B3,
which included a classical bulge from the start, are able to fit most
of the observational kinematic data to a surprising degree.  Indeed
the details of the X-shape provide more of a distinction than do the
kinematics.  Unfortunately therefore, while it is clear that the bulge
of the Milky Way has been sculpted by the bar, the current kinematic
data cast no light on whether the bulge is a pre-existing structure (a
classical bulge) or one forged by secular evolution (a pseudo bulge).
We emphasize that this is a result of the still sparse sampling of
velocities with relatively large uncertainties.  In particular we
  note that the kinematic signature of the X-shape in model B3 is
  carried largely by the disc, not by the classical bulge.  Moreover
  the classical bulge itself has a boxy shape, not an X-shape.  These
  properties may aid future studies to determine the contribution of a
  classical bulge to the Milky Way's bulge.

The available kinematic data however provide strong constraints on the
orientation of the bar relative to the line-of-sight to the Galactic
Centre. For both models (R1 and B3), the best-fitting bar angle,
compared with the BRAVA data, is consistently 15\degr..  This value is between the 13\degr\ value from
\cite{Robin2012} and the 20\degr\ value from \cite{Shen2012}. It is
unclear whether this is driven by the selection functions, or is a
real result based on the best fit to BRAVA data.  Our models show that
the relative distributions of near versus far line-of-sight velocities
in the region of the X-shape can provide stronger constraints on the
bar angle.

Finally we showed that it should be possible to measure the distance
to the Galactic Centre to quite high accuracy using the X-shape.  By
tracing the location of the density peaks in the arms along the
line-of-sight, and fitting lines to them, we were able to measure the
distance to the Galactic Centre in the models to accuracy of at least
5 per cent 96\% of the time. This requires measurements of the
positions of the density peaks along each arm with uncertainties of
$20\%$. Measurements with {\it Gaia} can constrain the distances to
the arms with comparable or better accuracy.  The major advantages of
this method are that it is purely geometric, works even if the X-shape
is not perfectly symmetric across the mid-plane, and looks at regions
of the bulge which are away from the heavily-extincted main plane of
the disc.  We anticipate that in the era of {\it Gaia} we shall be
able to make high accuracy measurements of the distance to the
Galactic Centre using this method.

\section*{Acknowledgements}

We would like to acknowledge Roger Fux, for letting us use his model.
We would also like to acknowledge Charlie Gatehouse, Peter Tipping and
Jessica Mowatt for their contributions. We thank the Nuffield
Foundation for supporting Charlie Gatehouse and Peter Tipping on their
internships during the summers of 2011 and 2012 respectively.  We
  thank Melissa Ness for suggesting that we examine the contribution
  of the bulge and the disc separately for model B3.  We thank the
  anonymous referee for comments that helped improve this paper.  We
acknowledge the support of the French Agence Nationale de la Recherche
under contract ANR-2010-BLAN-0508-01OTP. V. P. D.  is supported by
STFC Consolidated grant \# ST/J001341/1.  BGM simulations were
executed on computers from the Utinam Institute of the Universit\'e de
Franche-Comt\'e, supported by the R\'egion de Franche-Comt\'e and
Institut des Sciences de l'Univers (INSU). Simulation HG1 was run at
the High Performance Computer Facility of the University of Central
Lancashire. EG thanks the Jeremiah
Horrocks Institute for their generous hospitality during parts of this
project. SV and MZ acknowledge support from Fondecyt Regular 1110393,
the BASAL CATA PFB-06, Proyecto Anillo ACT-86 and by the Chilean
Ministry for the Economy, Development, and Tourism's Programa
Iniciativa Cientifica Milenio through grant P07-021-F awarded to the
Milky Way Millennium Nucleus.

\bibliographystyle{mn2e}
\bibliography{N-body_simulation_insights_into_the_X-shaped_bulge_of_the_Milky_Way}

\begin{thebibliography}{}

\bibitem[\protect\citeauthoryear{{Athanassoula}}{{Athanassoula}}{2003}]{Athanassoula2003}
{Athanassoula} E.,  2003, \mnras, 341, 1179

\bibitem[\protect\citeauthoryear{{Athanassoula}}{{Athanassoula}}{2005}]{Athanassoula2005}
{Athanassoula} E.,  2005, \mnras, 358, 1477

\bibitem[\protect\citeauthoryear{{Baugh}, {Cole} \& {Frenk}}{{Baugh}
  et~al.}{1996}]{Baugh1996}
{Baugh} C.~M.,  {Cole} S.,    {Frenk} C.~S.,  1996, \mnras, 283, 1361

\bibitem[\protect\citeauthoryear{{Bettoni} \& {Galletta}}{{Bettoni} \&
  {Galletta}}{1994}]{Bettoni1994}
{Bettoni} D.,  {Galletta} G.,  1994, \aap, 281, 1

\bibitem[\protect\citeauthoryear{{Binney}, {Gerhard} \& {Spergel}}{{Binney}
  et~al.}{1997}]{Binney1997}
{Binney} J.,  {Gerhard} O.,    {Spergel} D.,  1997, MNRAS, 288, 365

\bibitem[\protect\citeauthoryear{{Bissantz} \& {Gerhard}}{{Bissantz} \&
  {Gerhard}}{2002}]{Bissantz2002}
{Bissantz} N.,  {Gerhard} O.,  2002, MNRAS, 330, 591

\bibitem[\protect\citeauthoryear{{Burbidge} \& {Burbidge}}{{Burbidge} \&
  {Burbidge}}{1959}]{Burbidge1959}
{Burbidge} E.~M.,  {Burbidge} G.~R.,  1959, \apj, 130, 20

\bibitem[\protect\citeauthoryear{{Bureau} \& {Freeman}}{{Bureau} \&
  {Freeman}}{1999}]{Bureau1999}
{Bureau} M.,  {Freeman} K.~C.,  1999, \aj, 118, 126

\bibitem[\protect\citeauthoryear{{Chung} \& {Bureau}}{{Chung} \&
  {Bureau}}{2004}]{Chung2004}
{Chung} A.,  {Bureau} M.,  2004, \aj, 127, 3192

\bibitem[\protect\citeauthoryear{{Clarkson} et~al.,}{{Clarkson}
  et~al.}{2008}]{Clarkson2008}
{Clarkson} W.  et~al., 2008, \apj, 684, 1110

\bibitem[\protect\citeauthoryear{{Clarkson} et~al.,}{{Clarkson}
  et~al.}{2011}]{Clarkson2011}
{Clarkson} W.~I.  et~al., 2011, \apj, 735, 37

\bibitem[\protect\citeauthoryear{{Combes}, {Debbasch}, {Friedli} \&
  {Pfenniger}}{{Combes} et~al.}{1990}]{Combes1990}
{Combes} F.,  {Debbasch} F.,  {Friedli} D.,    {Pfenniger} D.,  1990, \aap,
  233, 82

\bibitem[\protect\citeauthoryear{{Combes} \& {Sanders}}{{Combes} \&
  {Sanders}}{1981}]{Combes1981}
{Combes} F.,  {Sanders} R.~H.,  1981, \aap, 96, 164

\bibitem[\protect\citeauthoryear{{Courteau}, {de Jong} \& {Broeils}}{{Courteau}
  et~al.}{1996}]{Courteau1996}
{Courteau} S.,  {de Jong} R.~S.,    {Broeils} A.~H.,  1996, \apjl, 457, L73

\bibitem[\protect\citeauthoryear{{De Propris} et~al.,}{{De Propris}
  et~al.}{2011}]{DePropris2011}
{De Propris} R.  et~al., 2011, \apjl, 732, L36

\bibitem[\protect\citeauthoryear{{Debattista}, {Carollo}, {Mayer} \&
  {Moore}}{{Debattista} et~al.}{2004}]{Debattista2004}
{Debattista} V.~P.,  {Carollo} C.~M.,  {Mayer} L.,    {Moore} B.,  2004, \apjl,
  604, L93

\bibitem[\protect\citeauthoryear{{Debattista}, {Carollo}, {Mayer} \&
  {Moore}}{{Debattista} et~al.}{2005}]{Debattista2005}
{Debattista} V.~P.,  {Carollo} C.~M.,  {Mayer} L.,    {Moore} B.,  2005, \apj,
  628, 678

\bibitem[\protect\citeauthoryear{{Debattista}, {Mayer}, {Carollo}, {Moore},
  {Wadsley} \& {Quinn}}{{Debattista} et~al.}{2006}]{Debattista2006}
{Debattista} V.~P.,  {Mayer} L.,  {Carollo} C.~M.,  {Moore} B.,  {Wadsley} J.,
    {Quinn} T.,  2006, \apj, 645, 209

\bibitem[\protect\citeauthoryear{{Driver} et~al.,}{{Driver}
  et~al.}{2011}]{GAMA}
{Driver} S.~P.  et~al., 2011, \mnras, 413, 971

\bibitem[\protect\citeauthoryear{{Drory} \& {Fisher}}{{Drory} \&
  {Fisher}}{2007}]{Drory2007}
{Drory} N.,  {Fisher} D.~B.,  2007, \apj, 664, 640

\bibitem[\protect\citeauthoryear{{Dwek} et~al.,}{{Dwek}
  et~al.}{1995}]{Dwek1995}
{Dwek} E.  et~al., 1995, \apj, 445, 716

\bibitem[\protect\citeauthoryear{{Eggen}, {Lynden-Bell} \& {Sandage}}{{Eggen}
  et~al.}{1962}]{Eggen1962}
{Eggen} O.~J.,  {Lynden-Bell} D.,    {Sandage} A.~R.,  1962, \apj, 136, 748

\bibitem[\protect\citeauthoryear{{Erwin}, {Beltr{\'a}n}, {Graham} \&
  {Beckman}}{{Erwin} et~al.}{2003}]{Erwin2003}
{Erwin} P.,  {Beltr{\'a}n} J.~C.~V.,  {Graham} A.~W.,    {Beckman} J.~E.,
  2003, \apj, 597, 929

\bibitem[\protect\citeauthoryear{{Erwin} \& {Debattista}}{{Erwin} \&
  {Debattista}}{2013}]{Erwin2013}
{Erwin} P.,  {Debattista} V.~P.,  2013, \mnras, 431, 3060

\bibitem[\protect\citeauthoryear{{Fukugita}, {Hogan} \& {Peebles}}{{Fukugita}
  et~al.}{1998}]{Fukugita1998}
{Fukugita} M.,  {Hogan} C.~J.,    {Peebles} P.~J.~E.,  1998, \apj, 503, 518

\bibitem[\protect\citeauthoryear{{Fux}}{{Fux}}{1999}]{Fux1999}
{Fux} R.,  1999, \aap, 345, 787

\bibitem[\protect\citeauthoryear{{Hopkins} et~al.,}{{Hopkins}
  et~al.}{2010}]{Hopkins2010}
{Hopkins} P.~F.  et~al., 2010, \apj, 715, 202

\bibitem[\protect\citeauthoryear{{Howard} et~al.,}{{Howard}
  et~al.}{2009}]{Howard2009}
{Howard} C.~D.  et~al., 2009, \apjl, 702, L153

\bibitem[\protect\citeauthoryear{{Kauffmann}, {White} \&
  {Guiderdoni}}{{Kauffmann} et~al.}{1993}]{Kauffmann1993}
{Kauffmann} G.,  {White} S.~D.~M.,    {Guiderdoni} B.,  1993, \mnras, 264, 201

\bibitem[\protect\citeauthoryear{{Kormendy} \& {Kennicutt} Jr.}{{Kormendy} \&
  {Kennicutt}}{2004}]{Kormendy2004}
{Kormendy} J.,  {Kennicutt} Jr. R.~C.,  2004, \araa, 42, 603

\bibitem[\protect\citeauthoryear{{Kuijken} \& {Merrifield}}{{Kuijken} \&
  {Merrifield}}{1995}]{Kuijken1995}
{Kuijken} K.,  {Merrifield} M.~R.,  1995, \apjl, 443, L13

\bibitem[\protect\citeauthoryear{{Kuijken} \& {Rich}}{{Kuijken} \&
  {Rich}}{2002}]{Kuijken2002}
{Kuijken} K.,  {Rich} R.~M.,  2002, \aj, 124, 2054

\bibitem[\protect\citeauthoryear{{Kunder} et~al.,}{{Kunder}
  et~al.}{2012}]{BRAVA}
{Kunder} A.  et~al., 2012, \aj, 143, 57

\bibitem[\protect\citeauthoryear{{Laurikainen}, {Salo}, {Buta} \&
  {Knapen}}{{Laurikainen} et~al.}{2011}]{Laurikainen2011}
{Laurikainen} E.,  {Salo} H.,  {Buta} R.,    {Knapen} J.~H.,  2011, \mnras,
  418, 1452

\bibitem[\protect\citeauthoryear{{Li} \& {Shen}}{{Li} \&
  {Shen}}{2012}]{Shen2012}
{Li} Z.-Y.,  {Shen} J.,  2012, \apjl, 757, L7

\bibitem[\protect\citeauthoryear{{Lindegren}, {Lammers}, {Hobbs}, {O'Mullane},
  {Bastian} \& {Hern{\'a}ndez}}{{Lindegren} et~al.}{2012}]{Lindegren2012}
{Lindegren} L.,  {Lammers} U.,  {Hobbs} D.,  {O'Mullane} W.,  {Bastian} U.,
  {Hern{\'a}ndez} J.,  2012, \aap, 538, A78

\bibitem[\protect\citeauthoryear{{L{\'o}pez-Corredoira}, {Cabrera-Lavers} \&
  {Gerhard}}{{L{\'o}pez-Corredoira} et~al.}{2005}]{Lopez2005}
{L{\'o}pez-Corredoira} M.,  {Cabrera-Lavers} A.,    {Gerhard} O.~E.,  2005,
  A\&A, 439, 107

\bibitem[\protect\citeauthoryear{{L{\"u}tticke}, {Dettmar} \&
  {Pohlen}}{{L{\"u}tticke} et~al.}{2000}]{Lutticke2000a}
{L{\"u}tticke} R.,  {Dettmar} R.-J.,    {Pohlen} M.,  2000, \aaps, 145, 405

\bibitem[\protect\citeauthoryear{{Marshall}, {Robin}, {Reyl{\'e}}, {Schultheis}
  \& {Picaud}}{{Marshall} et~al.}{2006}]{Marshall2006}
{Marshall} D.~J.,  {Robin} A.~C.,  {Reyl{\'e}} C.,  {Schultheis} M.,
  {Picaud} S.,  2006, \aap, 453, 635

\bibitem[\protect\citeauthoryear{{McWilliam}}{{McWilliam}}{1997}]{McWilliam97}
{McWilliam} A.,  1997, \araa, 35, 503

\bibitem[\protect\citeauthoryear{{McWilliam} \& {Zoccali}}{{McWilliam} \&
  {Zoccali}}{2010}]{McWilliam2010}
{McWilliam} A.,  {Zoccali} M.,  2010, \apj, 724, 1491

\bibitem[\protect\citeauthoryear{{M{\'e}ndez-Abreu}, {Corsini}, {Debattista},
  {De Rijcke}, {Aguerri} \& {Pizzella}}{{M{\'e}ndez-Abreu}
  et~al.}{2008}]{MendezAbreu2008}
{M{\'e}ndez-Abreu} J.,  {Corsini} E.~M.,  {Debattista} V.~P.,  {De Rijcke} S.,
  {Aguerri} J.~A.~L.,    {Pizzella} A.,  2008, \apjl, 679, L73

\bibitem[\protect\citeauthoryear{{Nataf}, {Udalski}, {Gould}, {Fouqu{\'e}} \&
  {Stanek}}{{Nataf} et~al.}{2010}]{Nataf2010}
{Nataf} D.~M.,  {Udalski} A.,  {Gould} A.,  {Fouqu{\'e}} P.,    {Stanek} K.~Z.,
   2010, \apjl, 721, L28

\bibitem[\protect\citeauthoryear{{Ness} et~al.,}{{Ness}
  et~al.}{2012}]{Ness2012}
{Ness} M.  et~al., 2012, \apj, 756, 22

\bibitem[\protect\citeauthoryear{{Norman}, {Sellwood} \& {Hasan}}{{Norman}
  et~al.}{1996}]{Norman1996}
{Norman} C.~A.,  {Sellwood} J.~A.,    {Hasan} H.,  1996, \apj, 462, 114

\bibitem[\protect\citeauthoryear{{Nowak}, {Thomas}, {Erwin}, {Saglia}, {Bender}
  \& {Davies}}{{Nowak} et~al.}{2010}]{Nowak2010}
{Nowak} N.,  {Thomas} J.,  {Erwin} P.,  {Saglia} R.~P.,  {Bender} R.,
  {Davies} R.~I.,  2010, \mnras, 403, 646

\bibitem[\protect\citeauthoryear{{Ortolani}, {Renzini}, {Gilmozzi}, {Marconi},
  {Barbuy}, {Bica} \& {Rich}}{{Ortolani} et~al.}{1995}]{Ortolani1995}
{Ortolani} S.,  {Renzini} A.,  {Gilmozzi} R.,  {Marconi} G.,  {Barbuy} B.,
  {Bica} E.,    {Rich} R.~M.,  1995, \nat, 377, 701

\bibitem[\protect\citeauthoryear{{Patsis}, {Athanassoula}, {Grosb{\o}l} \&
  {Skokos}}{{Patsis} et~al.}{2002}]{Patsis2002a}
{Patsis} P.~A.,  {Athanassoula} E.,  {Grosb{\o}l} P.,    {Skokos} C.,  2002,
  \mnras, 335, 1049

\bibitem[\protect\citeauthoryear{{Patsis}, {Skokos} \& {Athanassoula}}{{Patsis}
  et~al.}{2002}]{Patsis2002b}
{Patsis} P.~A.,  {Skokos} C.,    {Athanassoula} E.,  2002, \mnras, 337, 578

\bibitem[\protect\citeauthoryear{{Persic} \& {Salucci}}{{Persic} \&
  {Salucci}}{1992}]{Persic1992}
{Persic} M.,  {Salucci} P.,  1992, \mnras, 258, 14P

\bibitem[\protect\citeauthoryear{{Pfenniger}}{{Pfenniger}}{1984}]{Pfenniger1984}
{Pfenniger} D.,  1984, \aap, 134, 373

\bibitem[\protect\citeauthoryear{{Pfenniger}}{{Pfenniger}}{1985}]{Pfenniger1985}
{Pfenniger} D.,  1985, \aap, 150, 112

\bibitem[\protect\citeauthoryear{{Pfenniger} \& {Friedli}}{{Pfenniger} \&
  {Friedli}}{1991}]{Pfenniger1991}
{Pfenniger} D.,  {Friedli} D.,  1991, \aap, 252, 75

\bibitem[\protect\citeauthoryear{{Quillen}, {Kuchinski}, {Frogel} \&
  {Depoy}}{{Quillen} et~al.}{1997}]{Quillen1997}
{Quillen} A.~C.,  {Kuchinski} L.~E.,  {Frogel} J.~A.,    {Depoy} D.~L.,  1997,
  \apj, 481, 179

\bibitem[\protect\citeauthoryear{{Raha}, {Sellwood}, {James} \& {Kahn}}{{Raha}
  et~al.}{1991}]{Raha1991}
{Raha} N.,  {Sellwood} J.~A.,  {James} R.~A.,    {Kahn} F.~D.,  1991, \nat,
  352, 411

\bibitem[\protect\citeauthoryear{{Rangwala}, {Williams} \& {Stanek}}{{Rangwala}
  et~al.}{2009}]{Rangwala2009}
{Rangwala} N.,  {Williams} T.~B.,    {Stanek} K.~Z.,  2009, \apj, 691, 1387

\bibitem[\protect\citeauthoryear{{Reyl{\'e}}, {Marshall}, {Schultheis} \&
  {Robin}}{{Reyl{\'e}} et~al.}{2008}]{Reyle2008}
{Reyl{\'e}} C.,  {Marshall} D.~J.,  {Schultheis} M.,    {Robin} A.~C.,  2008,
  in {Charbonnel} C.,  {Combes} F.,   {Samadi} R.,  eds, SF2A-2008. p.~29

\bibitem[\protect\citeauthoryear{{Robin}, {Marshall}, {Schultheis} \&
  {Reyl{\'e}}}{{Robin} et~al.}{2012}]{Robin2012}
{Robin} A.~C.,  {Marshall} D.~J.,  {Schultheis} M.,    {Reyl{\'e}} C.,  2012,
  \aap, 538, A106

\bibitem[\protect\citeauthoryear{{Robin}, {Reyl{\'e}}, {Picaud} \&
  {Schultheis}}{{Robin} et~al.}{2005}]{Robin2005}
{Robin} A.~C.,  {Reyl{\'e}} C.,  {Picaud} S.,    {Schultheis} M.,  2005, \aap,
  430, 129

\bibitem[\protect\citeauthoryear{{Ro{\v s}kar}, {Debattista}, {Stinson},
  {Quinn}, {Kaufmann} \& {Wadsley}}{{Ro{\v s}kar} et~al.}{2008}]{Roskar2008}
{Ro{\v s}kar} R.,  {Debattista} V.~P.,  {Stinson} G.~S.,  {Quinn} T.~R.,
  {Kaufmann} T.,    {Wadsley} J.,  2008, \apjl, 675, L65

\bibitem[\protect\citeauthoryear{{Saha}, {Martinez-Valpuesta} \&
  {Gerhard}}{{Saha} et~al.}{2012}]{Saha2012}
{Saha} K.,  {Martinez-Valpuesta} I.,    {Gerhard} O.,  2012, \mnras, 421, 333

\bibitem[\protect\citeauthoryear{{Saito}, {Zoccali}, {McWilliam}, {Minniti},
  {Gonzalez} \& {Hill}}{{Saito} et~al.}{2011}]{Saito2011}
{Saito} R.~K.,  {Zoccali} M.,  {McWilliam} A.,  {Minniti} D.,  {Gonzalez}
  O.~A.,    {Hill} V.,  2011, \aj, 142, 76

\bibitem[\protect\citeauthoryear{{Schlegel}, {Finkbeiner} \&
  {Davis}}{{Schlegel} et~al.}{1998}]{Schlegel1998}
{Schlegel} D.~J.,  {Finkbeiner} D.~P.,    {Davis} M.,  1998, \apj, 500, 525

\bibitem[\protect\citeauthoryear{{Sch{\"o}nrich}}{{Sch{\"o}nrich}}{2012}]{Schonrich2012}
{Sch{\"o}nrich} R.,  2012, \mnras, 427, 274

\bibitem[\protect\citeauthoryear{{Schultheis} et~al.,}{{Schultheis}
  et~al.}{1999}]{Schultheis1999}
{Schultheis} M.  et~al., 1999, \aap, 349, L69

\bibitem[\protect\citeauthoryear{{Searle} \& {Zinn}}{{Searle} \&
  {Zinn}}{1978}]{Searle1978}
{Searle} L.,  {Zinn} R.,  1978, \apj, 225, 357

\bibitem[\protect\citeauthoryear{{Shaw}}{{Shaw}}{1987}]{Shaw1987}
{Shaw} M.~A.,  1987, \mnras, 229, 691

\bibitem[\protect\citeauthoryear{{Shen}, {Rich}, {Kormendy}, {Howard}, {De
  Propris} \& {Kunder}}{{Shen} et~al.}{2010}]{Shen2010}
{Shen} J.,  {Rich} R.~M.,  {Kormendy} J.,  {Howard} C.~D.,  {De Propris} R.,
  {Kunder} A.,  2010, \apjl, 720, L72

\bibitem[\protect\citeauthoryear{{Skrutskie} et~al.,}{{Skrutskie}
  et~al.}{2006}]{2MASS}
{Skrutskie} M.~F.  et~al., 2006, \aj, 131, 1163

\bibitem[\protect\citeauthoryear{{Tremaine}, {Ostriker} \& {Spitzer}
  Jr.}{{Tremaine} et~al.}{1975}]{Tremaine1975}
{Tremaine} S.~D.,  {Ostriker} J.~P.,    {Spitzer} Jr. L.,  1975, \apj, 196, 407

\bibitem[\protect\citeauthoryear{{Udalski} et~al.,}{{Udalski}
  et~al.}{2002}]{Udalski2002}
{Udalski} A.  et~al., 2002, \actaa, 52, 217

\bibitem[\protect\citeauthoryear{{Udalski}, {Szymanski}, {Soszynski} \&
  {Poleski}}{{Udalski} et~al.}{2008}]{Udalski2008}
{Udalski} A.,  {Szymanski} M.~K.,  {Soszynski} I.,    {Poleski} R.,  2008,
  \actaa, 58, 69

\bibitem[\protect\citeauthoryear{{Uttenthaler}, {Schultheis}, {Nataf}, {Robin},
  {Lebzelter} \& {Chen}}{{Uttenthaler} et~al.}{2012}]{Uttenthaler2012}
{Uttenthaler} S.,  {Schultheis} M.,  {Nataf} D.~M.,  {Robin} A.~C.,
  {Lebzelter} T.,    {Chen} B.,  2012, \aap, 546, A57

\bibitem[\protect\citeauthoryear{{Valenti}, {Fallon} \&
  {Johns-Krull}}{{Valenti} et~al.}{2003}]{Valenti2003}
{Valenti} J.~A.,  {Fallon} A.~A.,    {Johns-Krull} C.~M.,  2003, \apjs, 147,
  305

\bibitem[\protect\citeauthoryear{{van den Bosch}}{{van den
  Bosch}}{1998}]{vandenBosch1998}
{van den Bosch} F.~C.,  1998, \apj, 507, 601

\bibitem[\protect\citeauthoryear{{van den Bosch} \& {van de Ven}}{{van den
  Bosch} \& {van de Ven}}{2009}]{vdBosch2009}
{van den Bosch} R.~C.~E.,  {van de Ven} G.,  2009, \mnras, 398, 1117

\bibitem[\protect\citeauthoryear{{Vanhollebeke}, {Groenewegen} \&
  {Girardi}}{{Vanhollebeke} et~al.}{2009}]{Vanhollebeke2009}
{Vanhollebeke} E.,  {Groenewegen} M.~A.~T.,    {Girardi} L.,  2009, A\&A, 498,
  95

\bibitem[\protect\citeauthoryear{{Vasquez} et~al.,}{{Vasquez}
  et~al.}{2013}]{Vasquez2013}
{Vasquez} S.  et~al., 2013, \aap, 555, A91

\bibitem[\protect\citeauthoryear{{Zoccali} et~al.,}{{Zoccali}
  et~al.}{2003}]{Zoccali2003}
{Zoccali} M.  et~al., 2003, \aap, 399, 931

\end{thebibliography}

\clearpage
\appendix
\section{Analysis of six unscaled models}
\label{appendix}

We concentrate here on the analysis of all six models, with a few
simplifications. The distance of the observer to the centre is set to
2 bar lengths, and the models are completely unscaled. The comparisons
of the models will be qualitative, instead of quantitative, as in the
main part of the paper.  All models have been rotated so that the bar
angle is $15 \degr$.

\begin{figure}
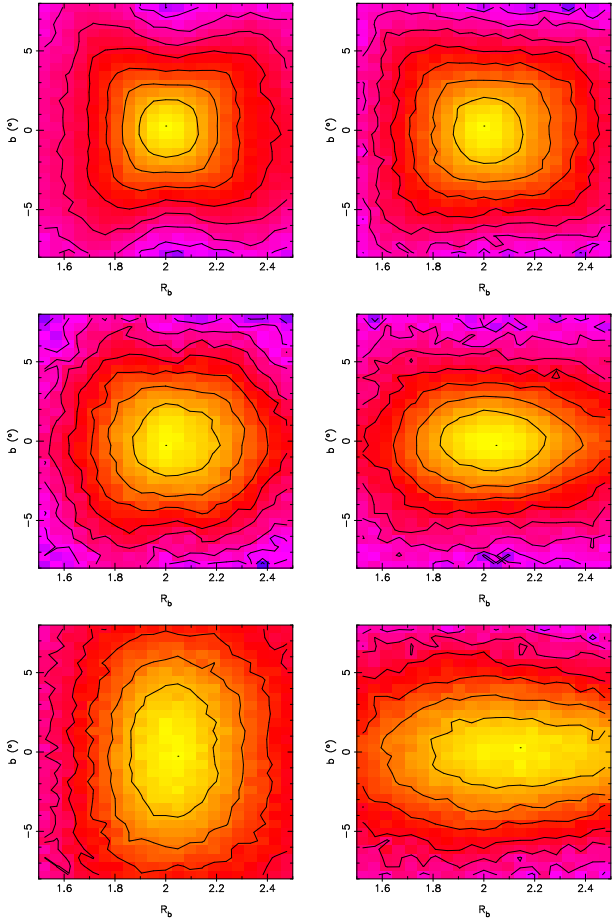

\begin{tabular}{lr}

\includegraphics[angle=-90,width=1.5in]{figA1.1.ps}&
\includegraphics[angle=-90,width=1.5in]{figA1.2.ps}\\
\includegraphics[angle=-90,width=1.5in]{figA1.3.ps} &
\includegraphics[angle=-90,width=1.5in]{figA1.4.ps}\\
\includegraphics[angle=-90,width=1.5in]{figA1.5.ps} &
\includegraphics[angle=-90,width=1.5in]{figA1.6.ps}

\end{tabular}

\caption{Density of models R1 (top-left), B3 (top-right), B2
(middle-left), R5 (middle-right), R2 (bottom-left) and R6
(bottom-right) in the $l=0\degr$ plane. The plot shows the density as
a function of latitude and distance, mimicking the projection in Fig.
4 of \protect\cite{Saito2011}.  The models are ordered by their
overall boxiness, with the last two exhibiting no signs of a {\it
visible} X-shape.}
\label{Ashapes1}
\end{figure}

Of the models in Fig. \ref{Ashapes1}, R1 and B3 exhibit strong B/P
shapes, B2 and R5 have weak B/P shapes, while R6 does not have an
X-shape at all. R2 is a special case where there is an X-shape, but it
requires a much larger observer distance to see it at $|b|<8\degr$.

\subsection{Velocities}

\begin{figure}
\includegraphics[width=\columnwidth]{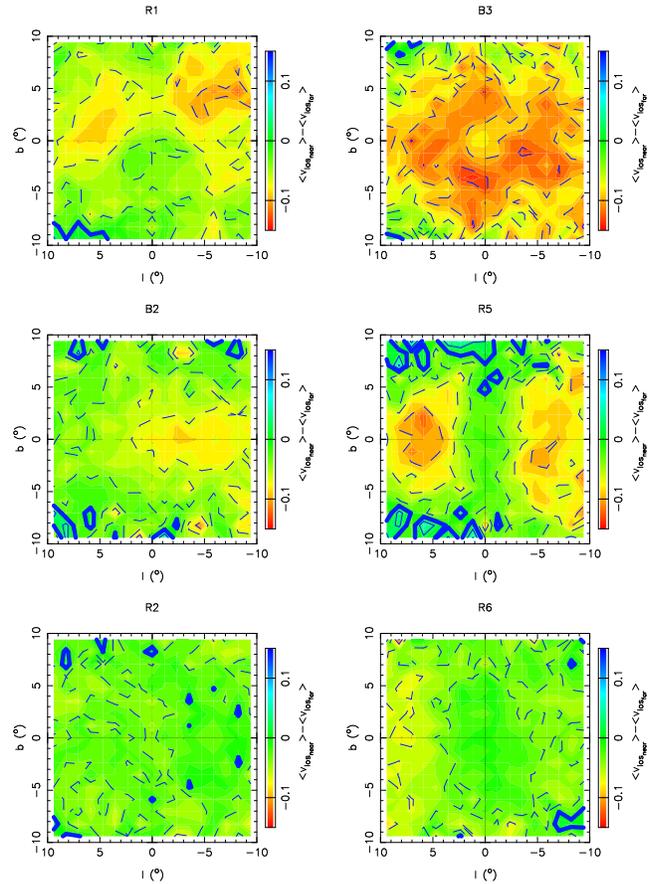}
\caption{The difference in mean line-of-sight velocity ($V_{los}$)
between the near and far sides of the bulge, for a bar angle of
15\degr, for each simulation. The dashed lines show negative contours
while the bold contour shows 0 km s$^{-1}$.}
\label{A02vlos}
\end{figure}

The presence of the X-shape is evident in the difference between mean
$V_{los}$ (Figs \ref{A02vlos} and \ref{A02svlos}).  The differences in
galactocentric $V_\phi$ and $\sigma_\phi$ (Figs \ref{A02vphi} and
\ref{A02svphi} show no evidence of the
imprint of an X-shape.  Both vertical velocity (Fig. \ref{A02vz}) and
vertical velocity dispersion (Fig. \ref{A02svz}) show no signs of
coherent differences between means on the near and far sides.

\begin{figure}
\includegraphics[width=\columnwidth]{figA3.ps}    
\caption{The difference in line-of-sight velocity dispersion
($\sigma_{los}$) between the near and far sides of the bulge, for a
bar angle of 15\degr, for each simulation. The dashed lines show
negative contours while the bold contour shows 0 km s$^{-1}$.}
\label{A02svlos}
\end{figure}

\begin{figure}
\includegraphics[width=\columnwidth]{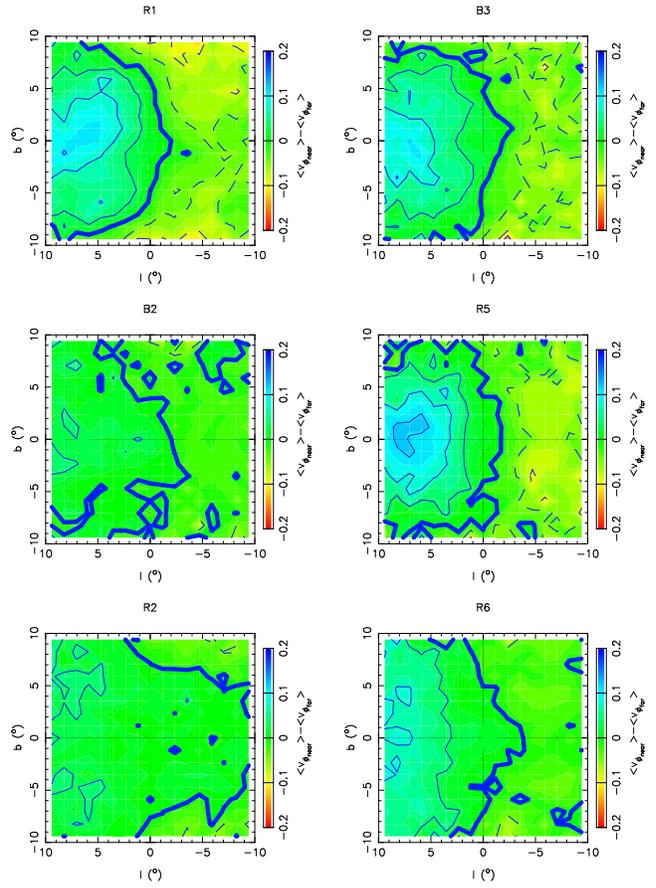}
\caption{Same as Fig. \protect\ref{A02vlos}, for galactocentric
azimuthal velocity (V$_\phi$).}
\label{A02vphi}
\end{figure}

\begin{figure}
\includegraphics[width=\columnwidth]{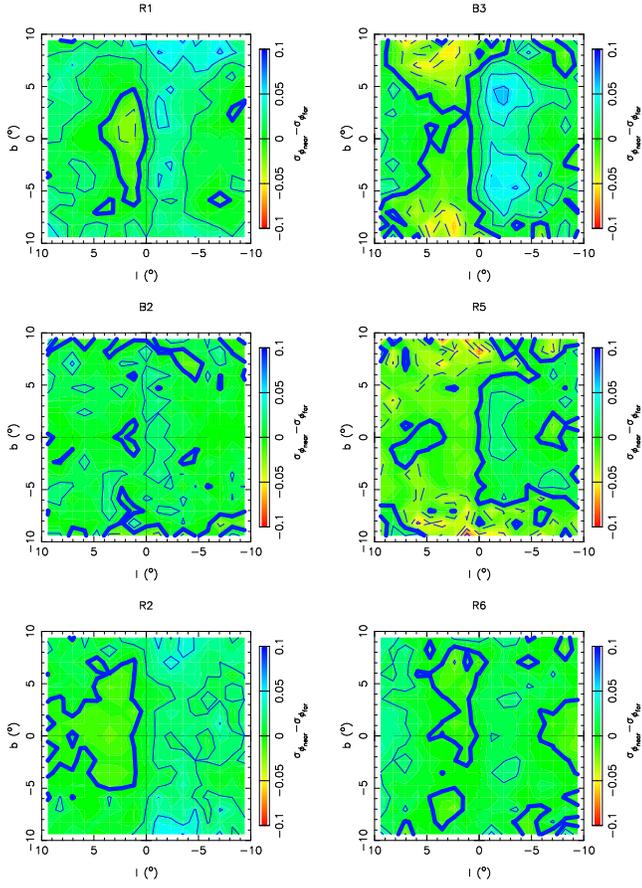}
\caption{Same as Fig. \protect\ref{A02svlos}, for galactocentric
azimuthal velocity dispersion ($\sigma_\phi$).}
\label{A02svphi}
\end{figure}

\begin{figure}
\includegraphics[width=\columnwidth]{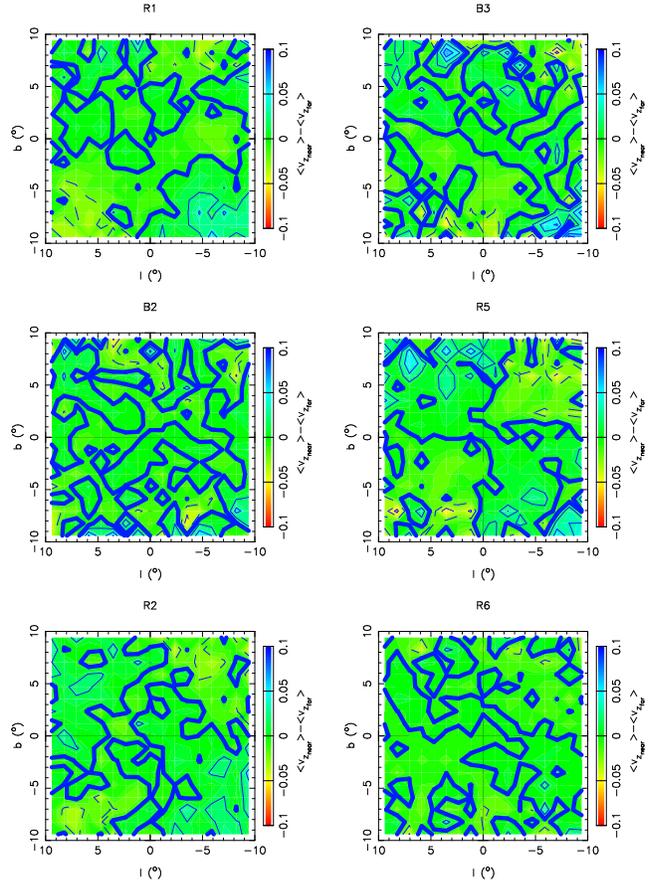}    
\caption{Same as Fig. \protect\ref{A02vlos}, for vertical velocity
(V$_z$).}
\label{A02vz}
\end{figure}

\begin{figure}
\includegraphics[width=\columnwidth]{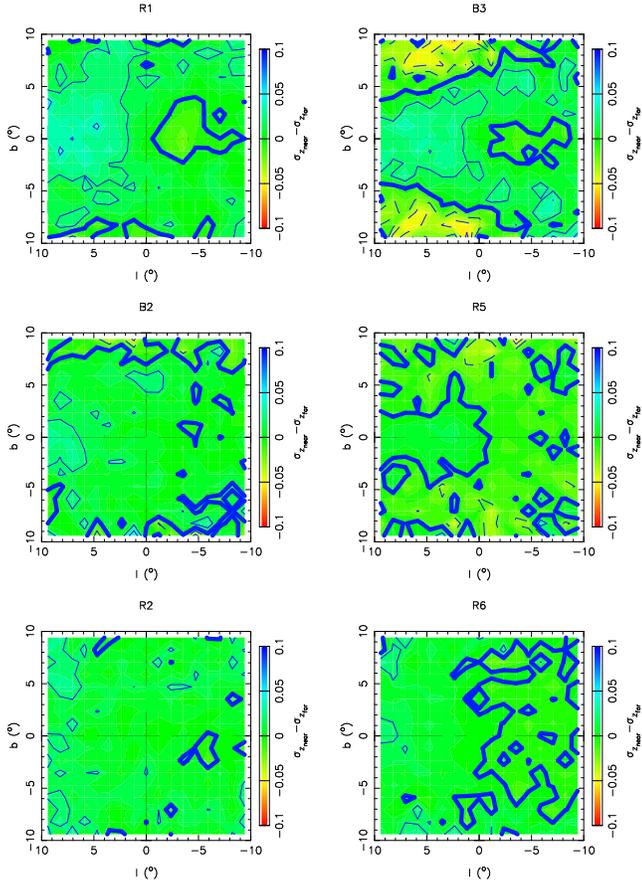}
\caption{Same as Fig. \protect\ref{A02svlos}, for vertical velocity
dispersion ($\sigma_z$).}
\label{A02svz}
\end{figure}

\subsection{The effect of bar angle on velocities}

\begin{figure}

\begin{tabular}{lccc}
  Angle:   15$^\circ$ & 25$^\circ$ & 35$^\circ$\\

    R1 
    \includegraphics[width=0.65in, angle=-90 ]{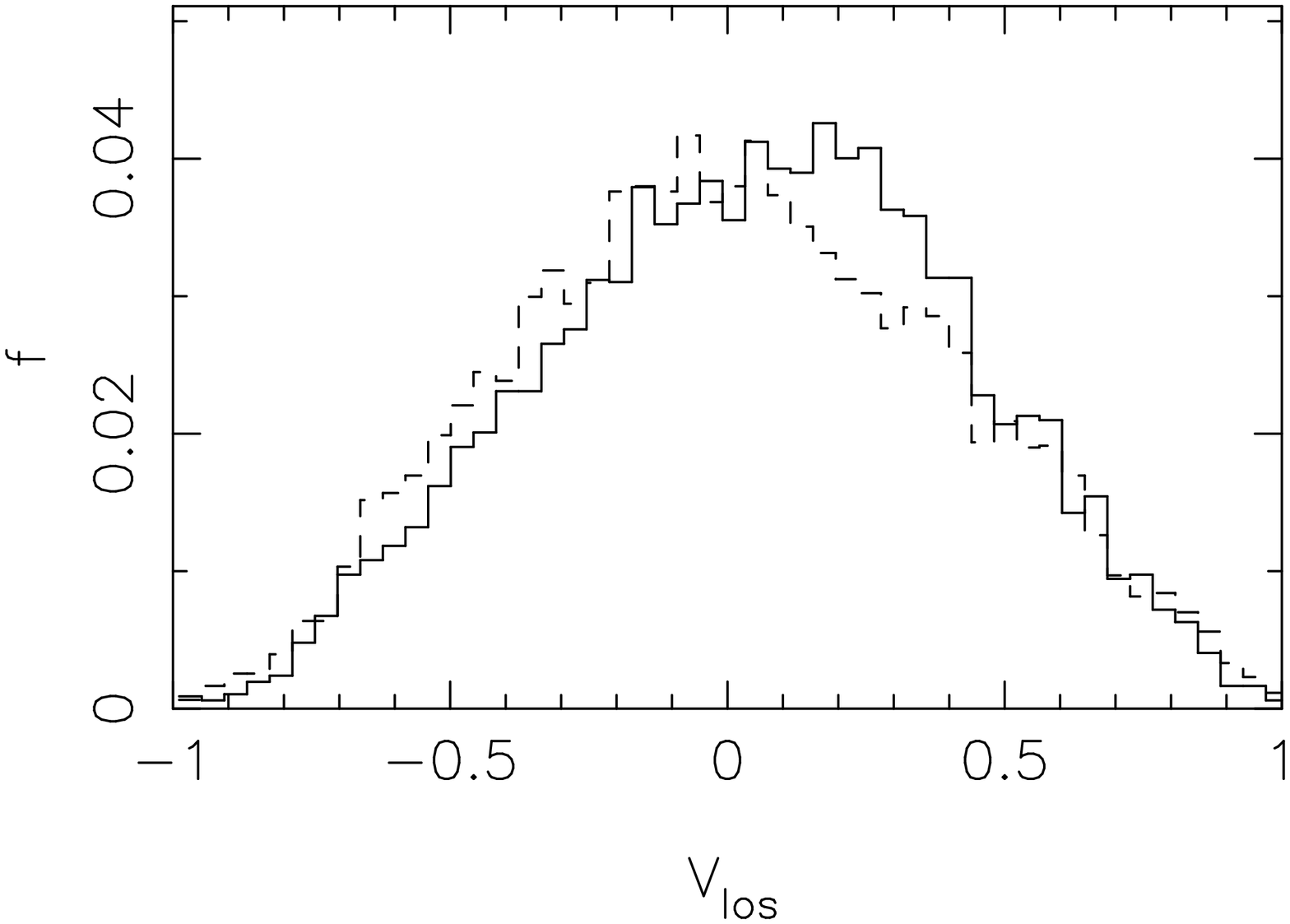}&
    \includegraphics[width=0.65in, angle=-90 ]{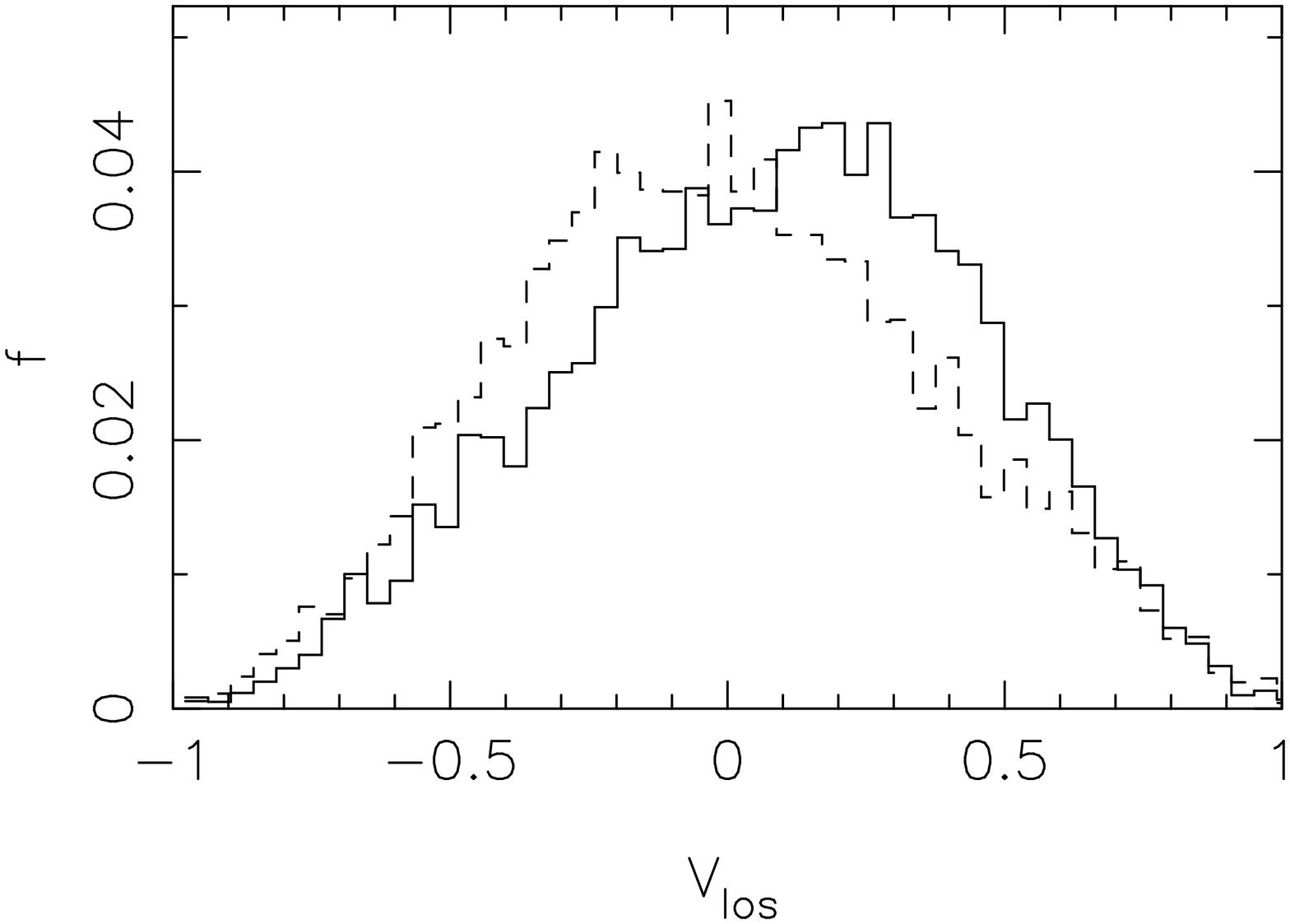}&
    \includegraphics[width=0.65in, angle=-90 ]{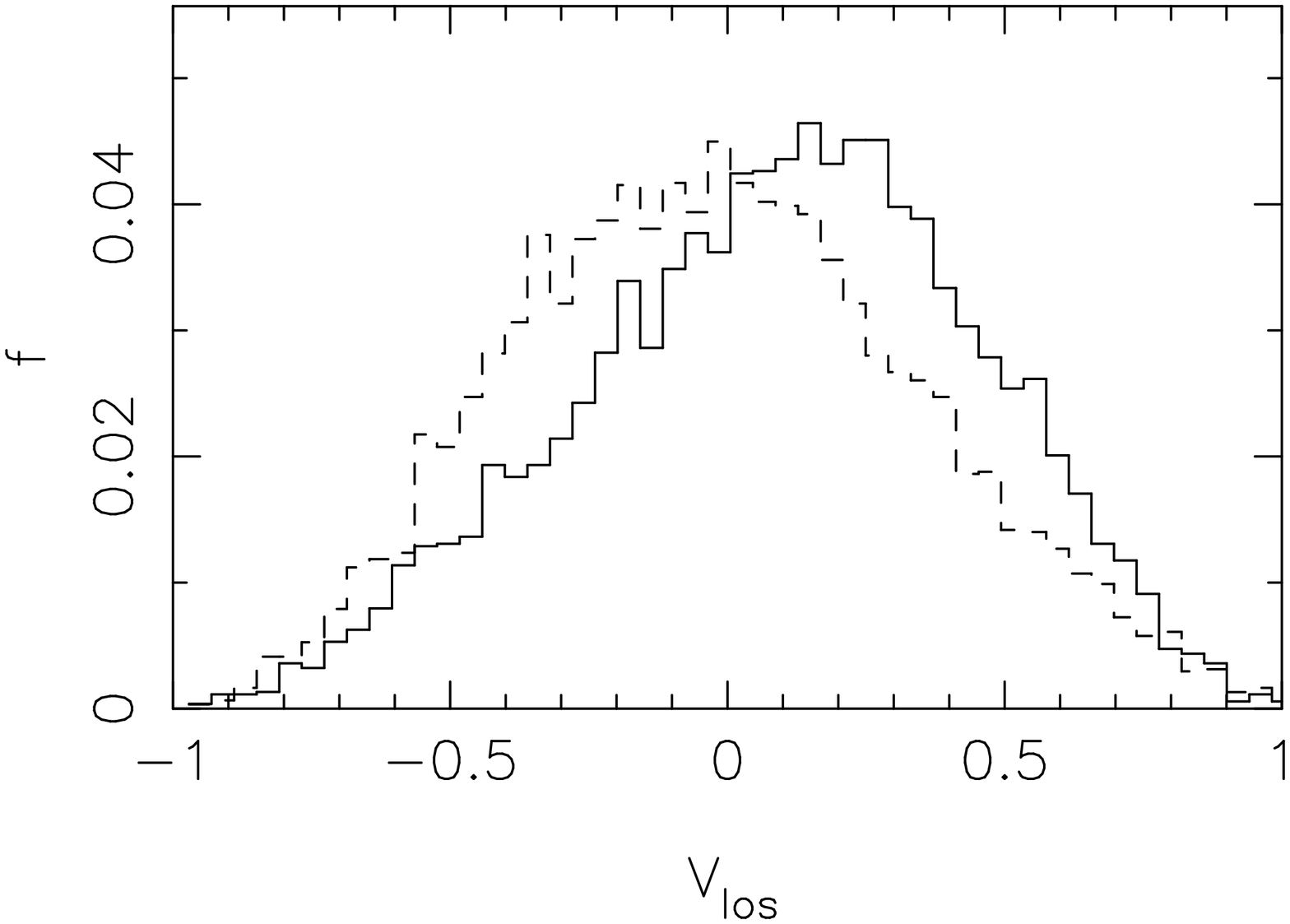}\\
    B3    
    \includegraphics[width=0.65in, angle=-90 ]{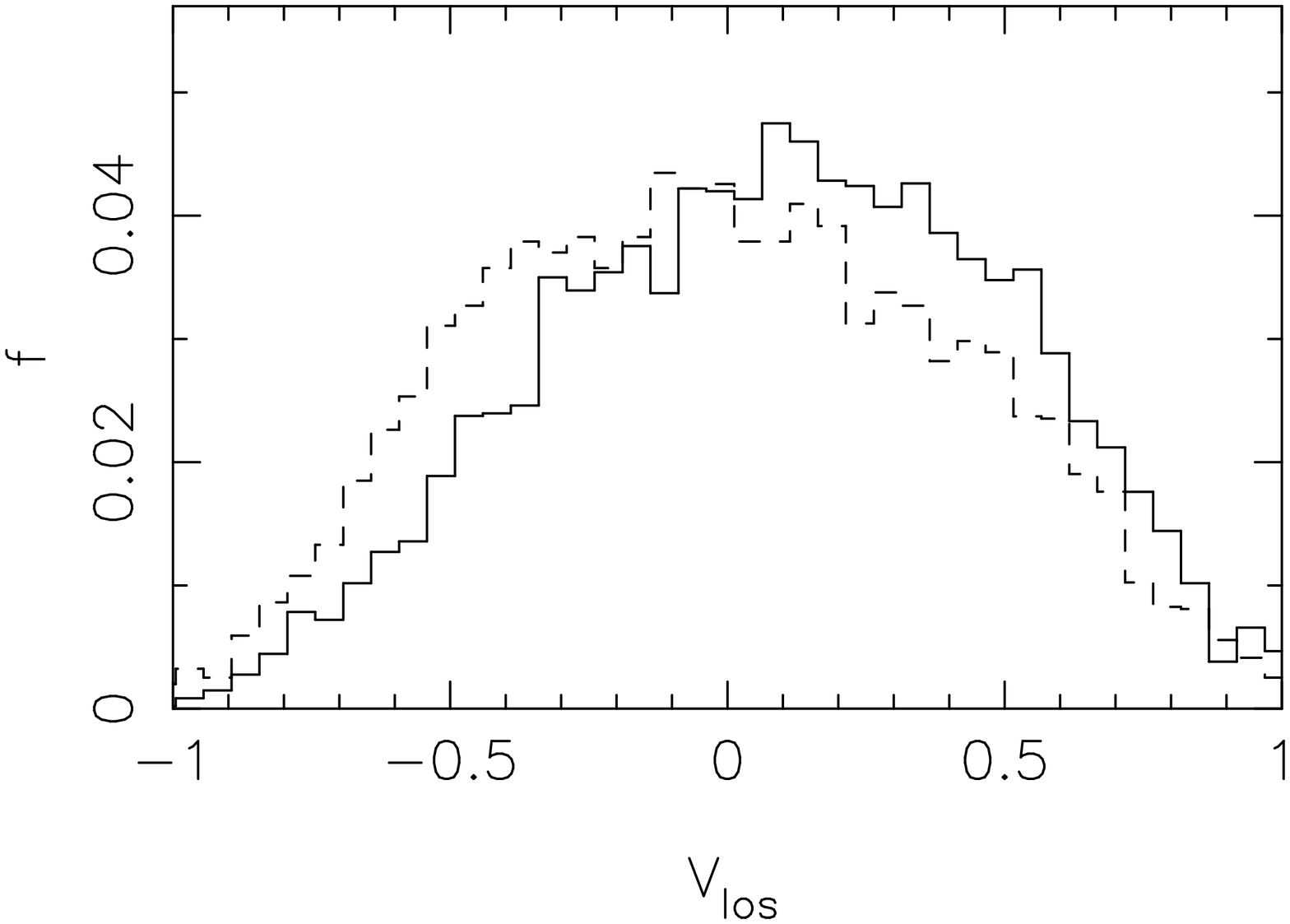}&
    \includegraphics[width=0.65in, angle=-90 ]{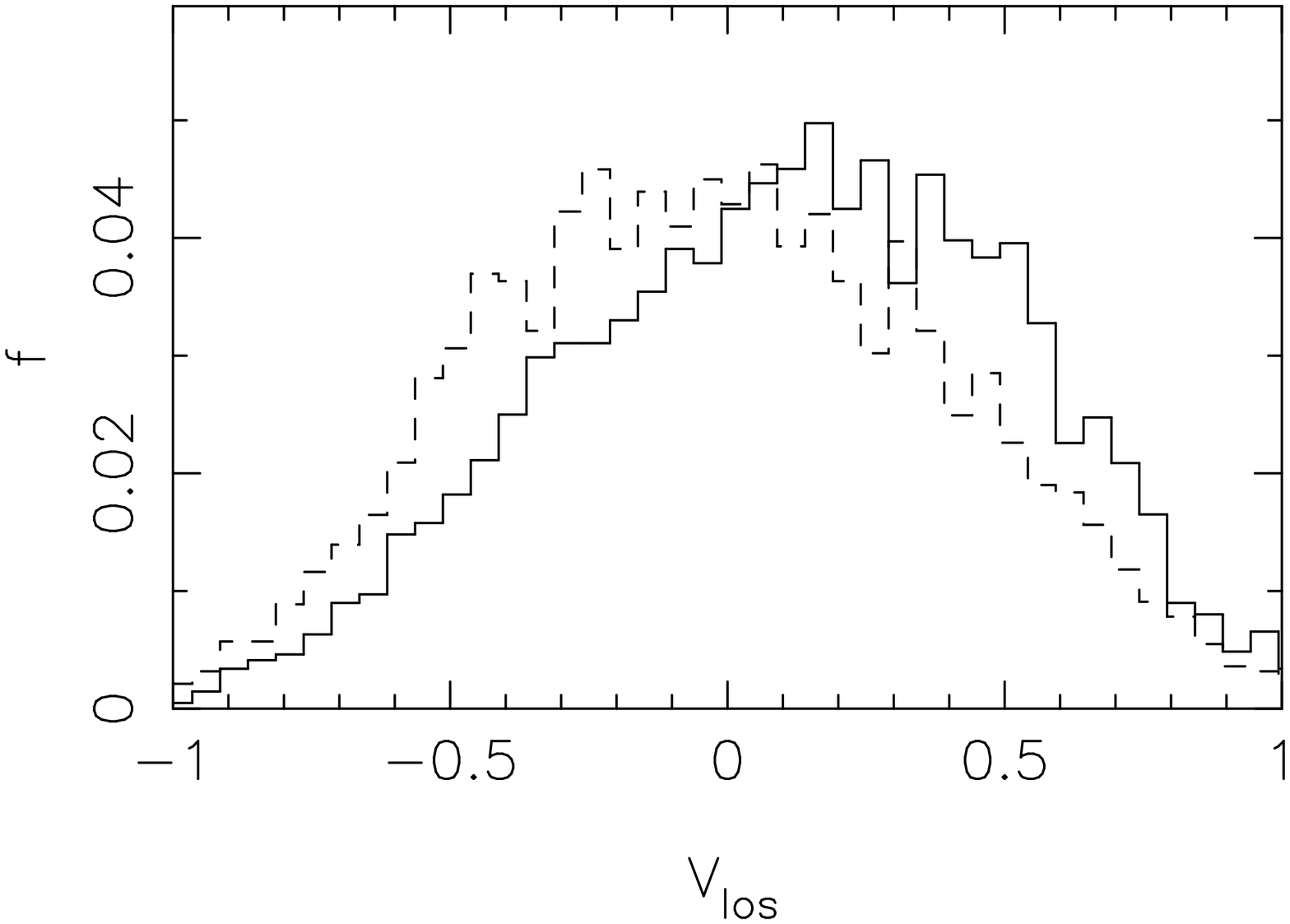}&
    \includegraphics[width=0.65in, angle=-90 ]{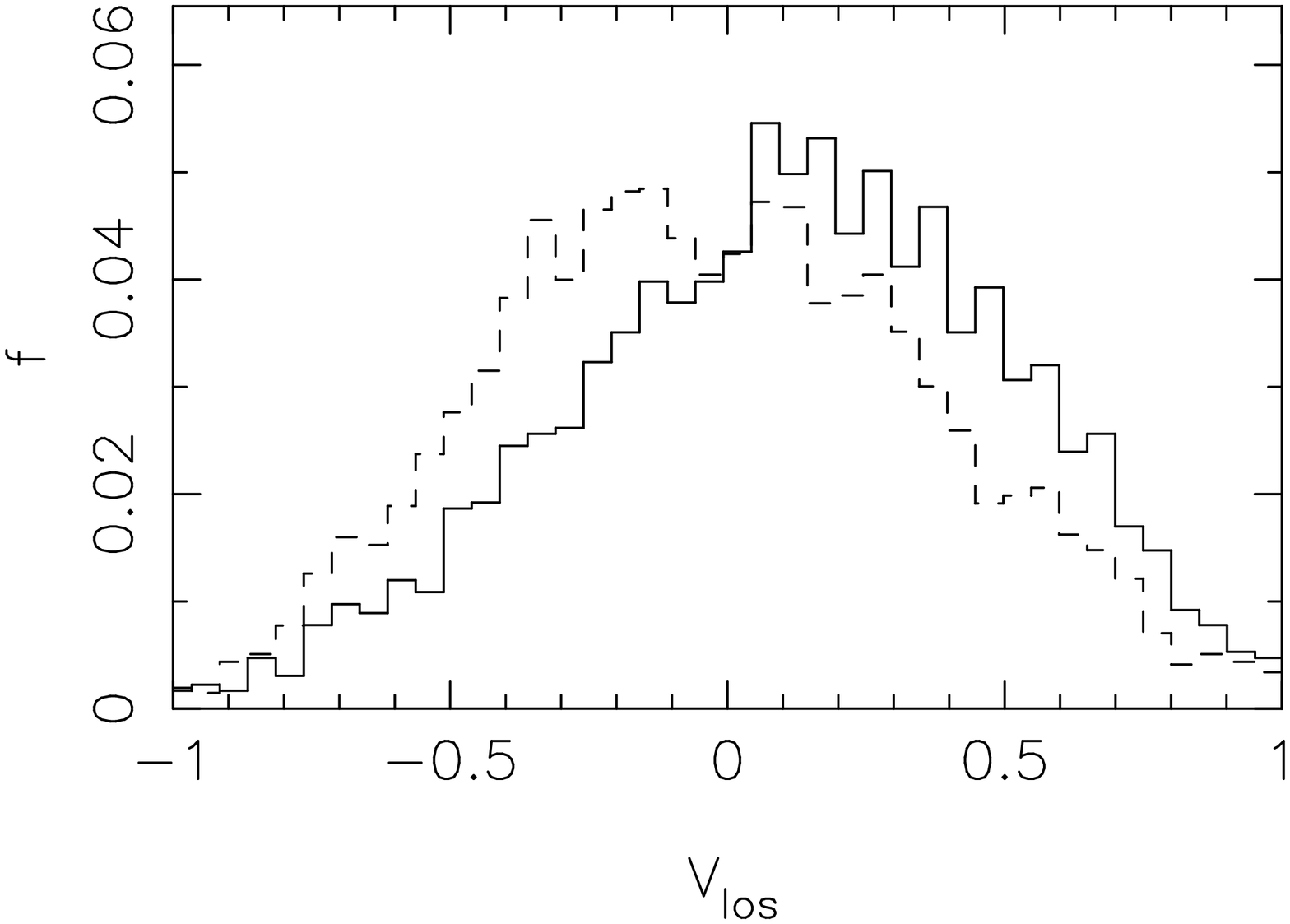}\\
    B2 
    \includegraphics[width=0.65in, angle=-90 ]{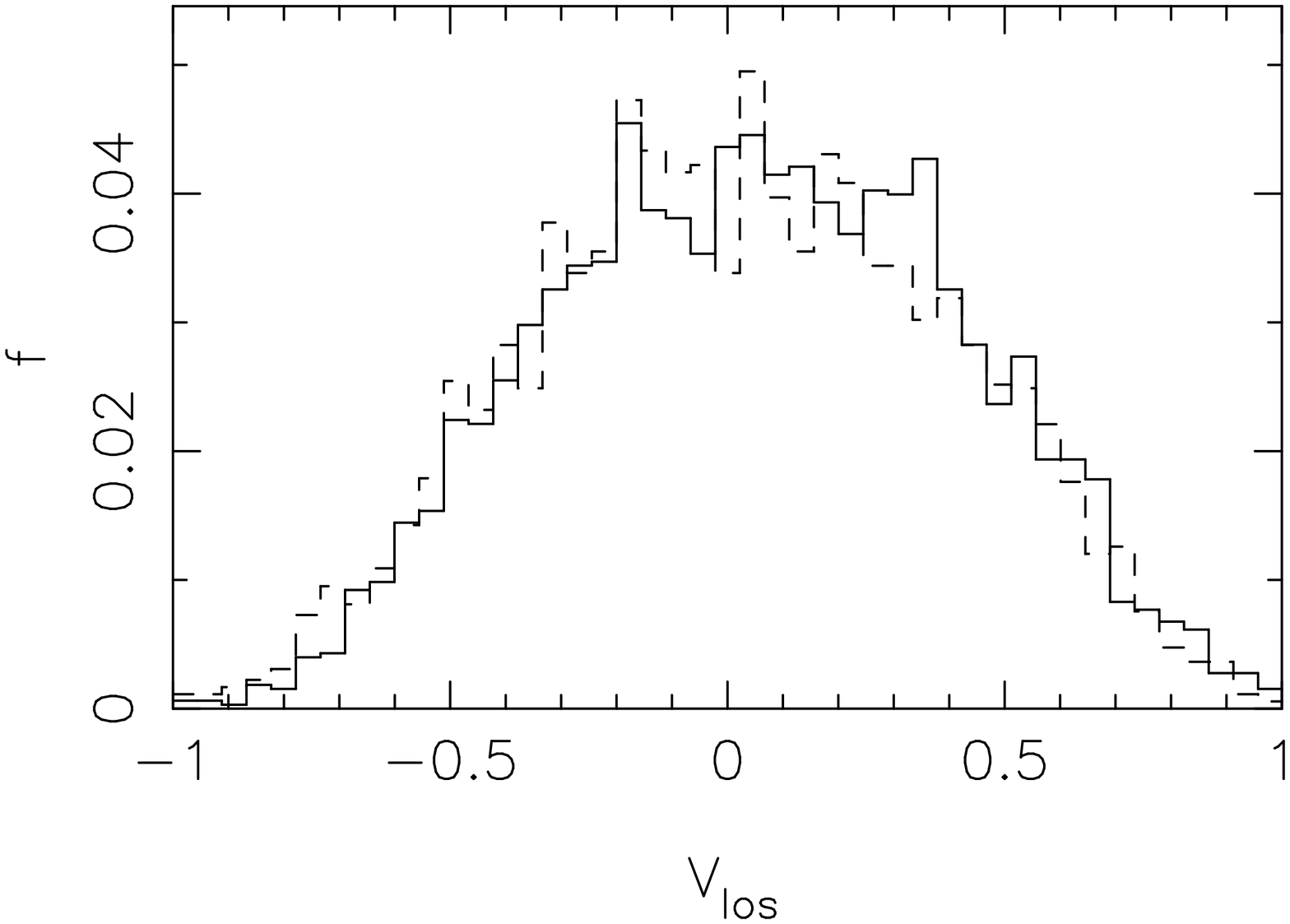}&
    \includegraphics[width=0.65in, angle=-90 ]{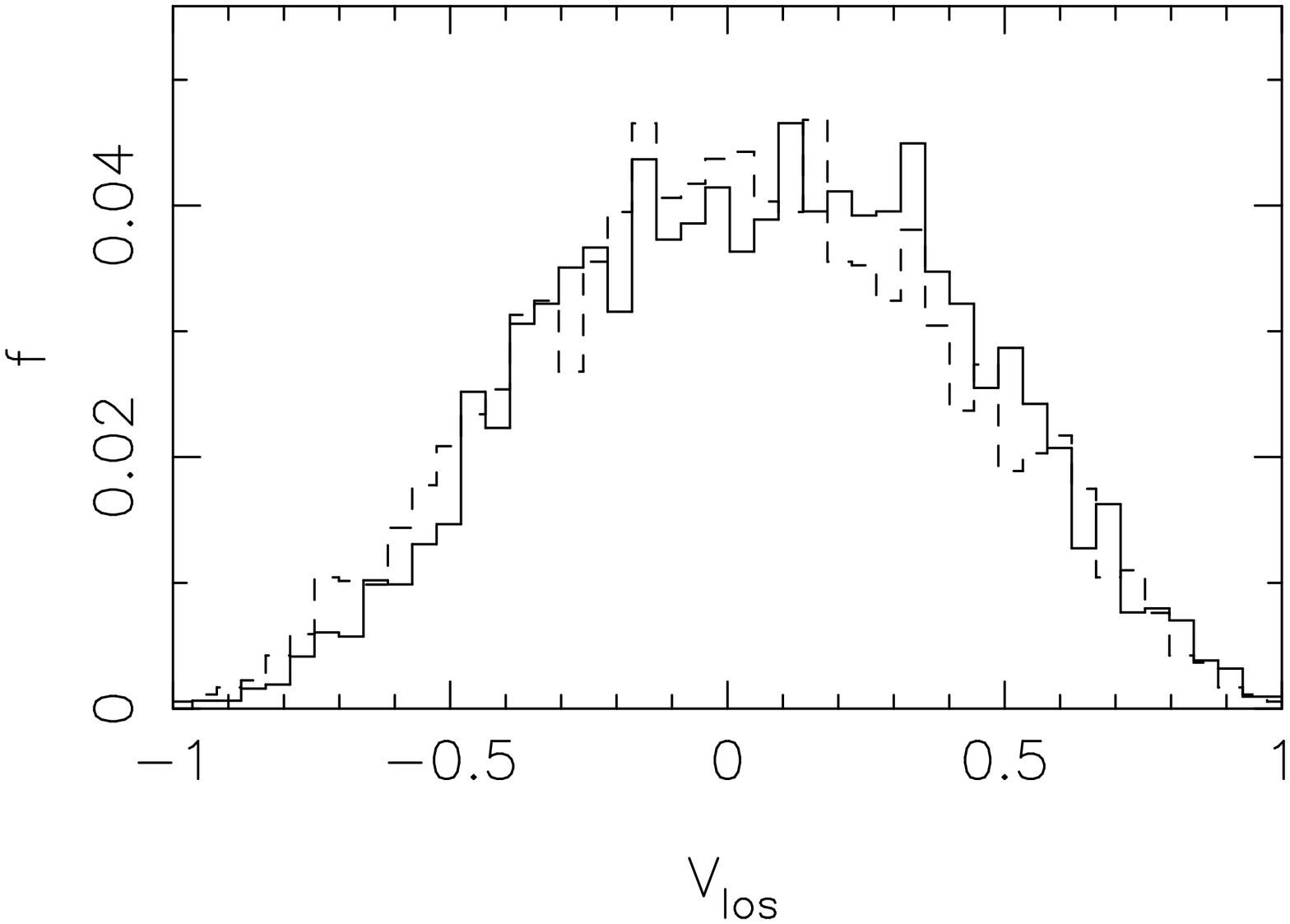}&
    \includegraphics[width=0.65in, angle=-90 ]{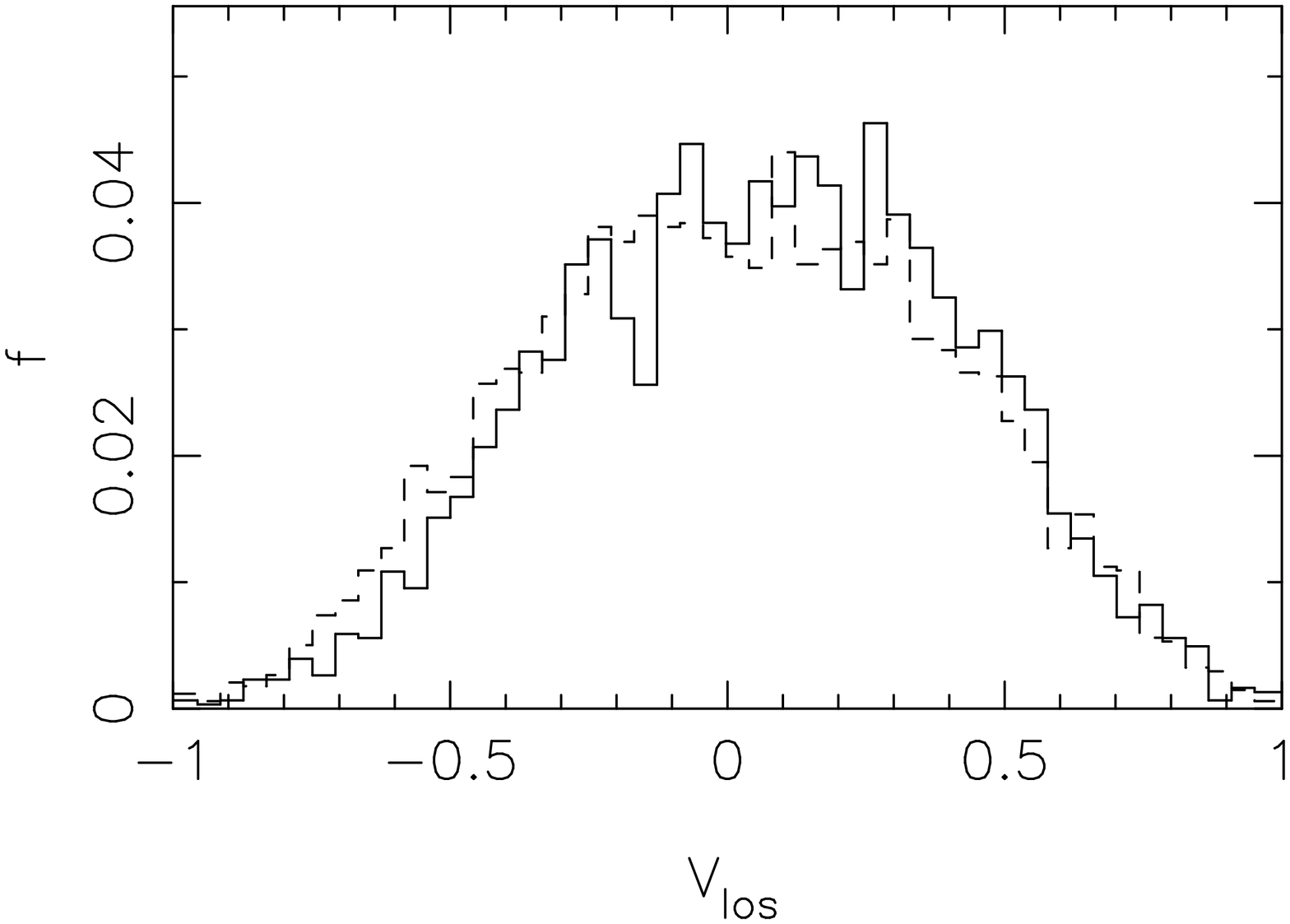}\\
    R5 
    \includegraphics[width=0.65in, angle=-90 ]{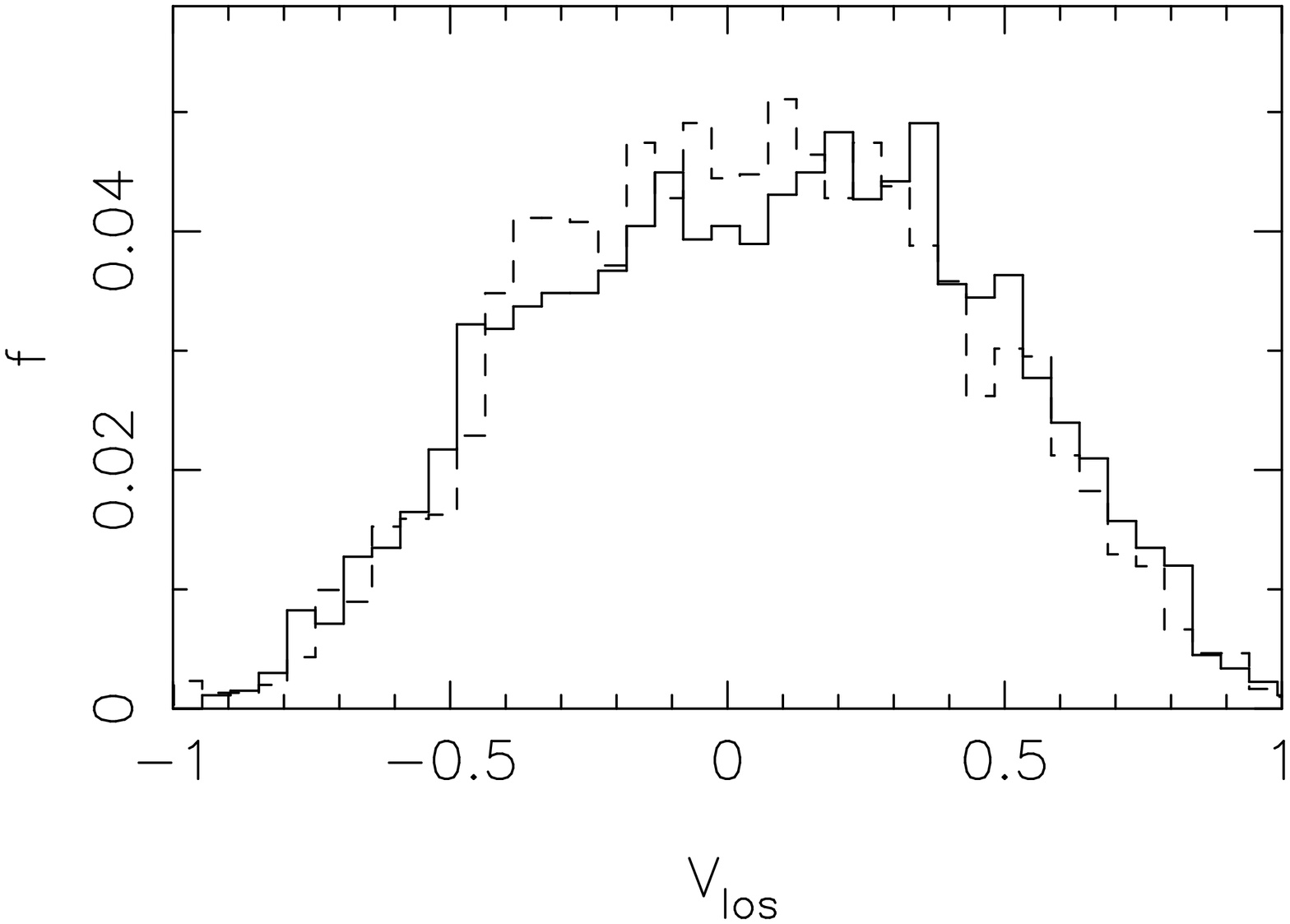}&
    \includegraphics[width=0.65in, angle=-90 ]{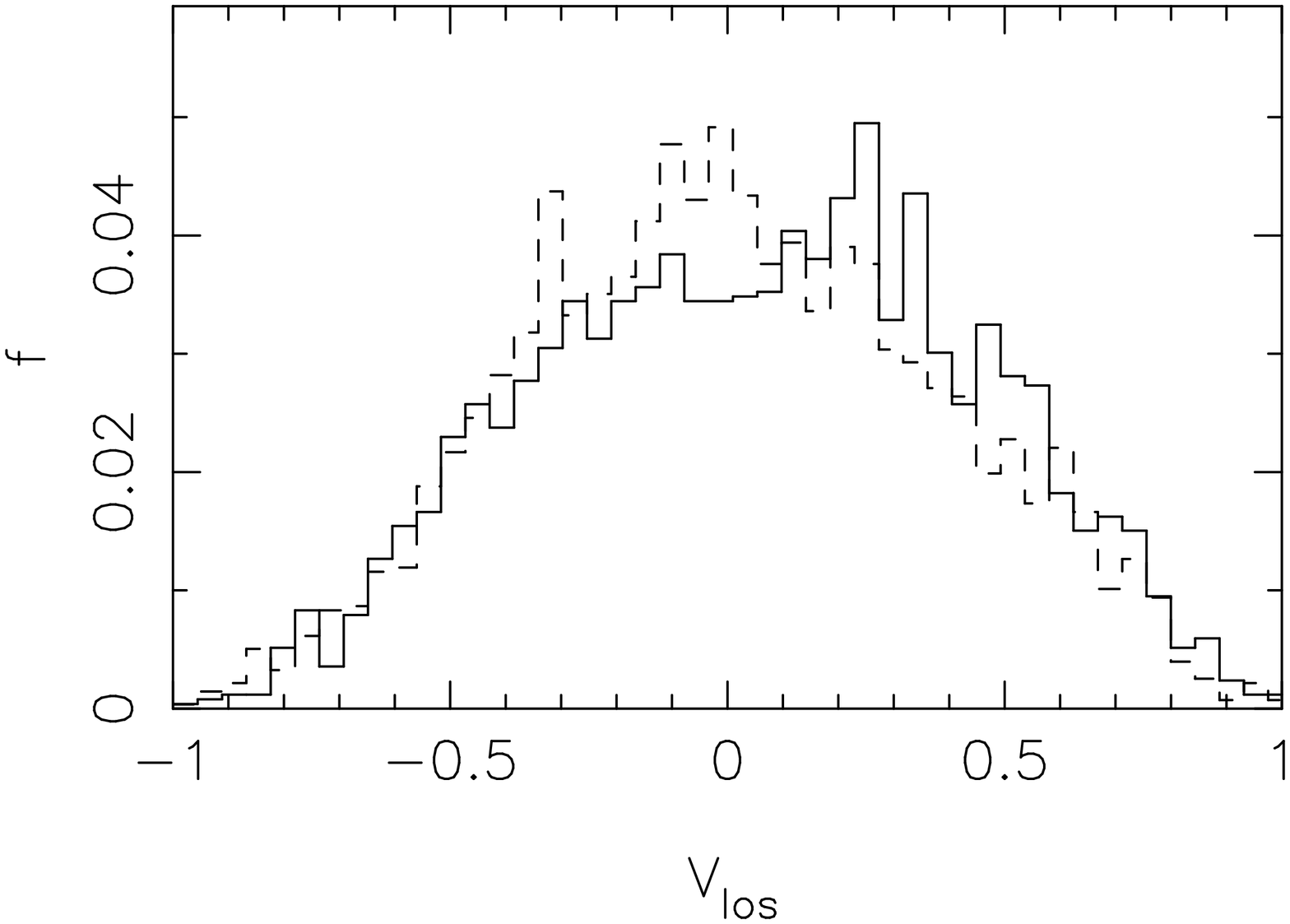}&
    \includegraphics[width=0.65in, angle=-90 ]{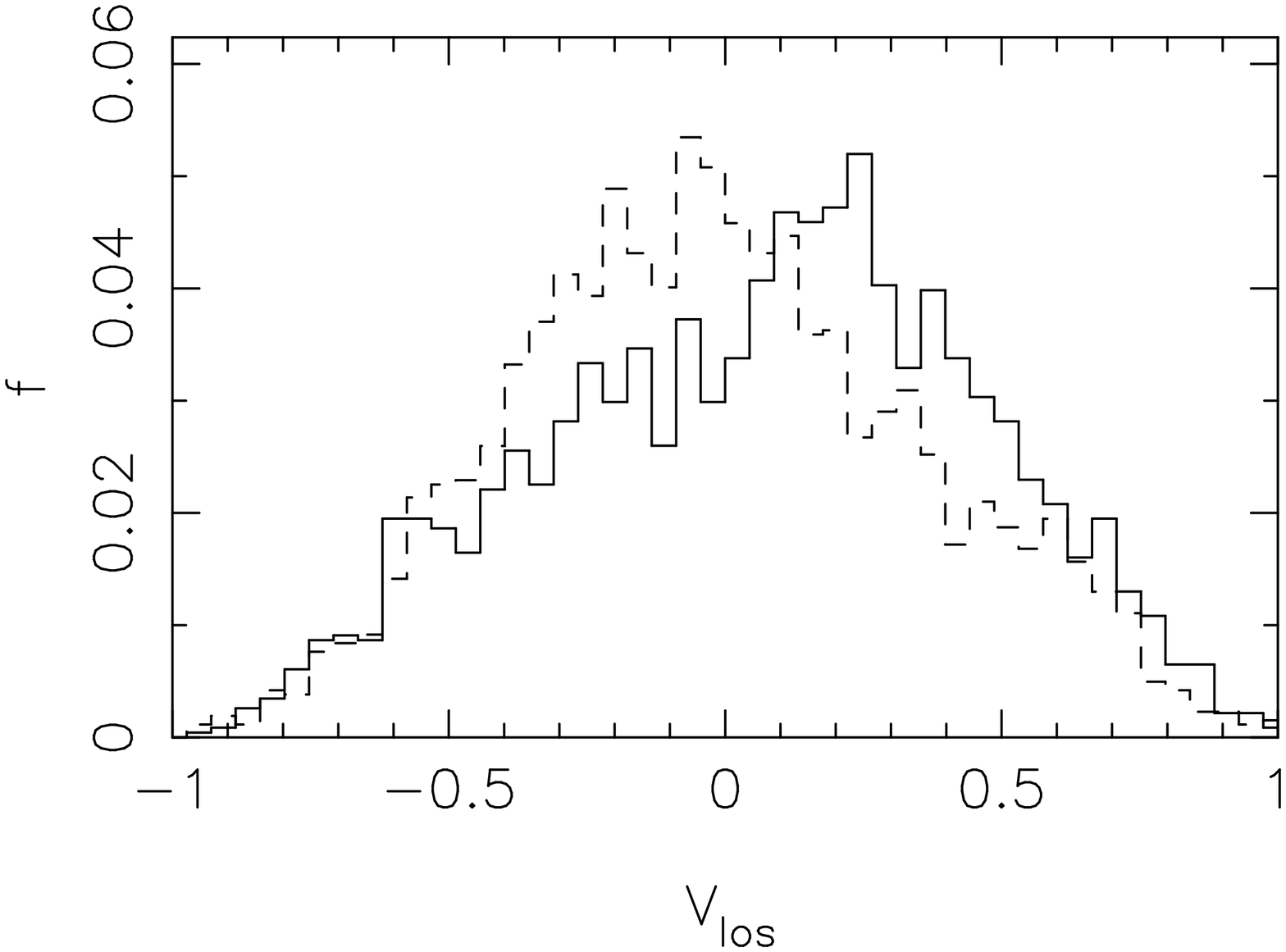}\\
    R2 
    \includegraphics[width=0.65in, angle=-90 ]{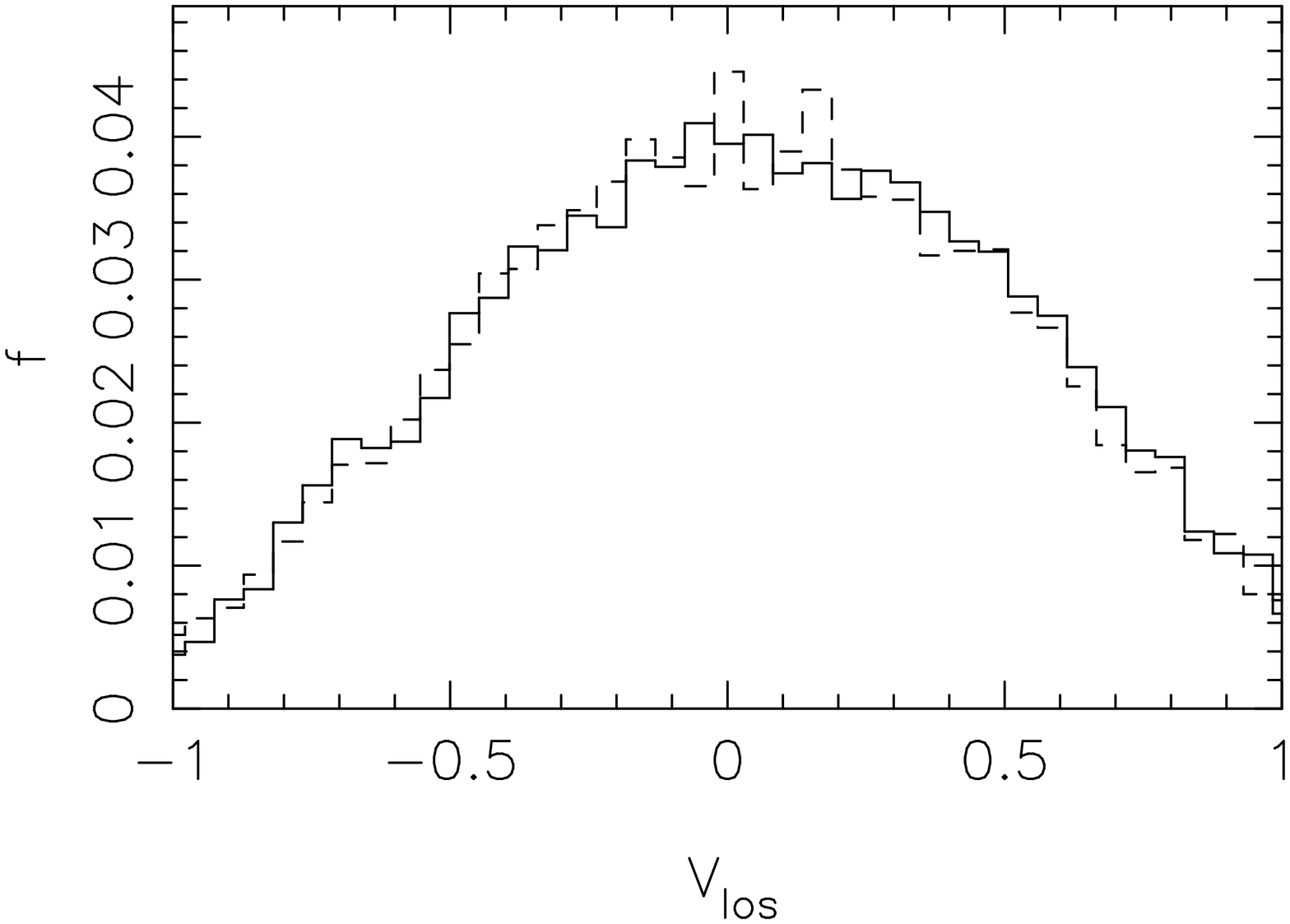}&
    \includegraphics[width=0.65in, angle=-90 ]{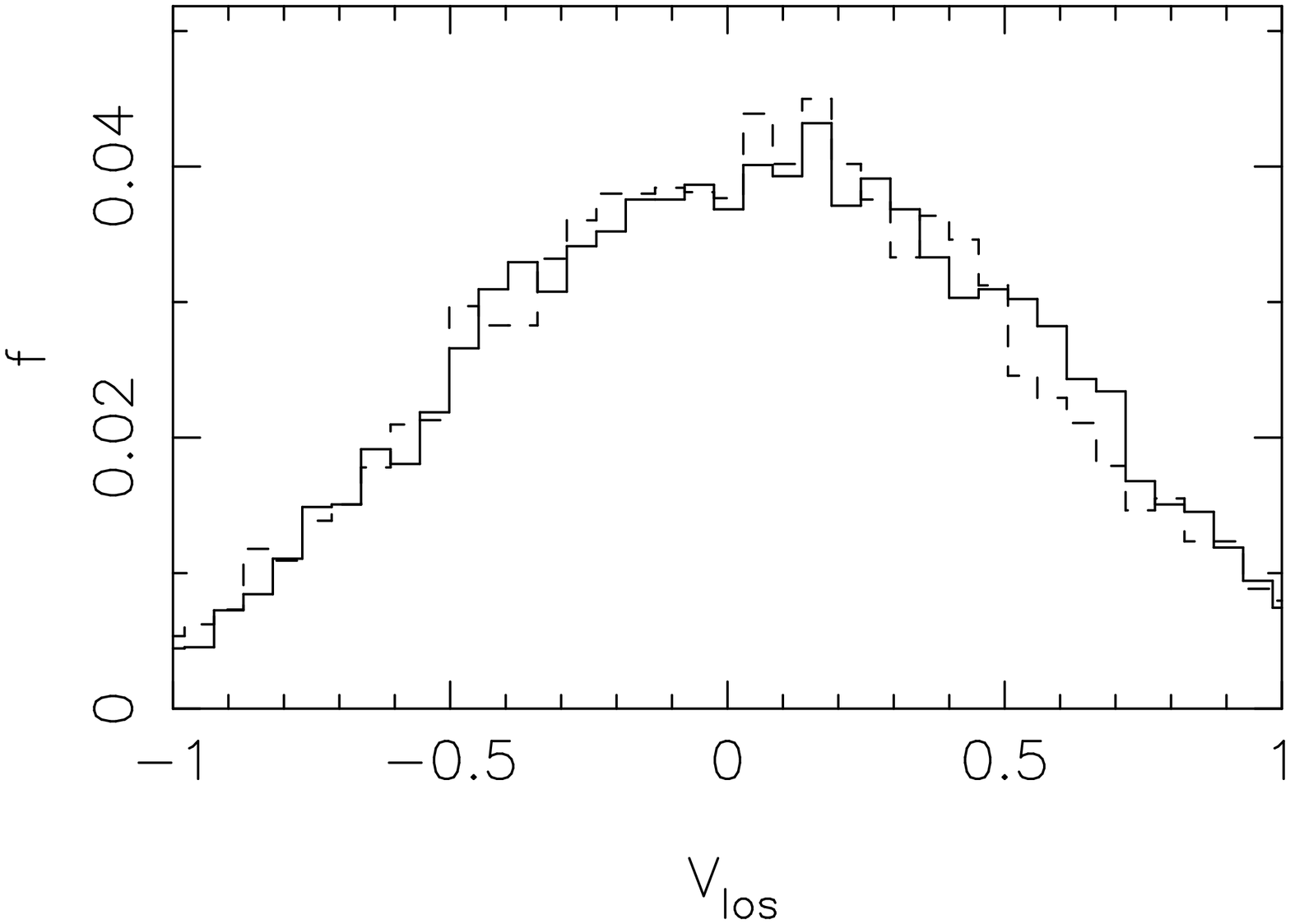}&
    \includegraphics[width=0.65in, angle=-90 ]{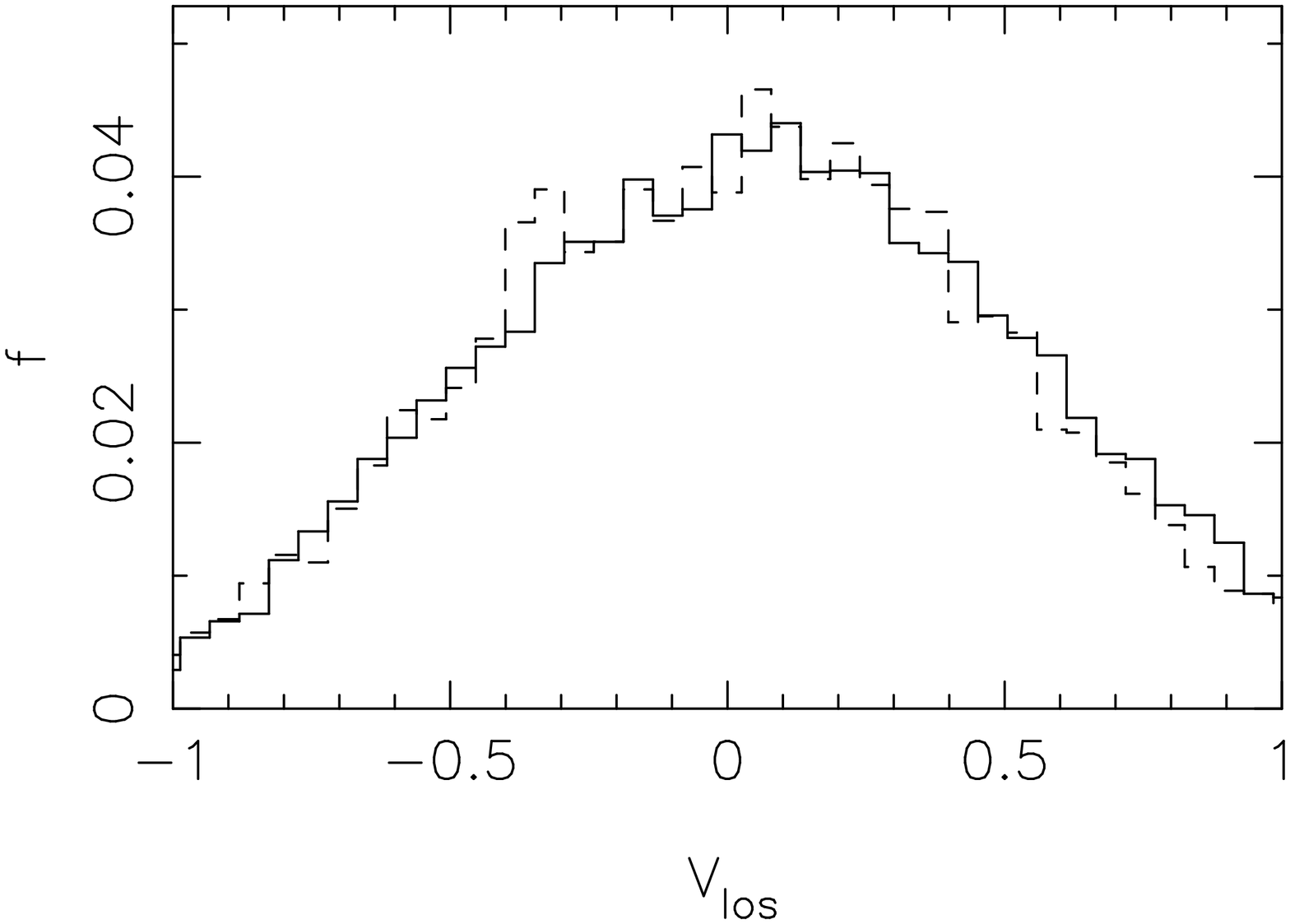}\\
    R6 
    \includegraphics[width=0.65in, angle=-90 ]{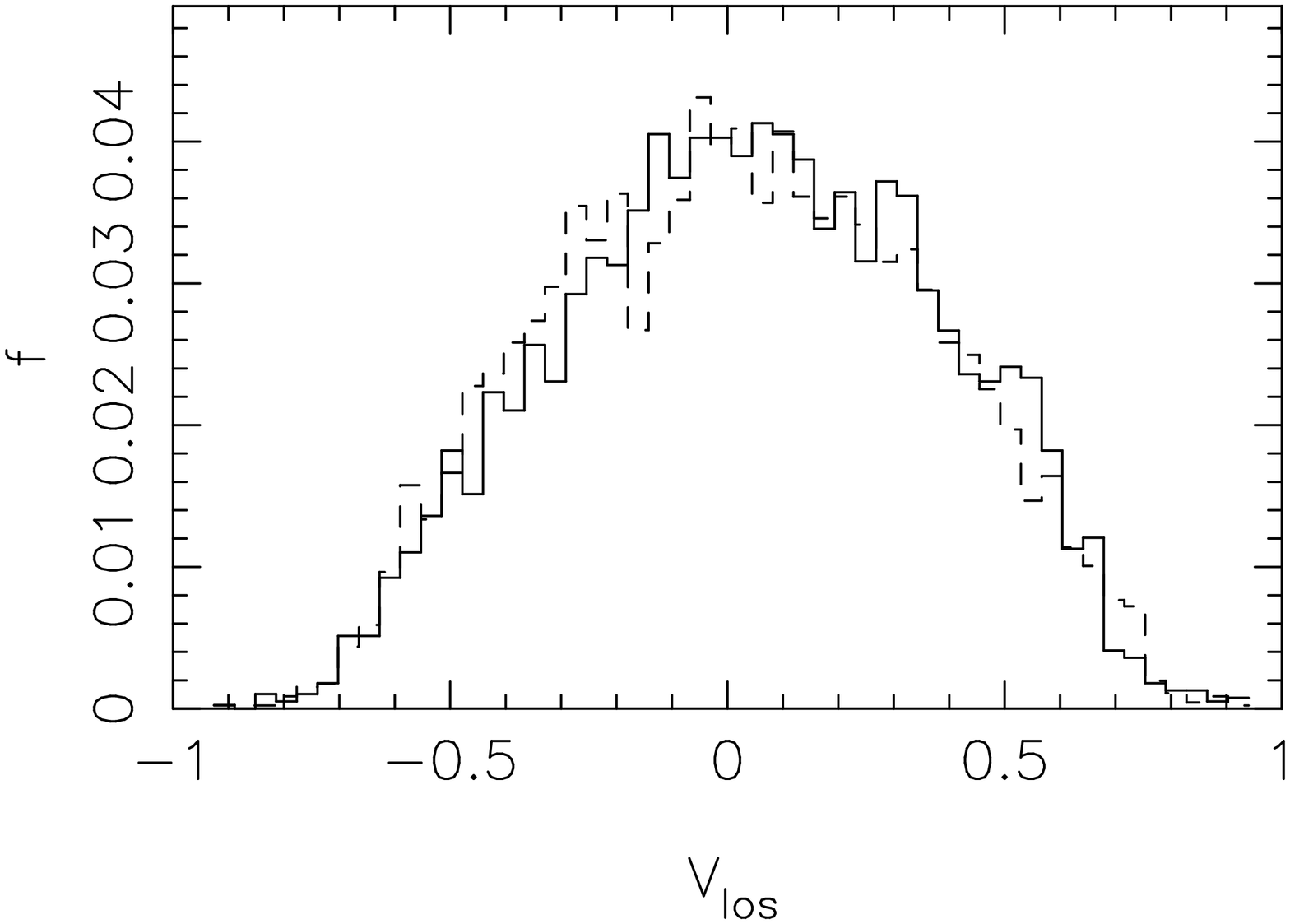}&
    \includegraphics[width=0.65in, angle=-90 ]{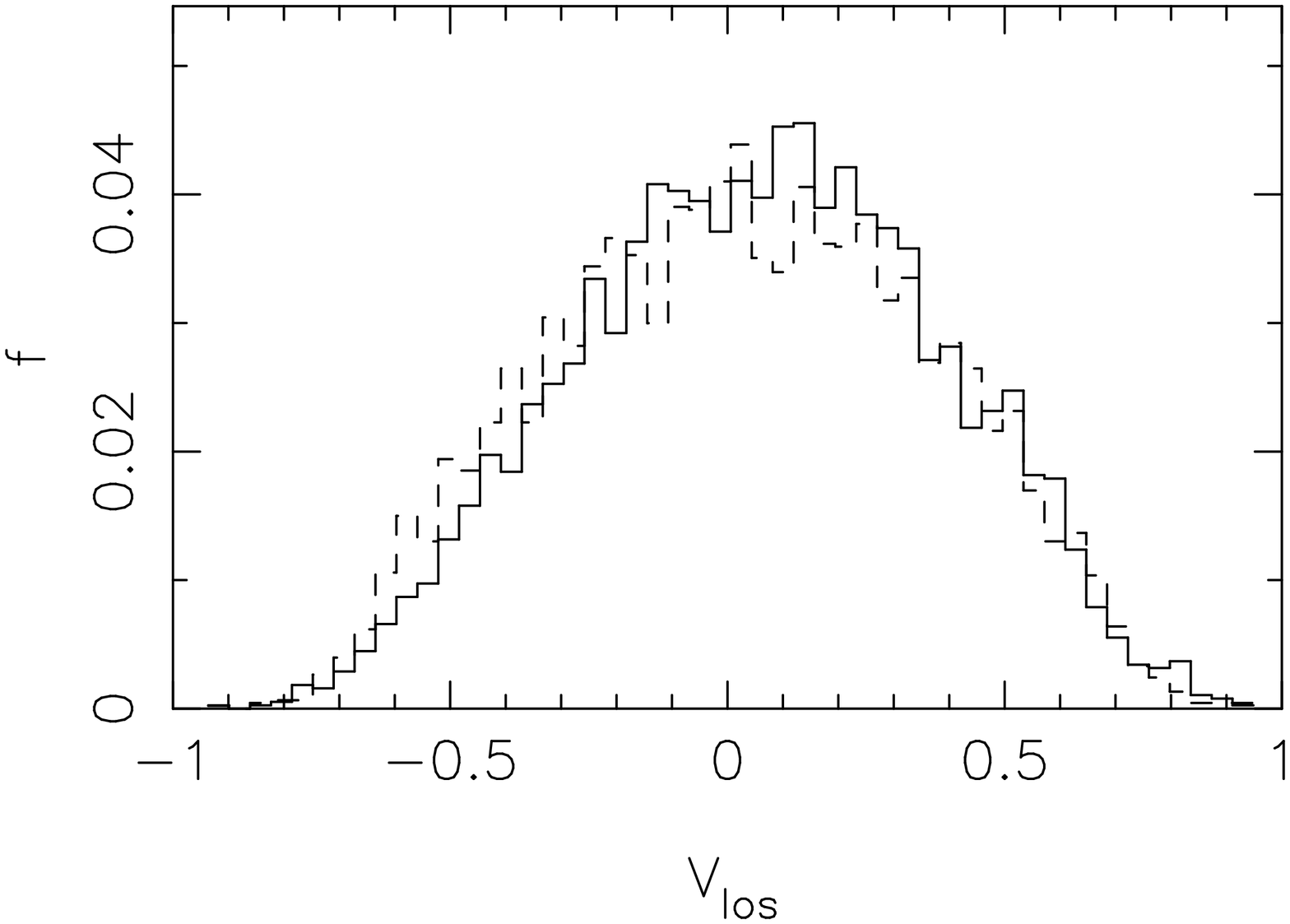}&
    \includegraphics[width=0.65in, angle=-90 ]{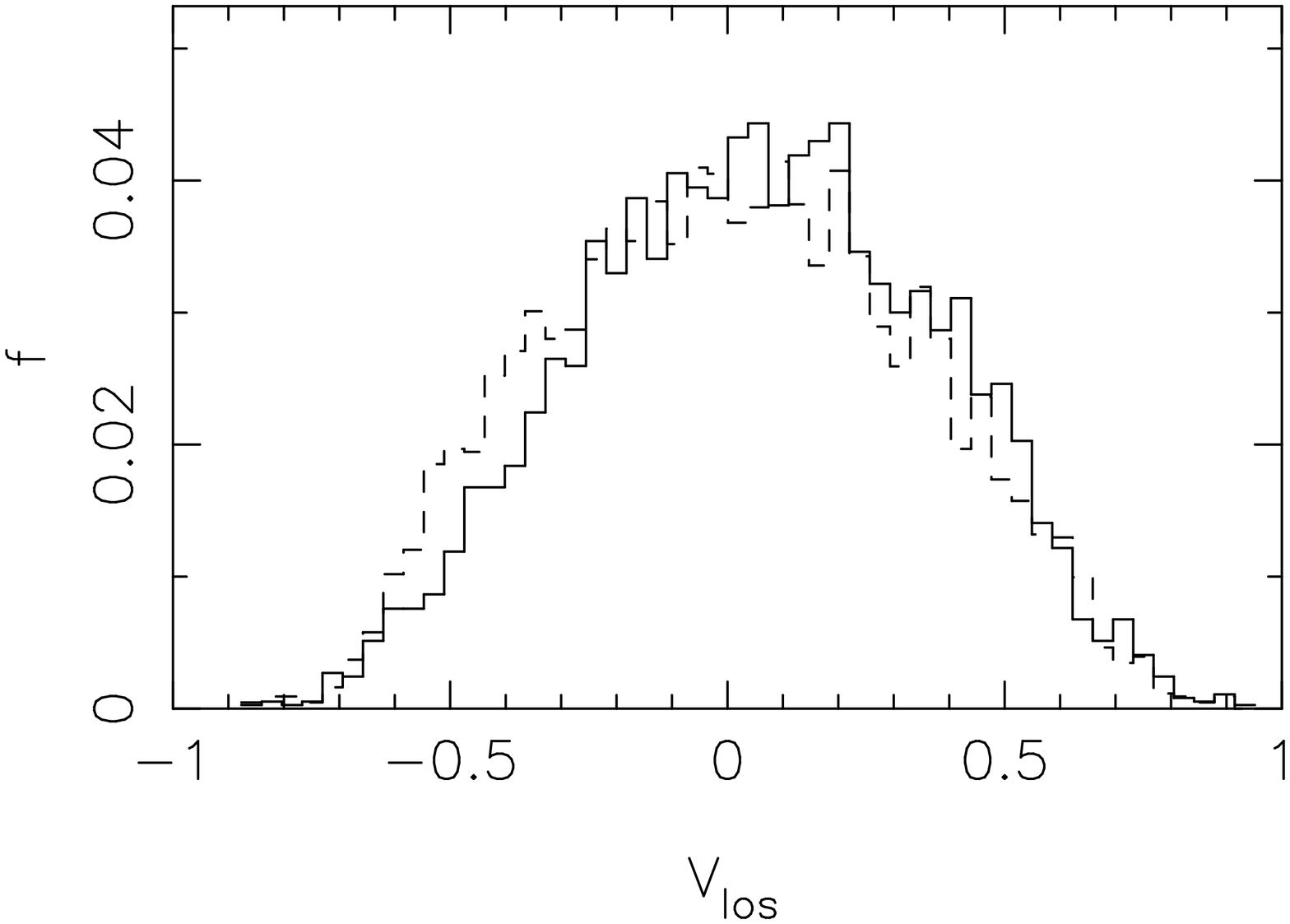}\\

\end{tabular}

\caption{Distribution of line-of-sight velocities V$_{los}$, at
($l,b$) = (0\degr,$-$6\degr), for varying bar angles in the six
simulations.  The solid line, in each panel, is the far side of the
model, while the dashed line is the near side of the model. The models
that exhibit an X-shape show near/far asymmetry, at all bar
angles. Those without an X-shape do not exhibit comparable asymmetry.}
\label{Avr}
\end{figure}

\begin{figure}

\begin{tabular}{lccc}
  Angle:  15$^\circ$ & 25$^\circ$ & 35$^\circ$\\
    R1 
    \includegraphics[width=0.65in, angle=-90 ]{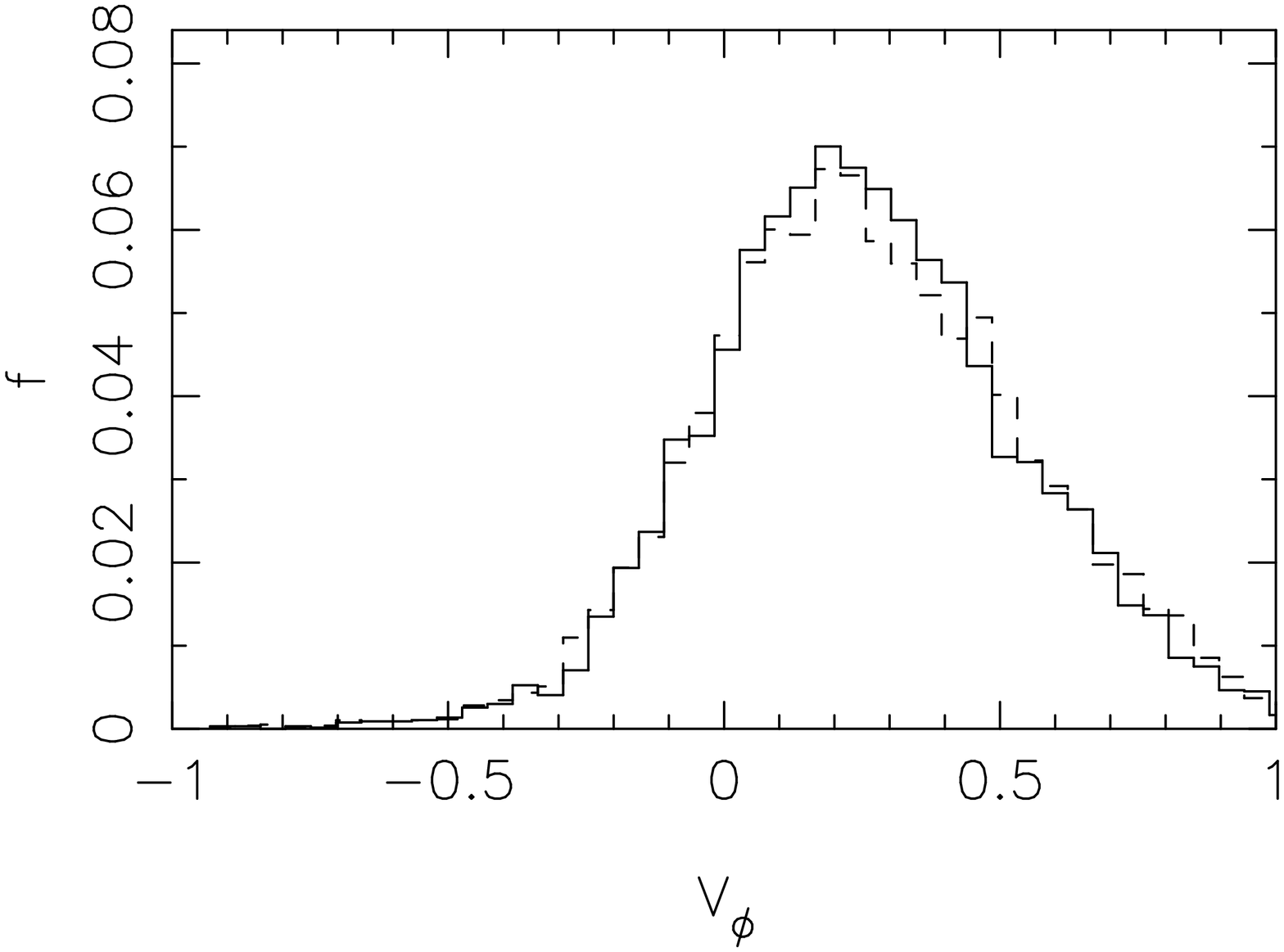}&
    \includegraphics[width=0.65in, angle=-90 ]{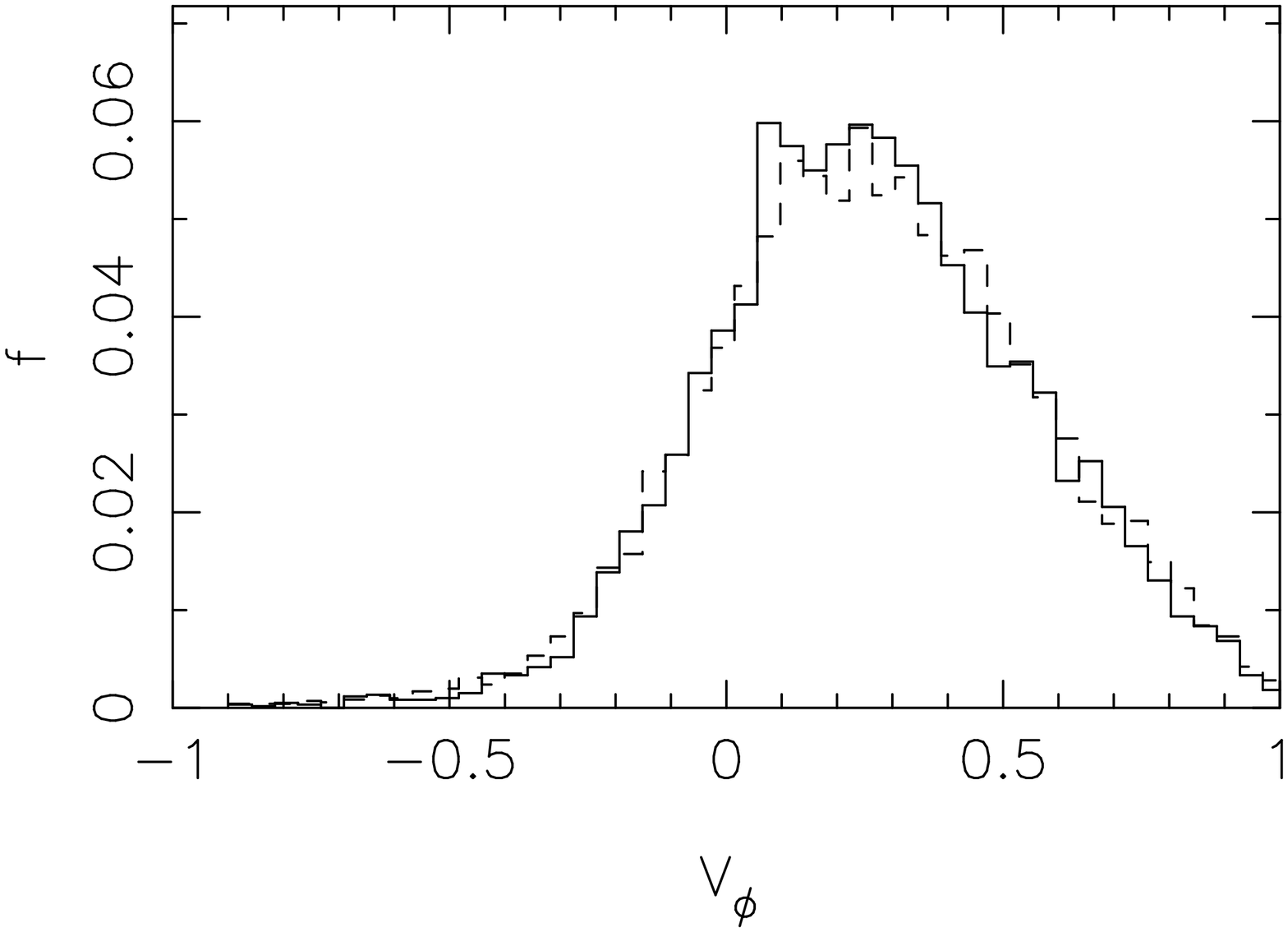}&
    \includegraphics[width=0.65in, angle=-90 ]{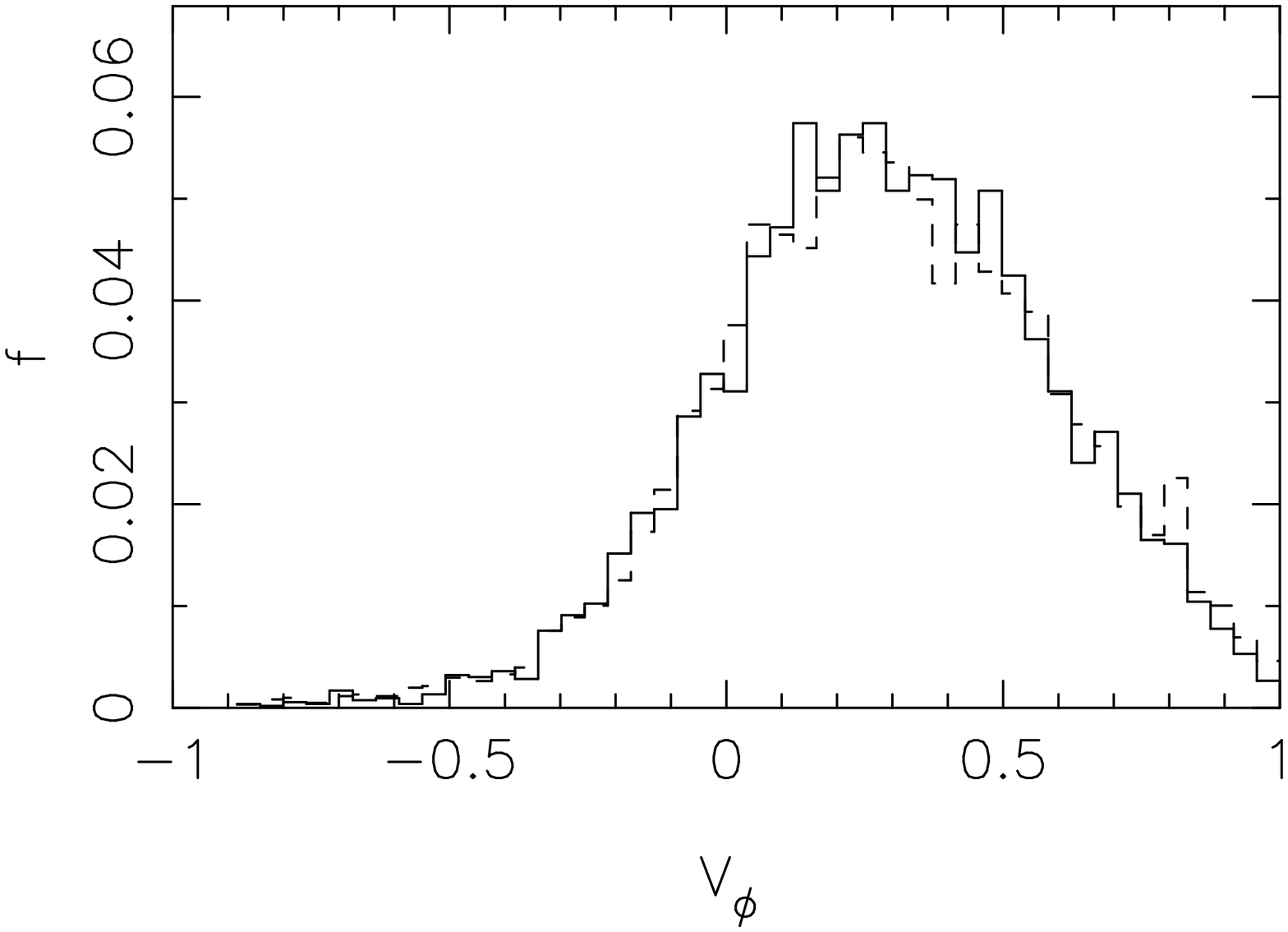}\\
    B3    
    \includegraphics[width=0.65in, angle=-90 ]{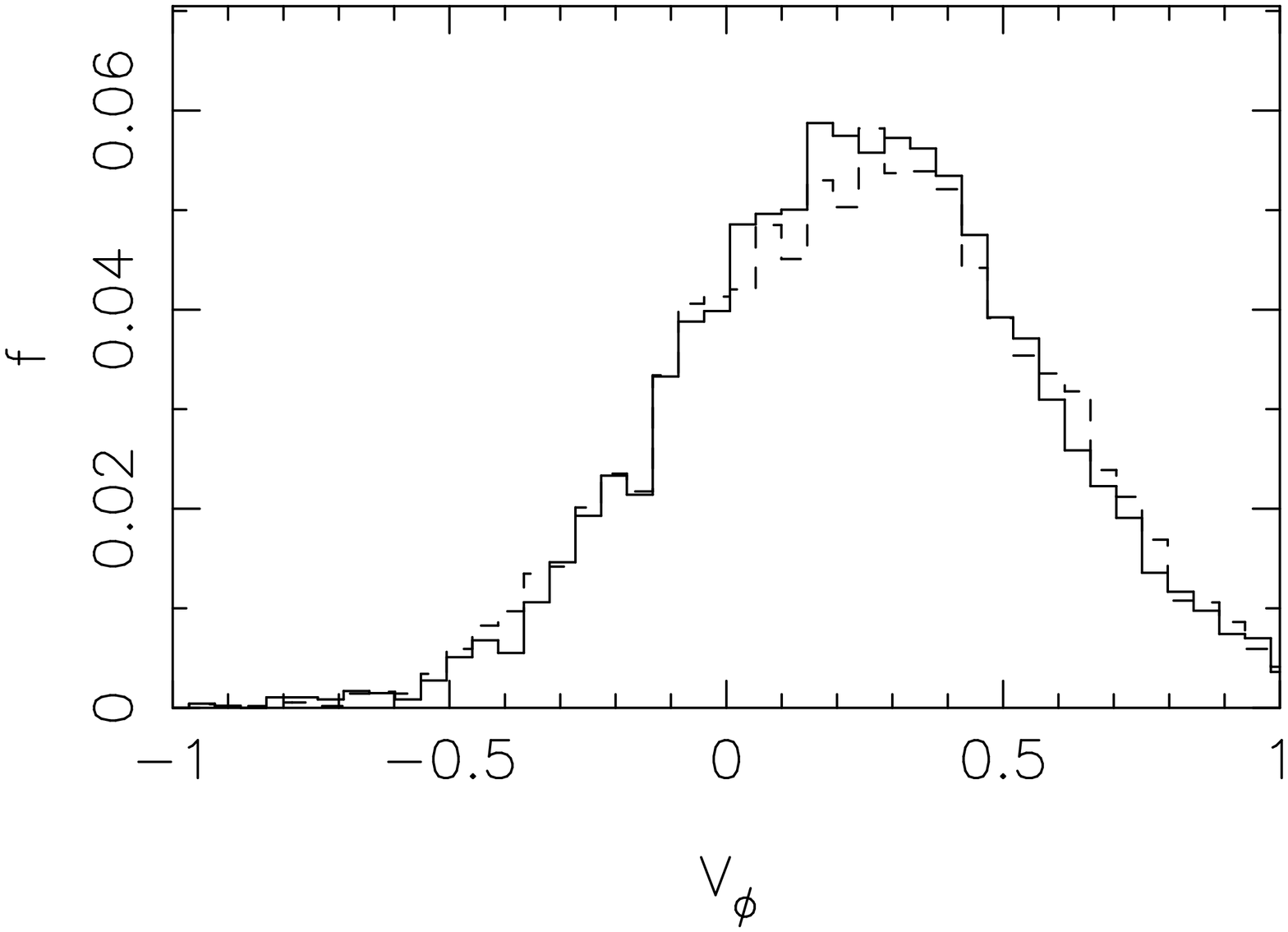}&
    \includegraphics[width=0.65in, angle=-90 ]{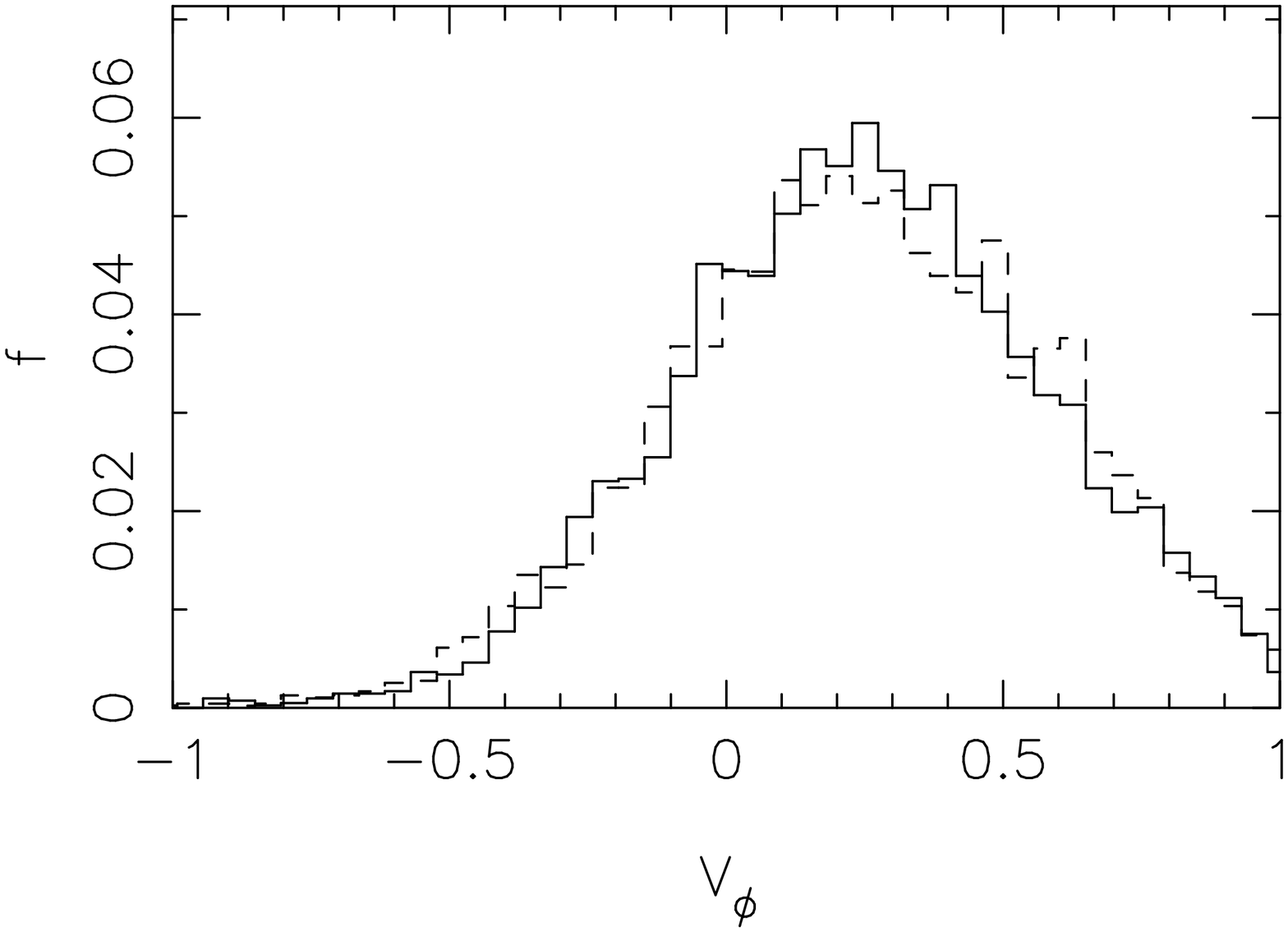}&
    \includegraphics[width=0.65in, angle=-90 ]{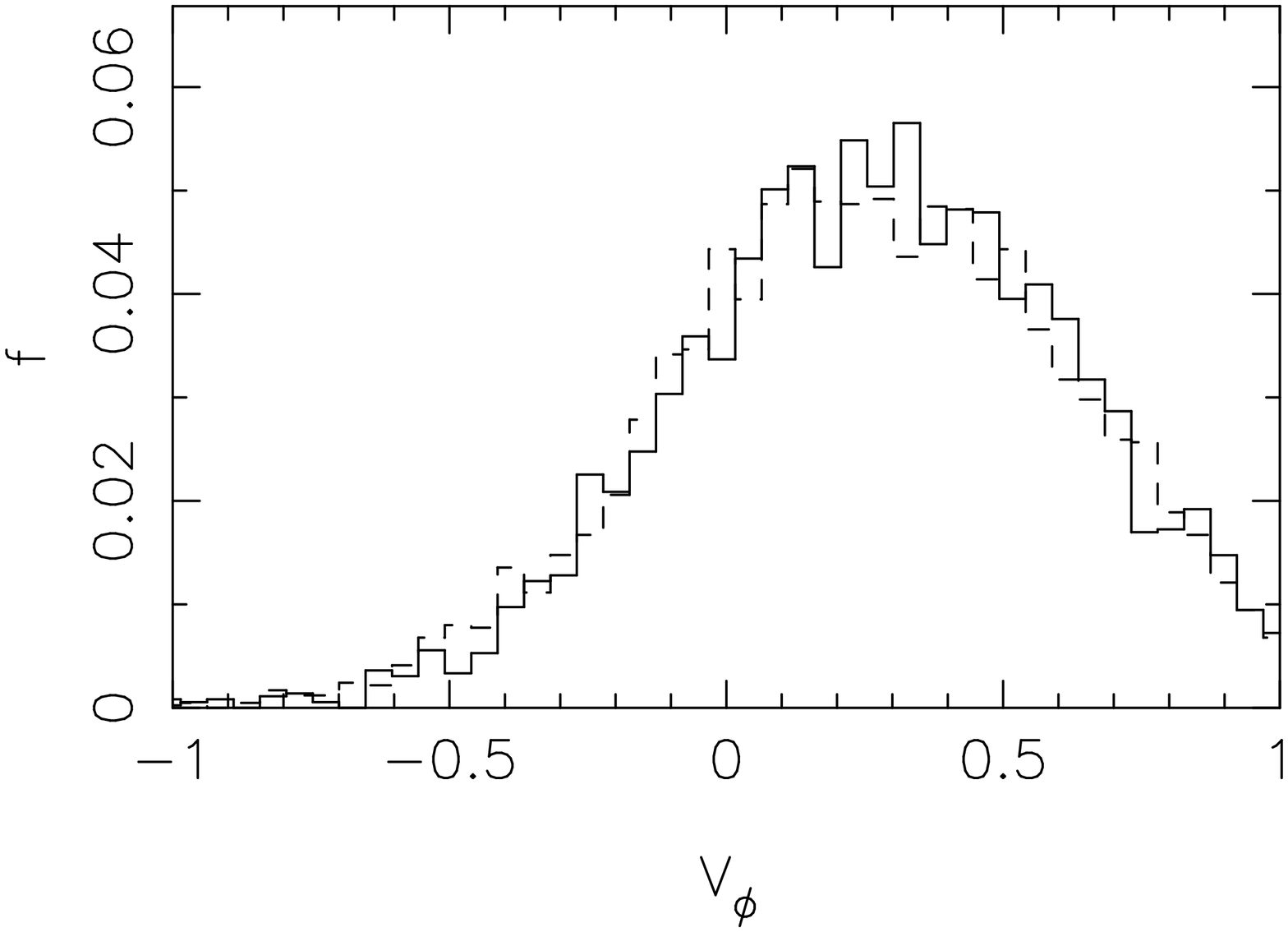}\\
    B2 
    \includegraphics[width=0.65in, angle=-90 ]{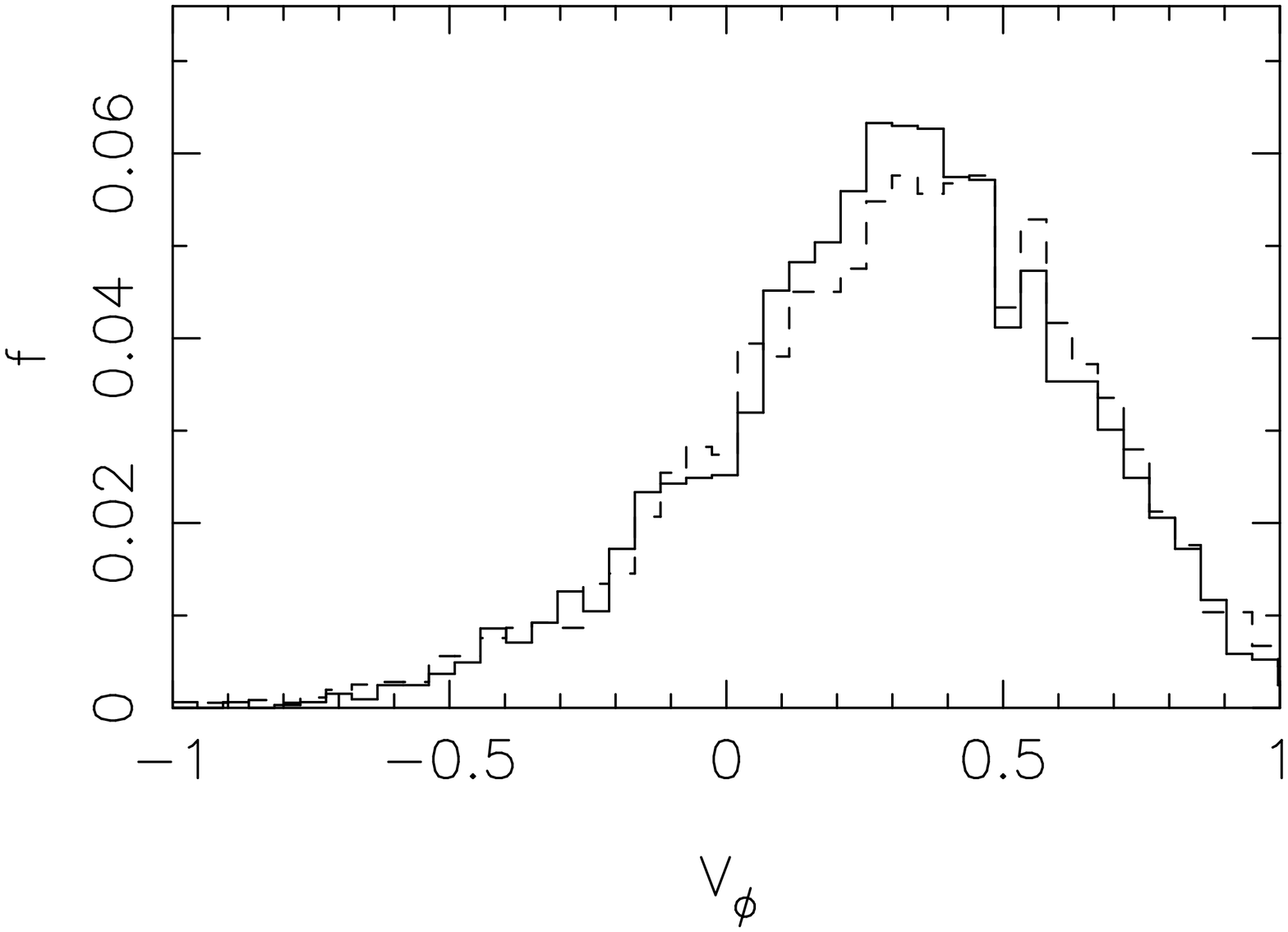}&
    \includegraphics[width=0.65in, angle=-90 ]{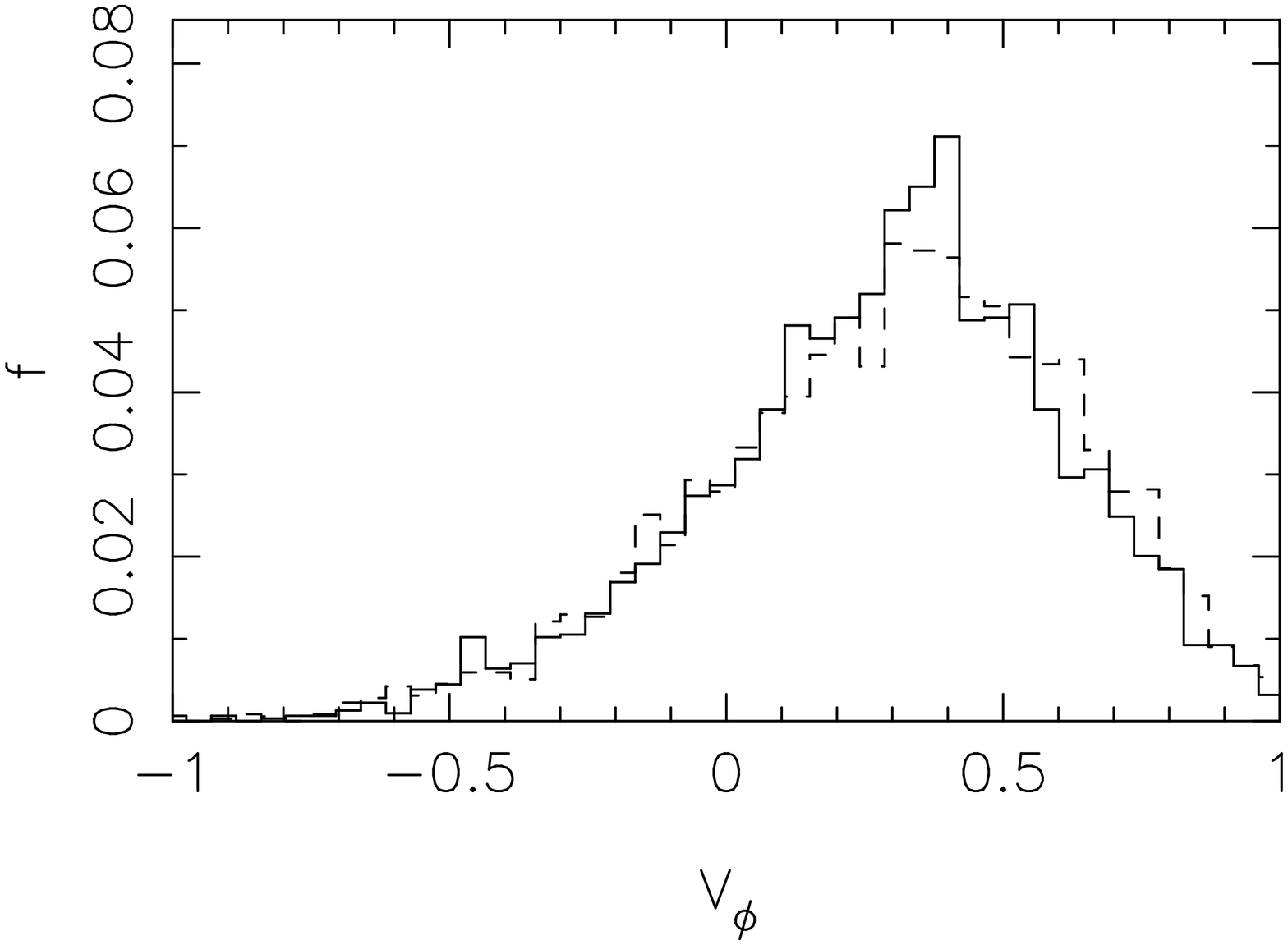}&
    \includegraphics[width=0.65in, angle=-90 ]{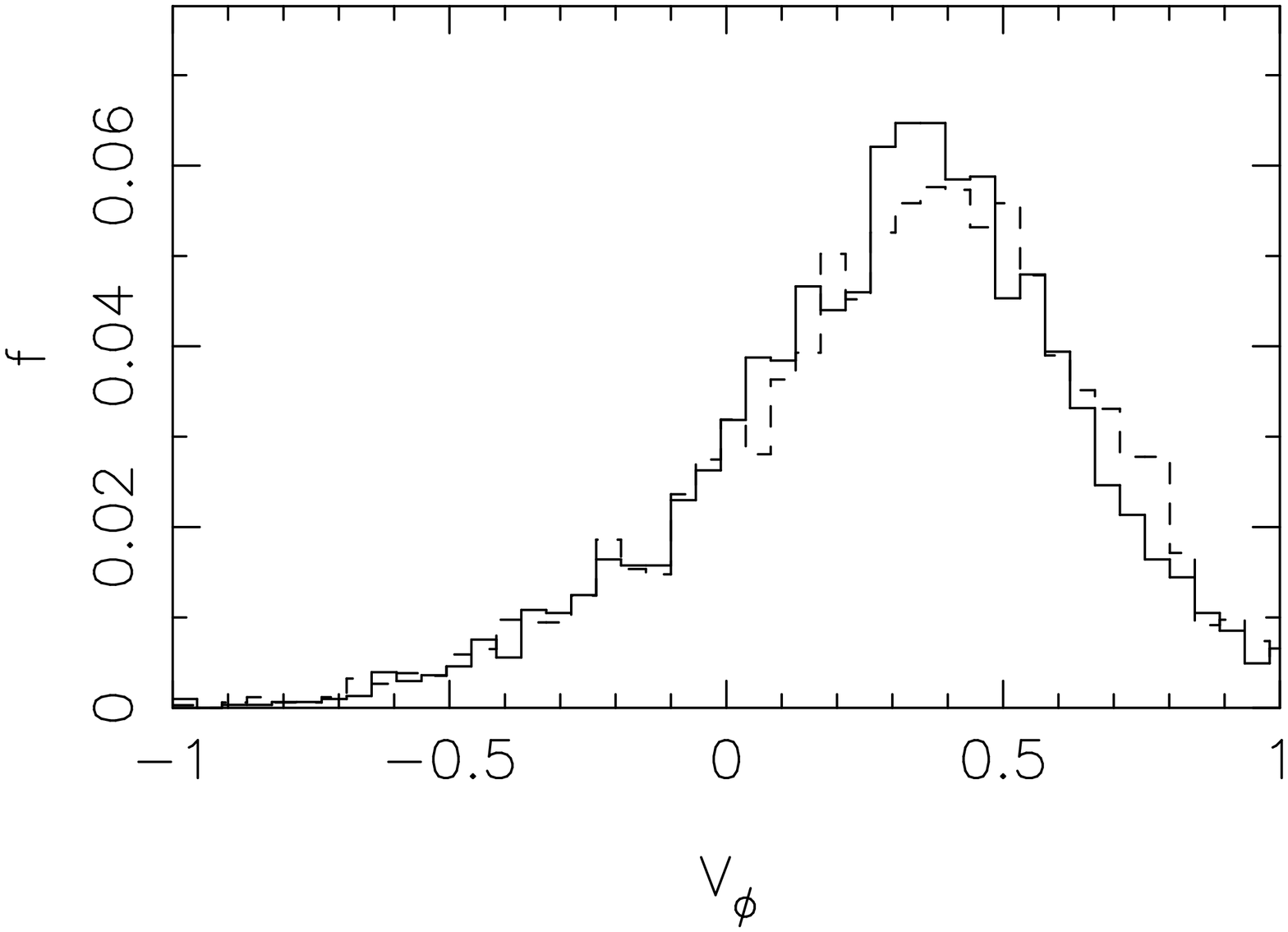}\\
    R5 
    \includegraphics[width=0.65in, angle=-90 ]{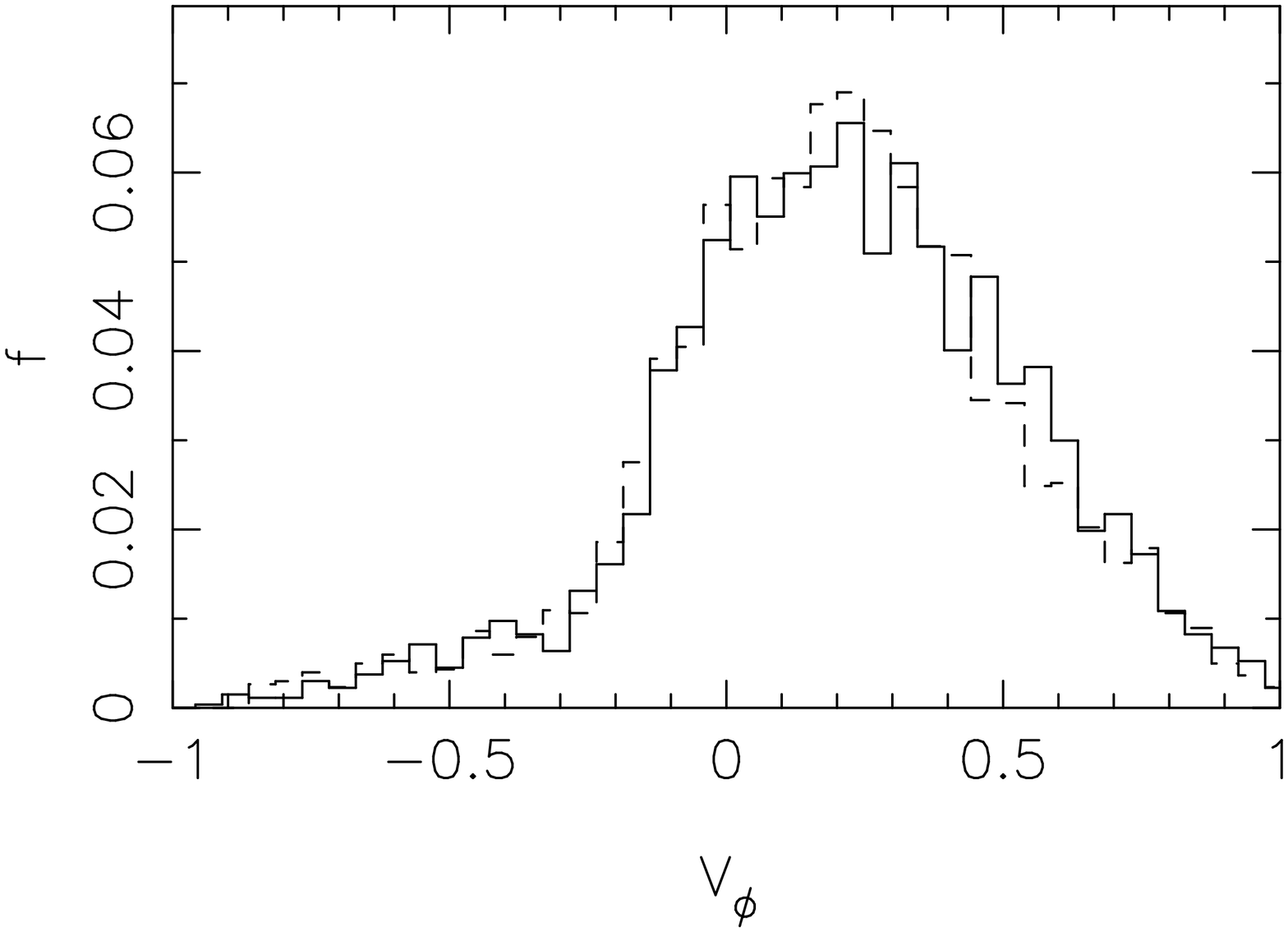}&
    \includegraphics[width=0.65in, angle=-90 ]{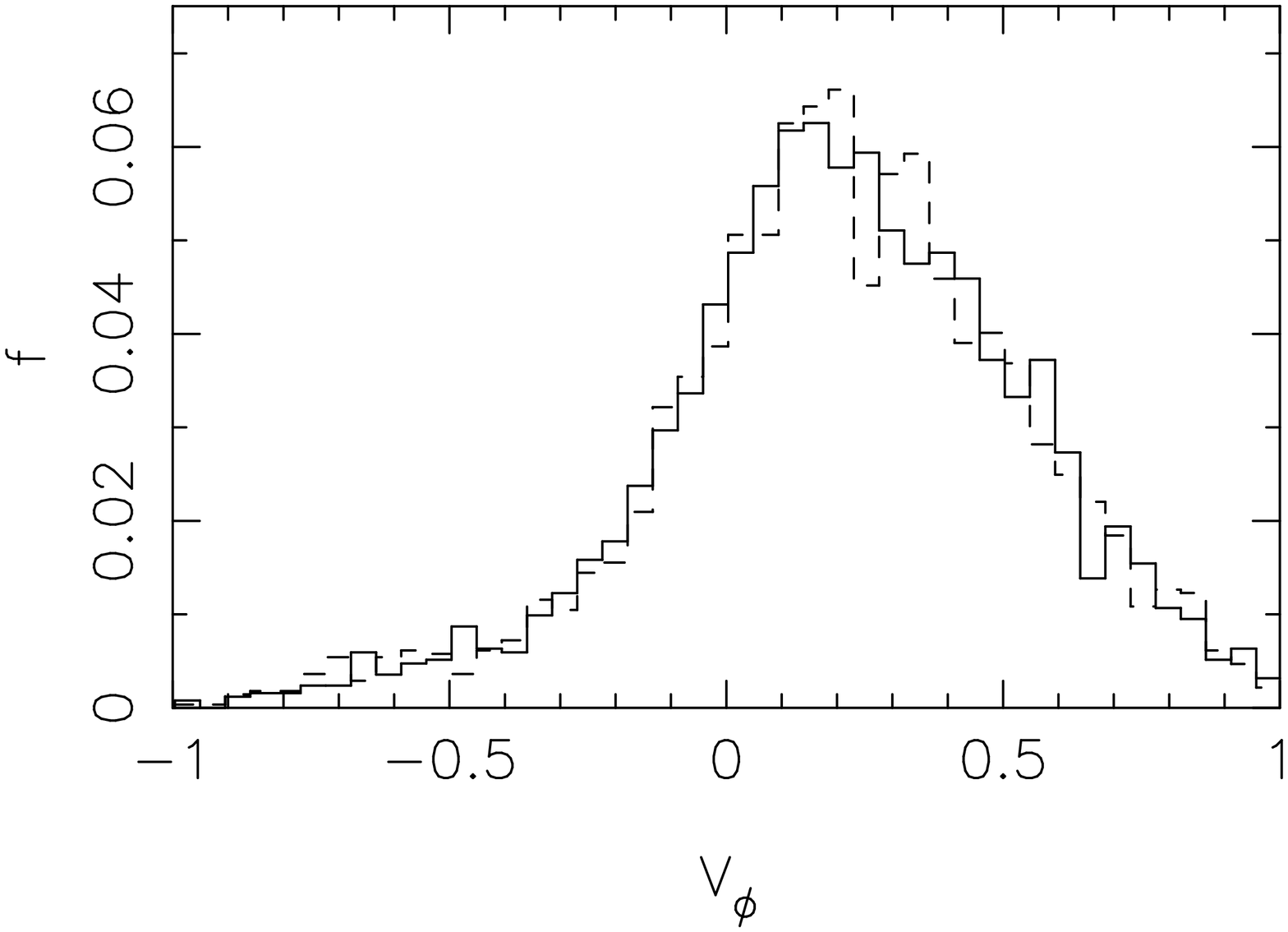}&
    \includegraphics[width=0.65in, angle=-90 ]{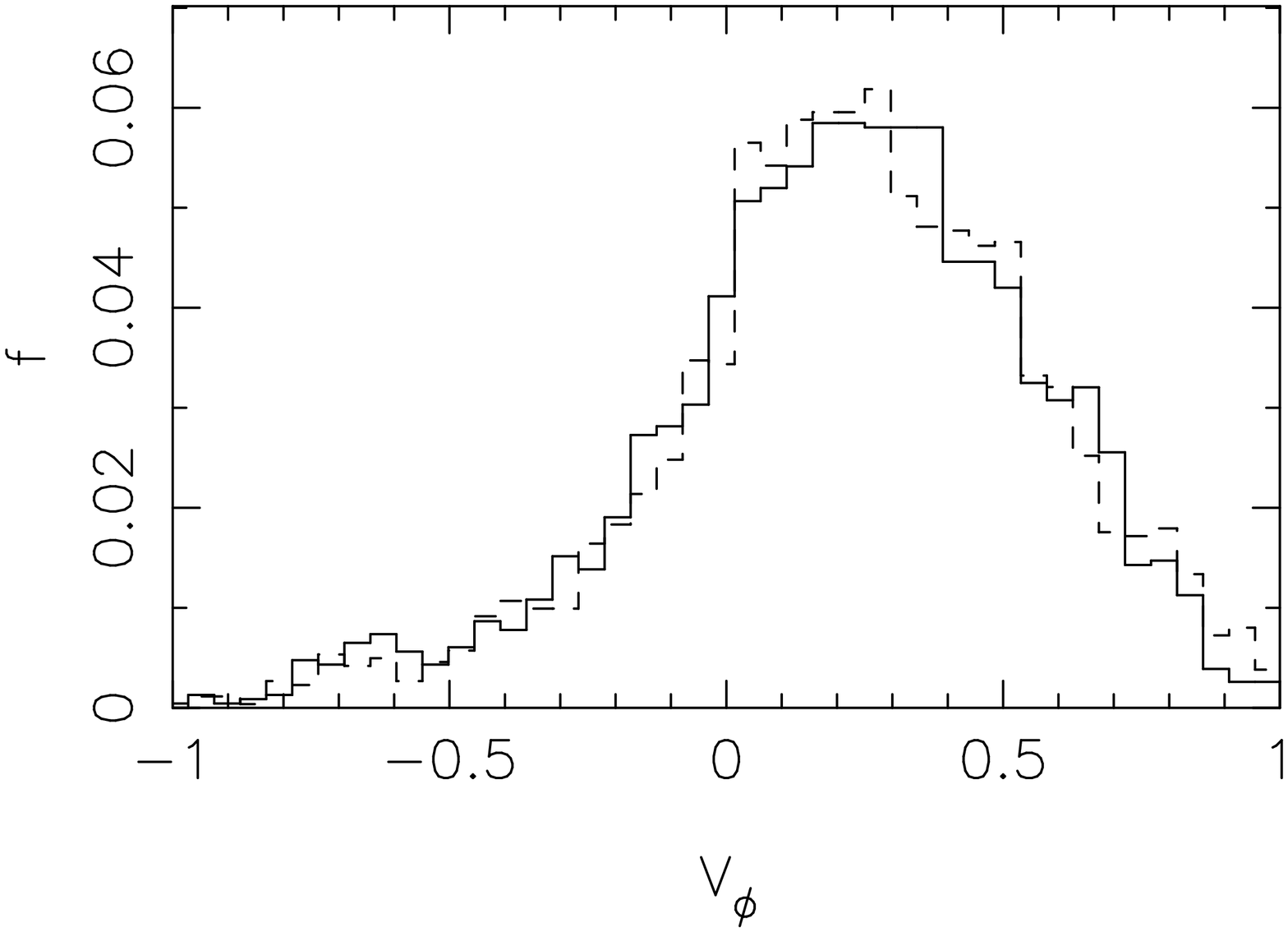}\\
    R2 
    \includegraphics[width=0.65in, angle=-90 ]{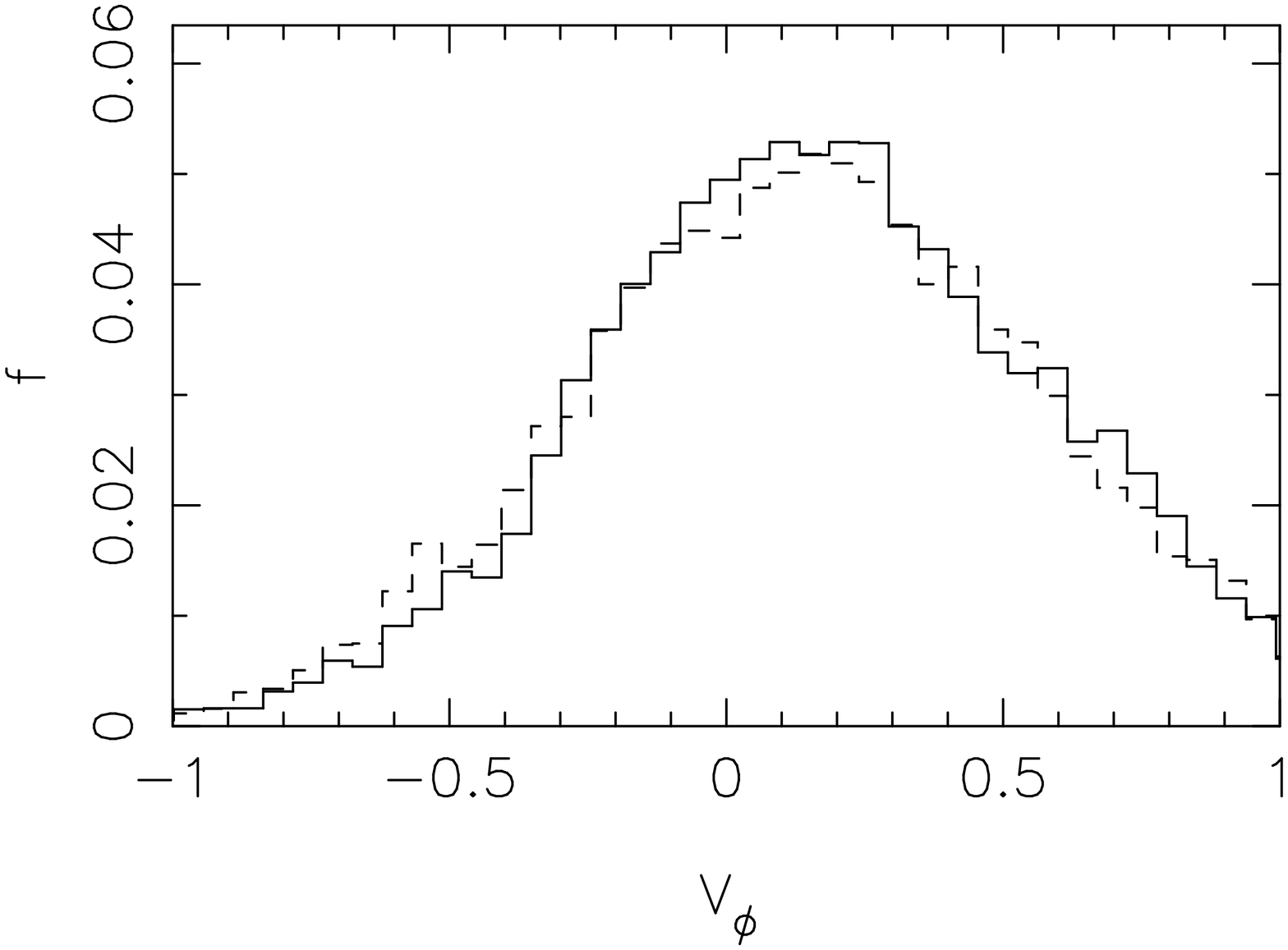}&
    \includegraphics[width=0.65in, angle=-90 ]{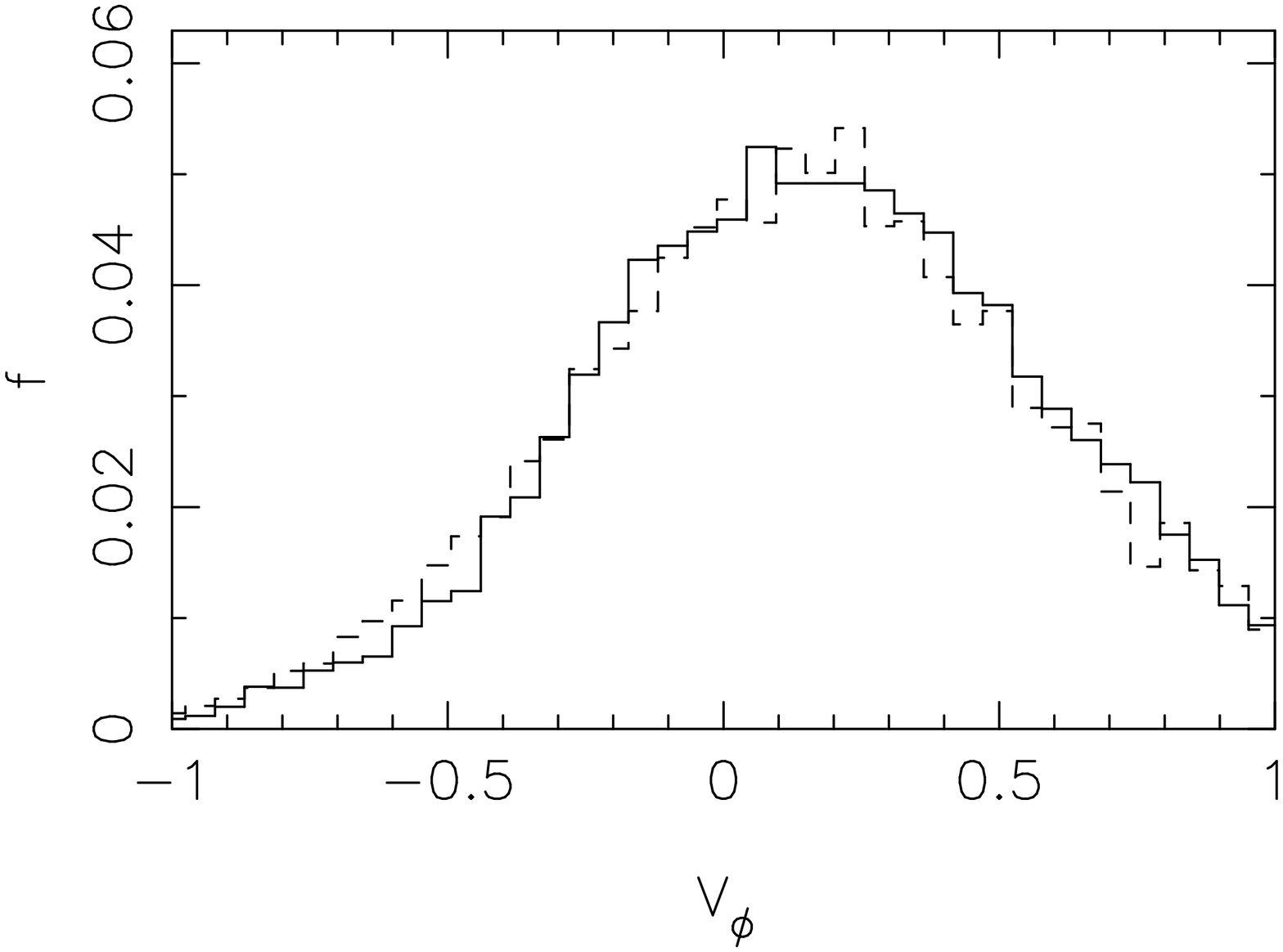}&
    \includegraphics[width=0.65in, angle=-90 ]{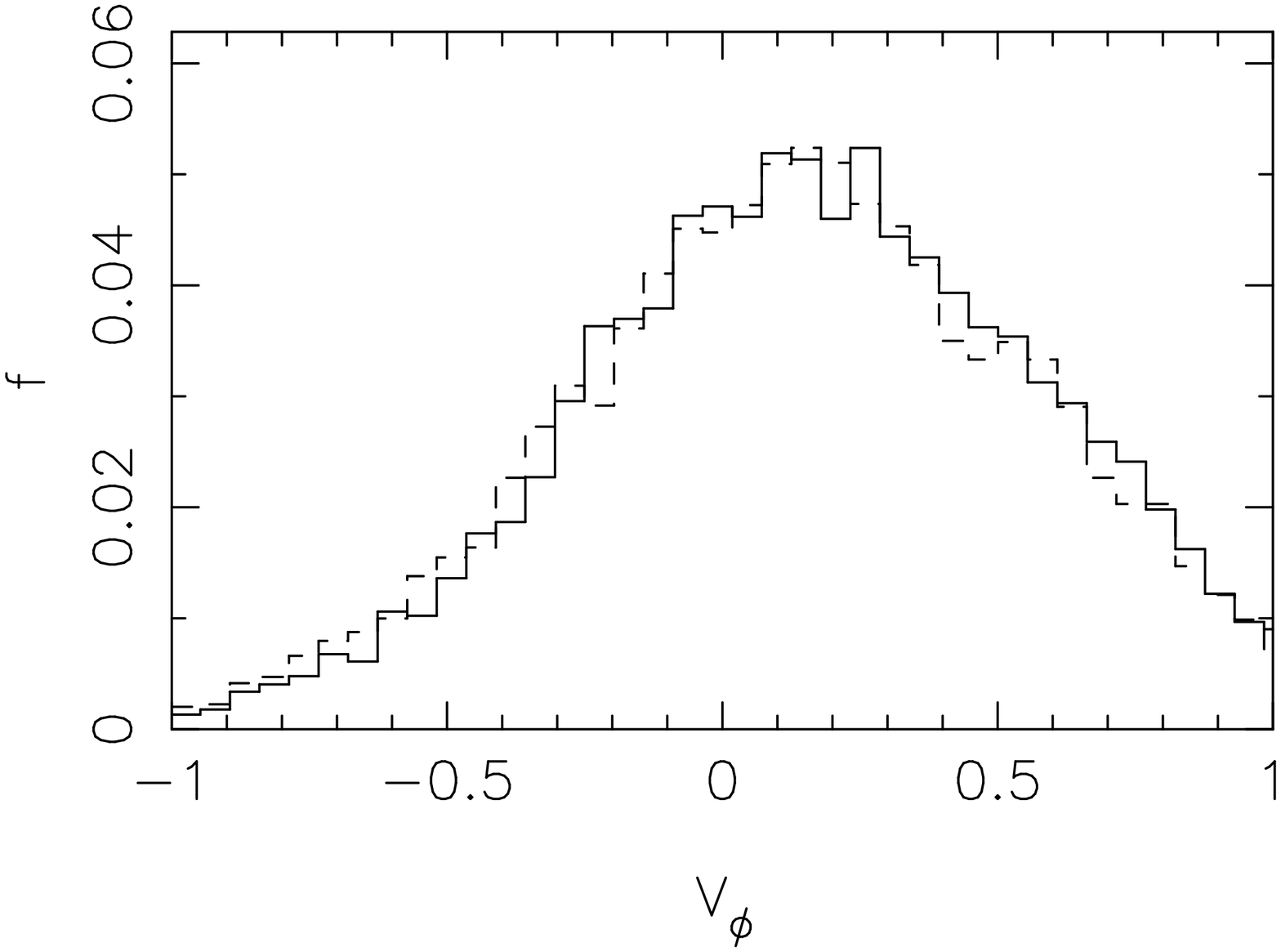}\\
    R6 
    \includegraphics[width=0.65in, angle=-90 ]{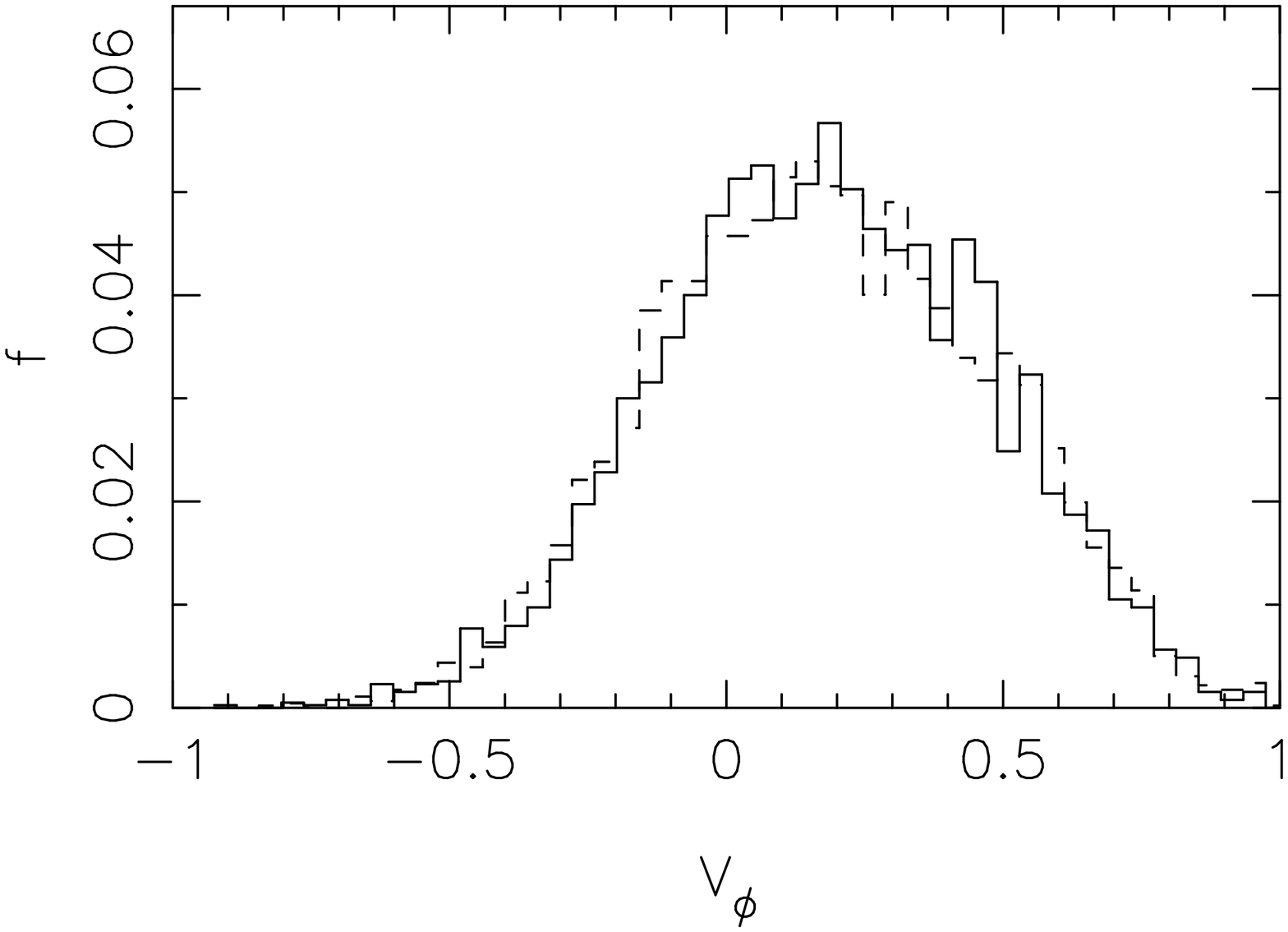}&
    \includegraphics[width=0.65in, angle=-90 ]{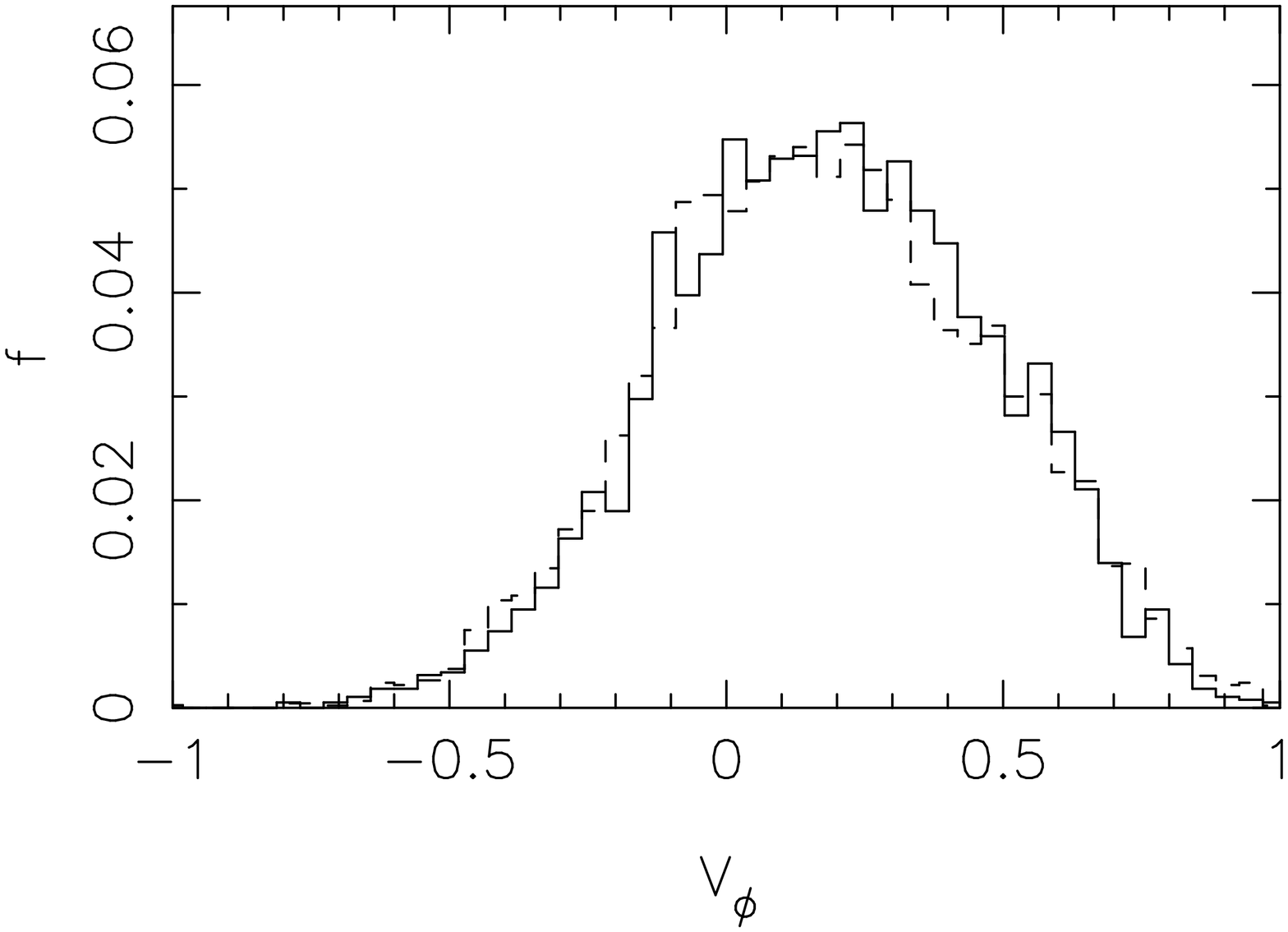}&
    \includegraphics[width=0.65in, angle=-90 ]{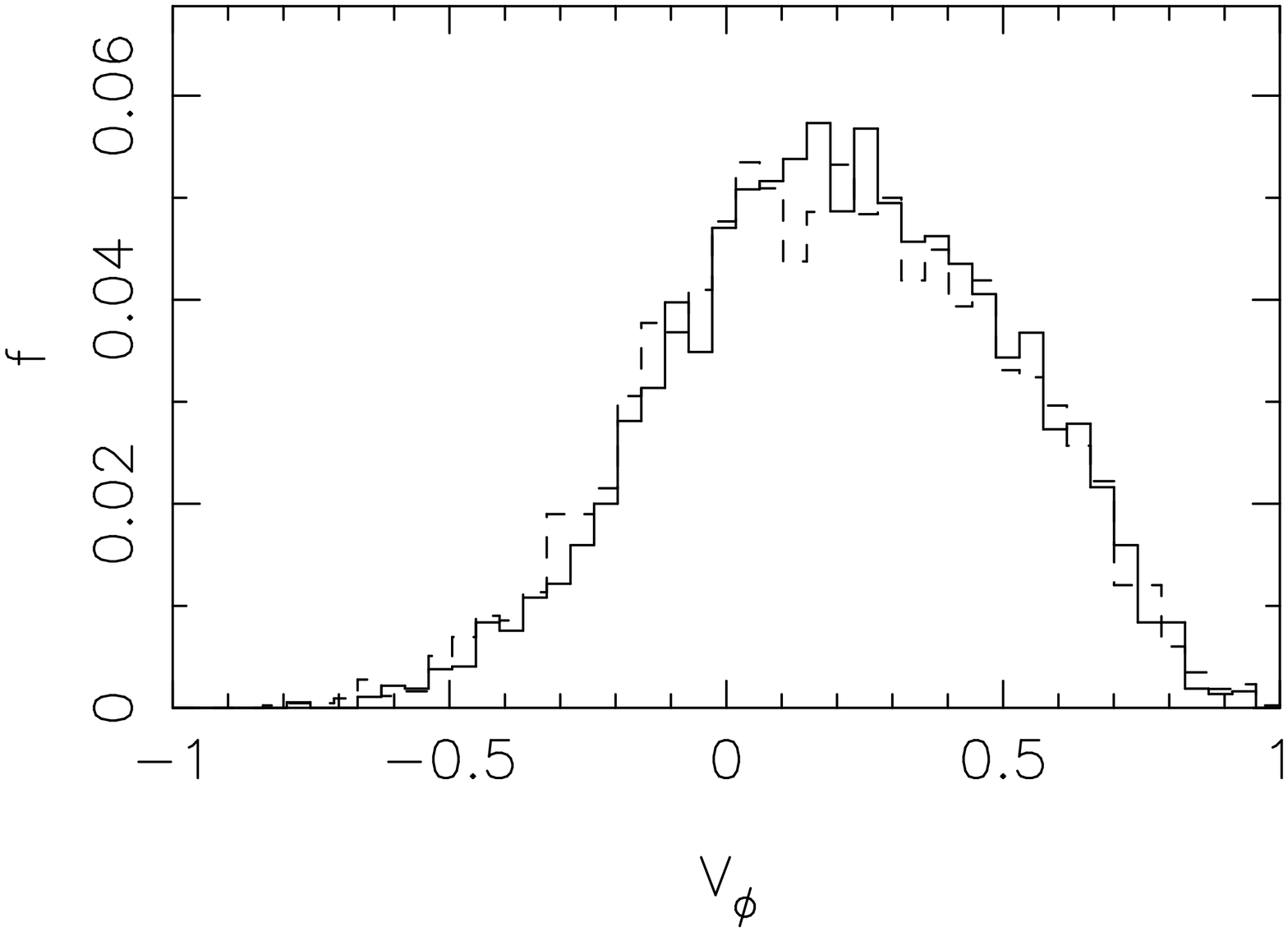}\\

\end{tabular}

\caption{Same as Fig. \ref{Avr} for galactocentric azimuthal
velocities, V$_\phi$. None of the models exhibit substantial near/far
asymmetry.}
\label{Avphi}
\end{figure}

\begin{figure}
\begin{tabular}{lccc}
  Angle: 15$^\circ$ & 25$^\circ$ & 35$^\circ$\\
    R1 
    \includegraphics[width=0.63in, angle=-90 ]{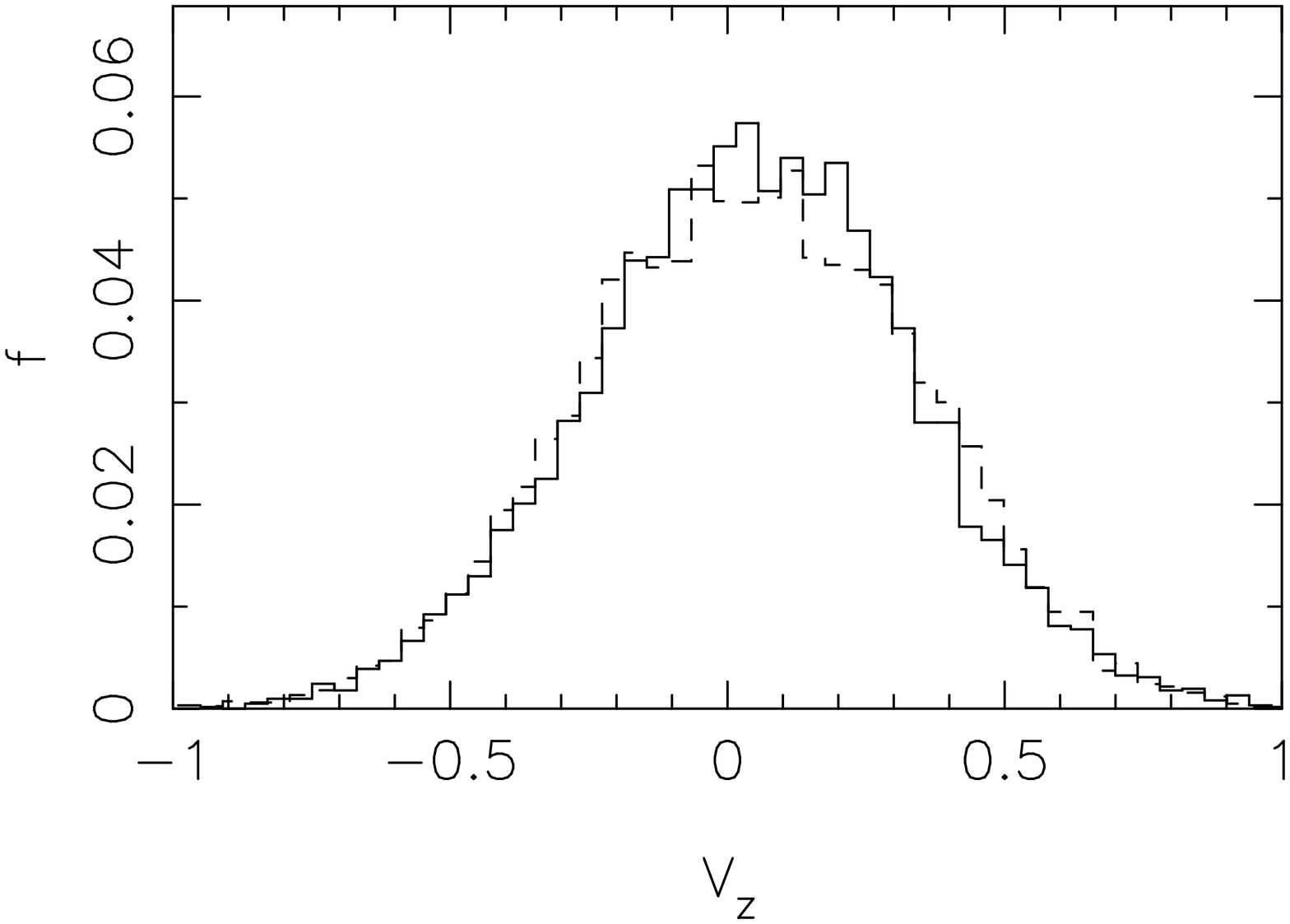}&
    \includegraphics[width=0.63in, angle=-90 ]{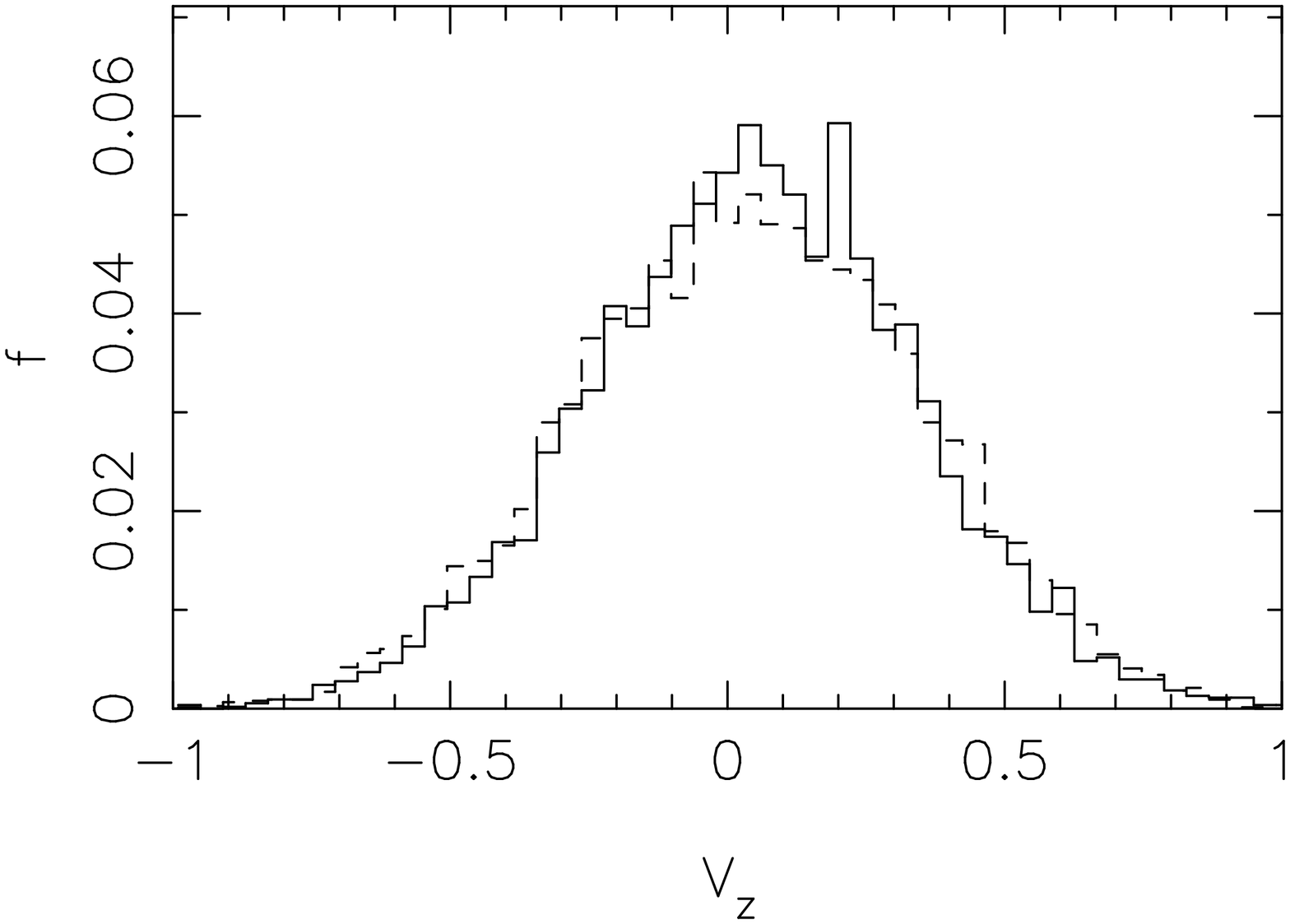}&
    \includegraphics[width=0.63in, angle=-90 ]{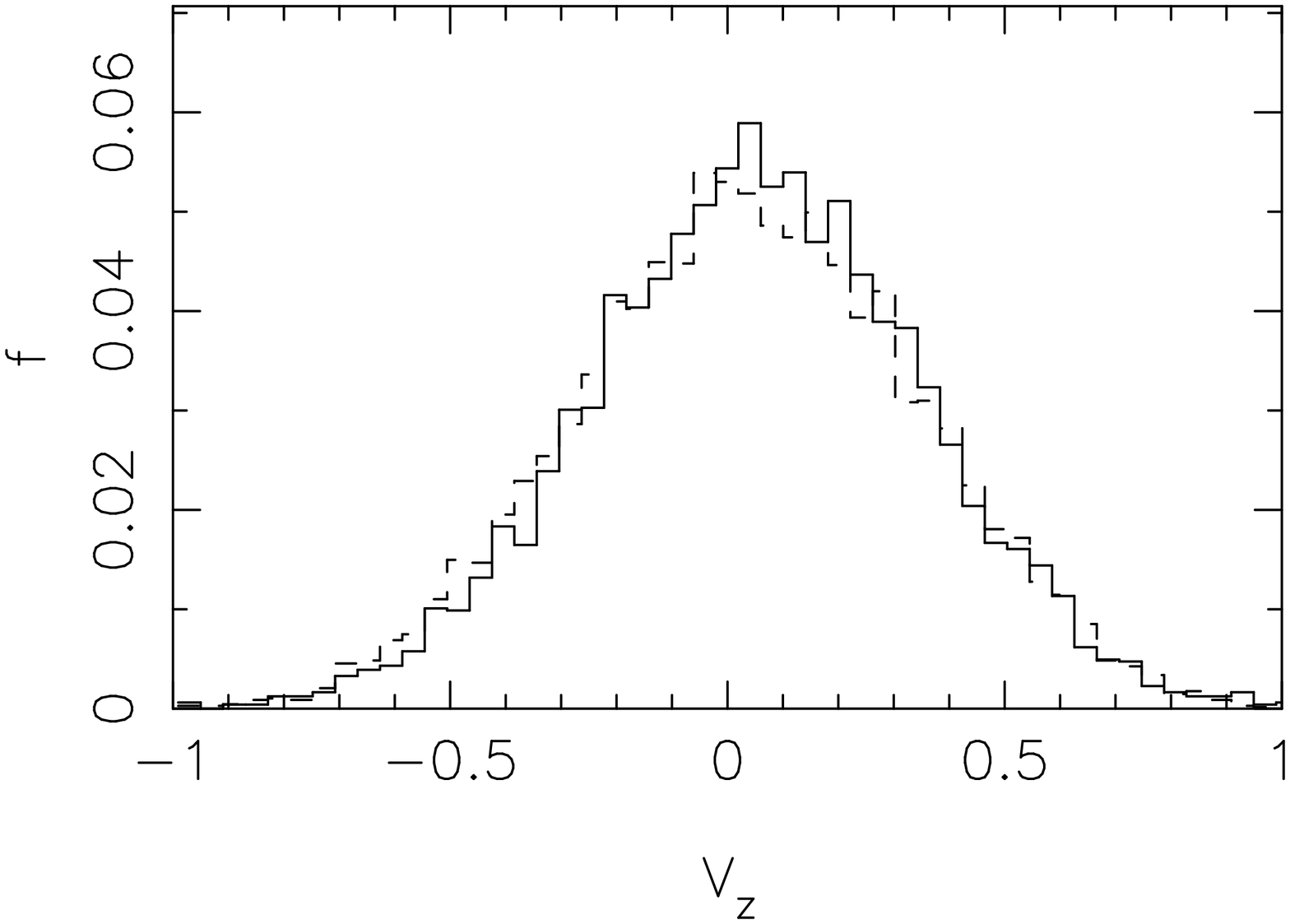}\\
    B3    
    \includegraphics[width=0.63in, angle=-90 ]{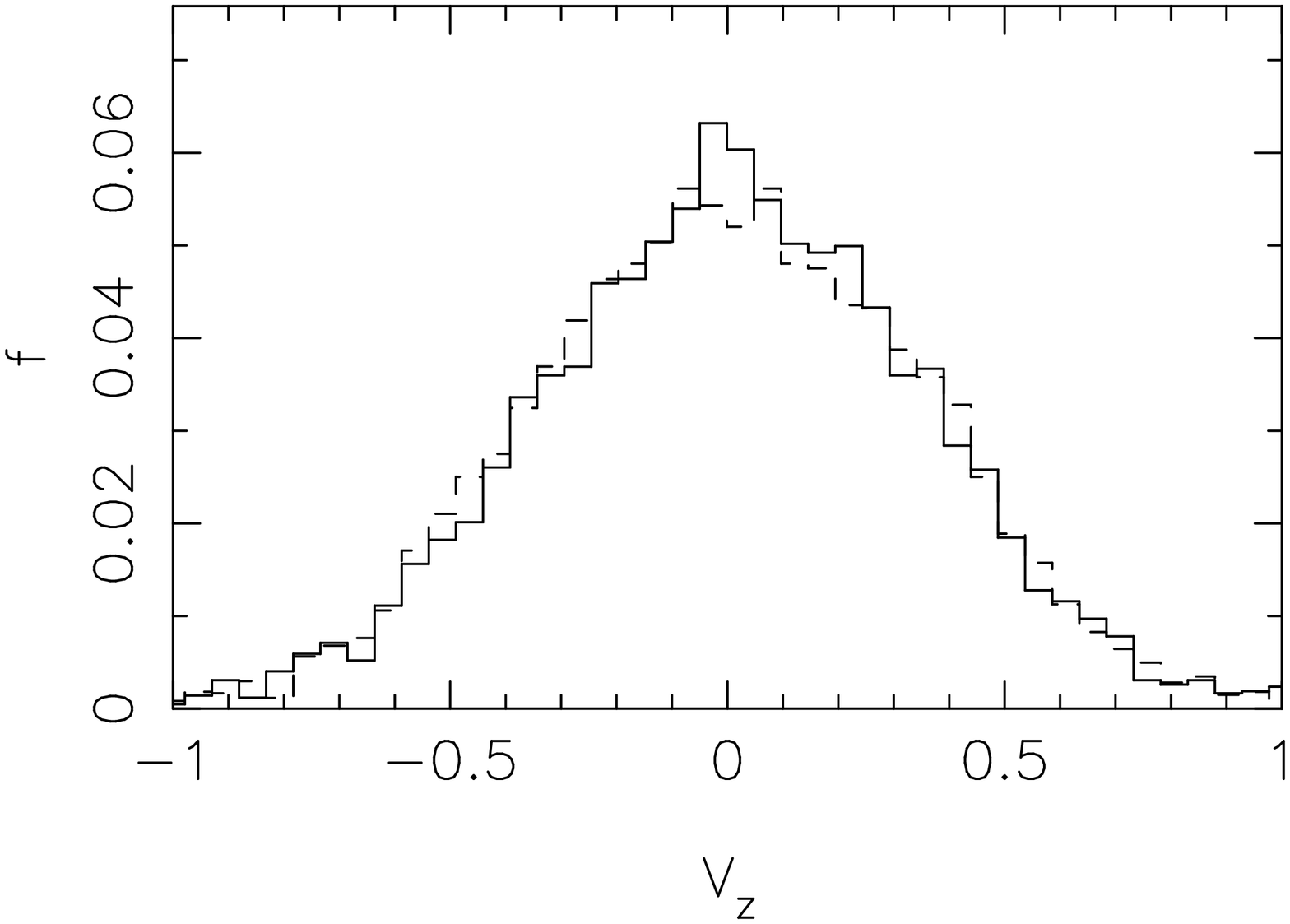}&
    \includegraphics[width=0.63in, angle=-90 ]{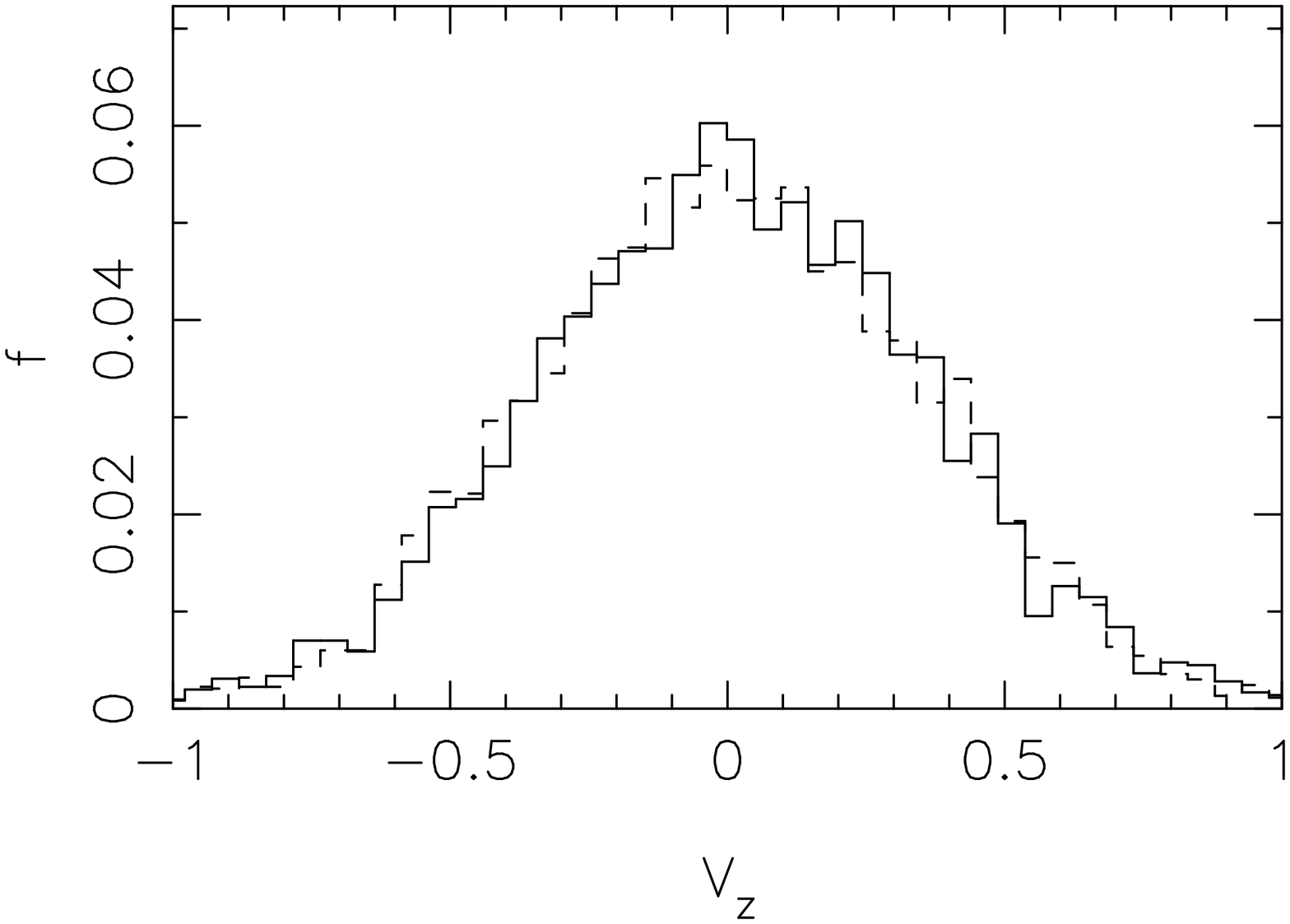}&
    \includegraphics[width=0.63in, angle=-90 ]{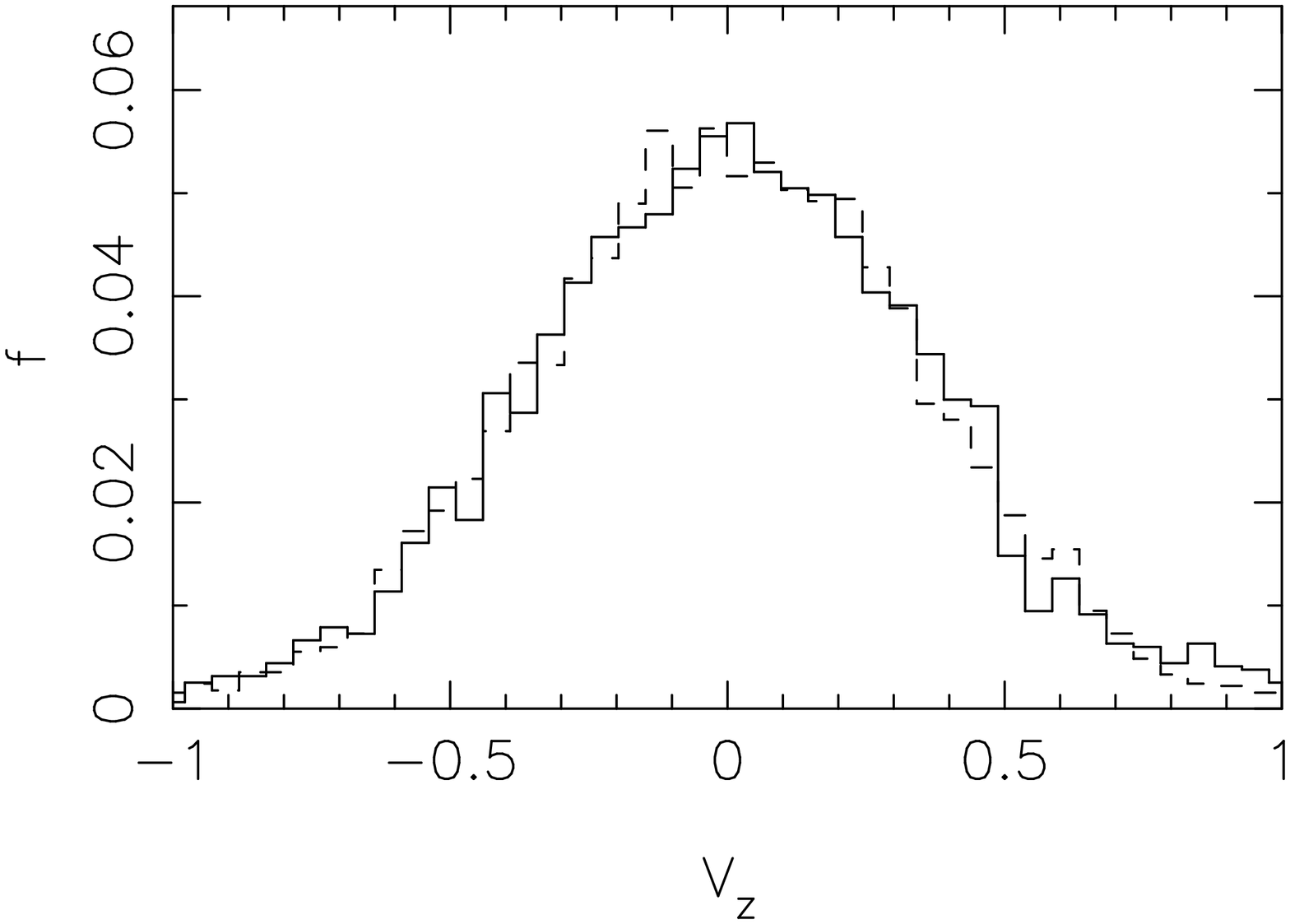}\\
    B2 
    \includegraphics[width=0.63in, angle=-90 ]{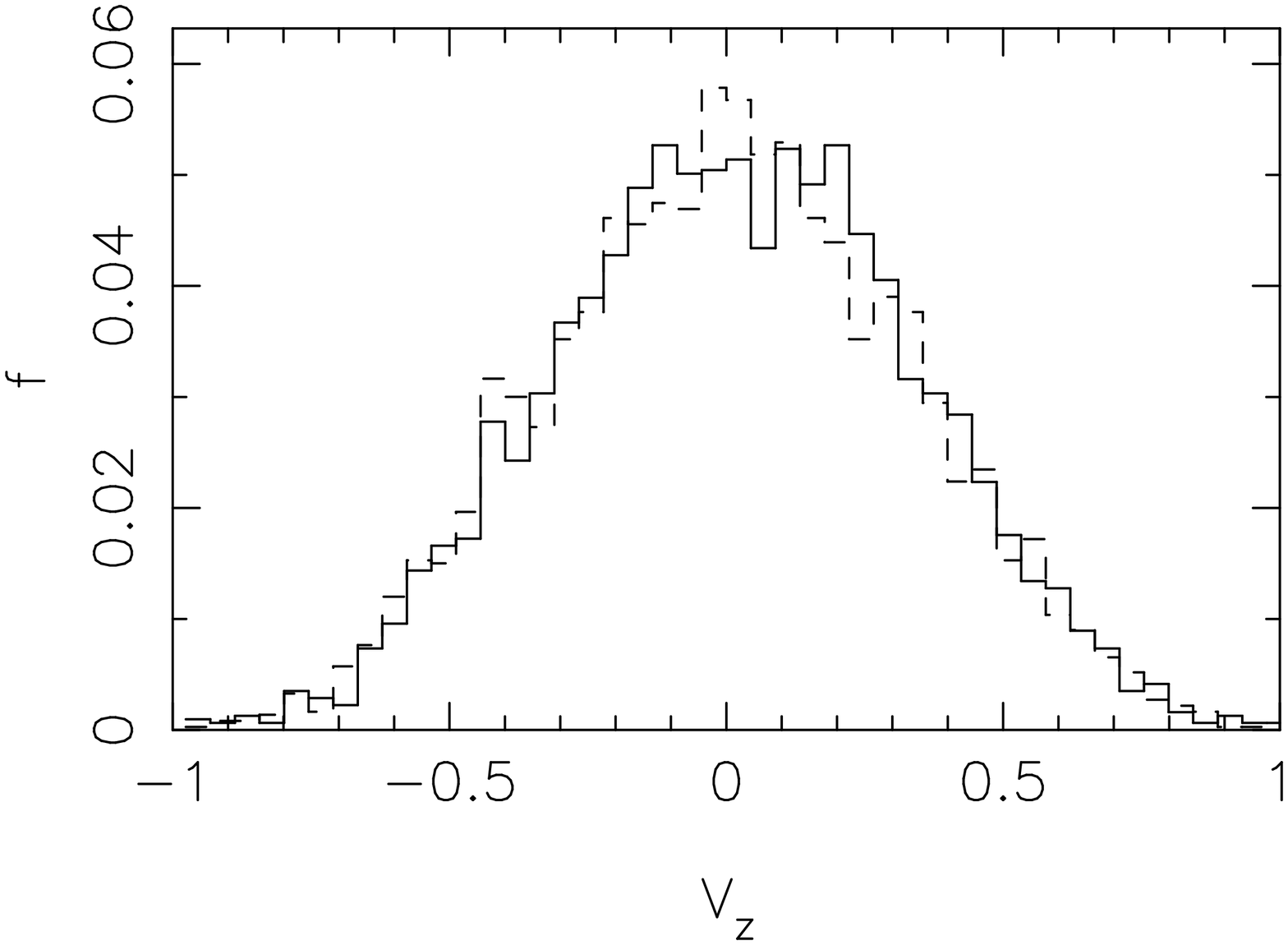}&
    \includegraphics[width=0.63in, angle=-90 ]{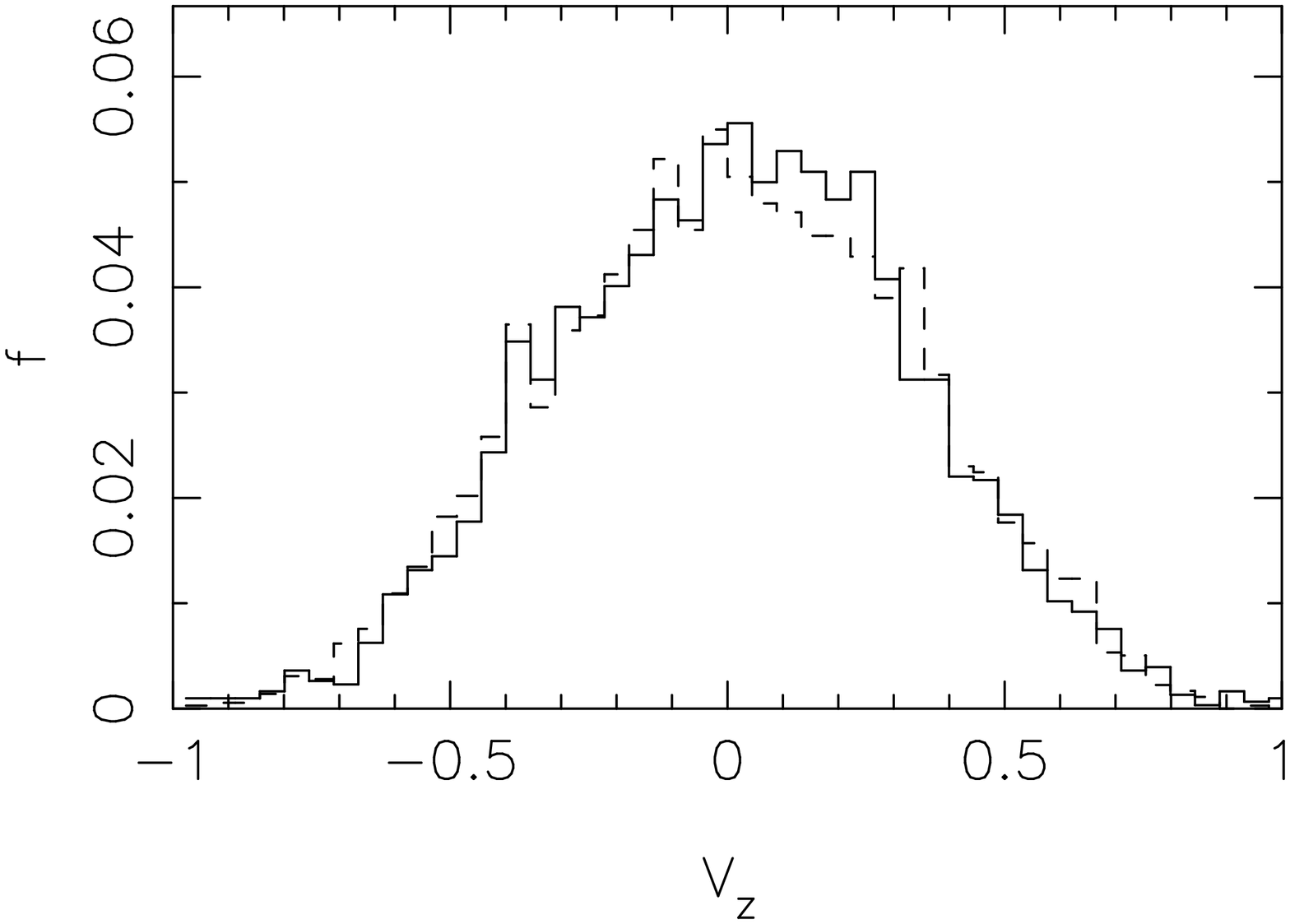}&
    \includegraphics[width=0.63in, angle=-90 ]{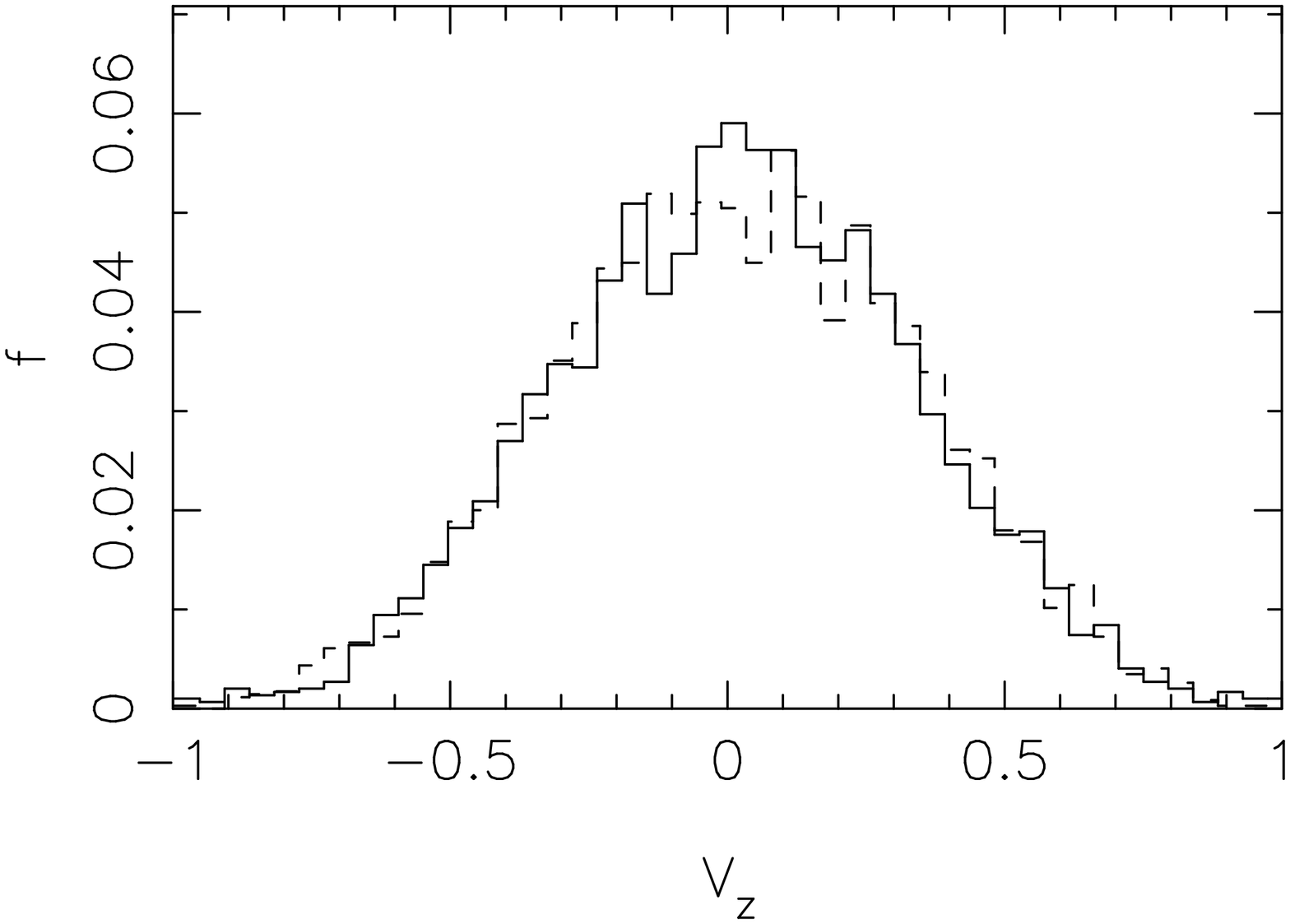}\\
    R5 
    \includegraphics[width=0.63in, angle=-90 ]{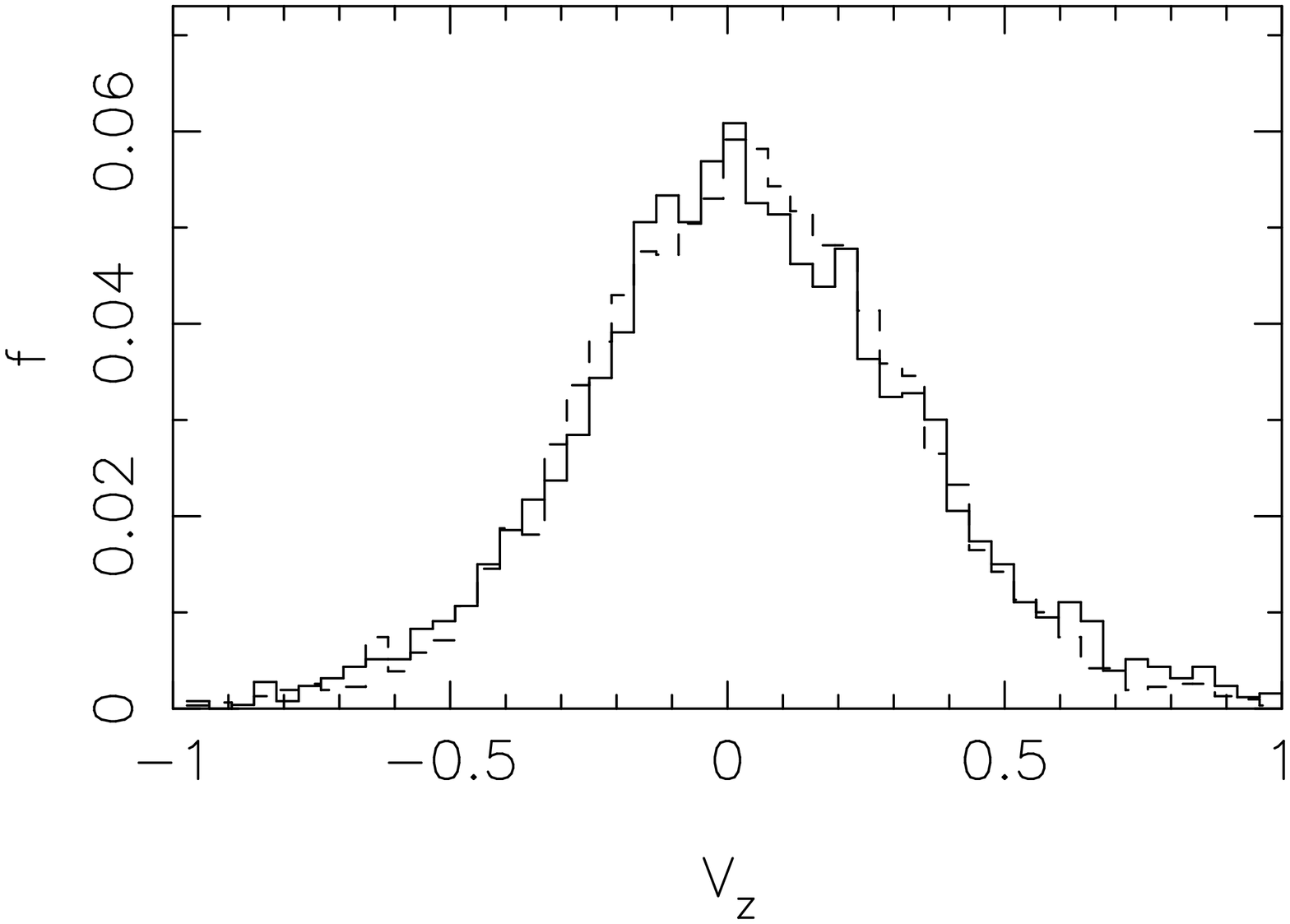}&
    \includegraphics[width=0.63in, angle=-90 ]{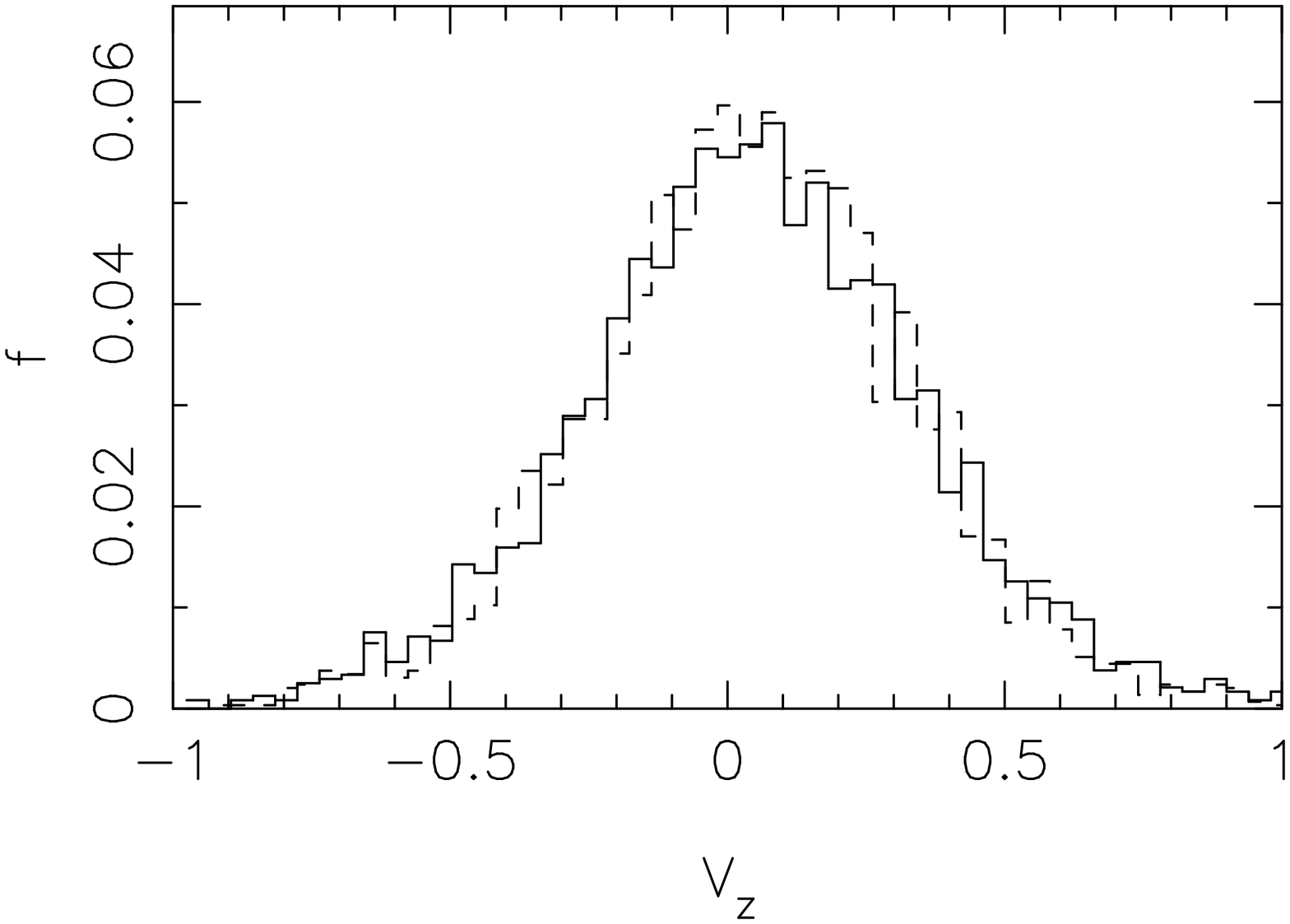}&
    \includegraphics[width=0.63in, angle=-90 ]{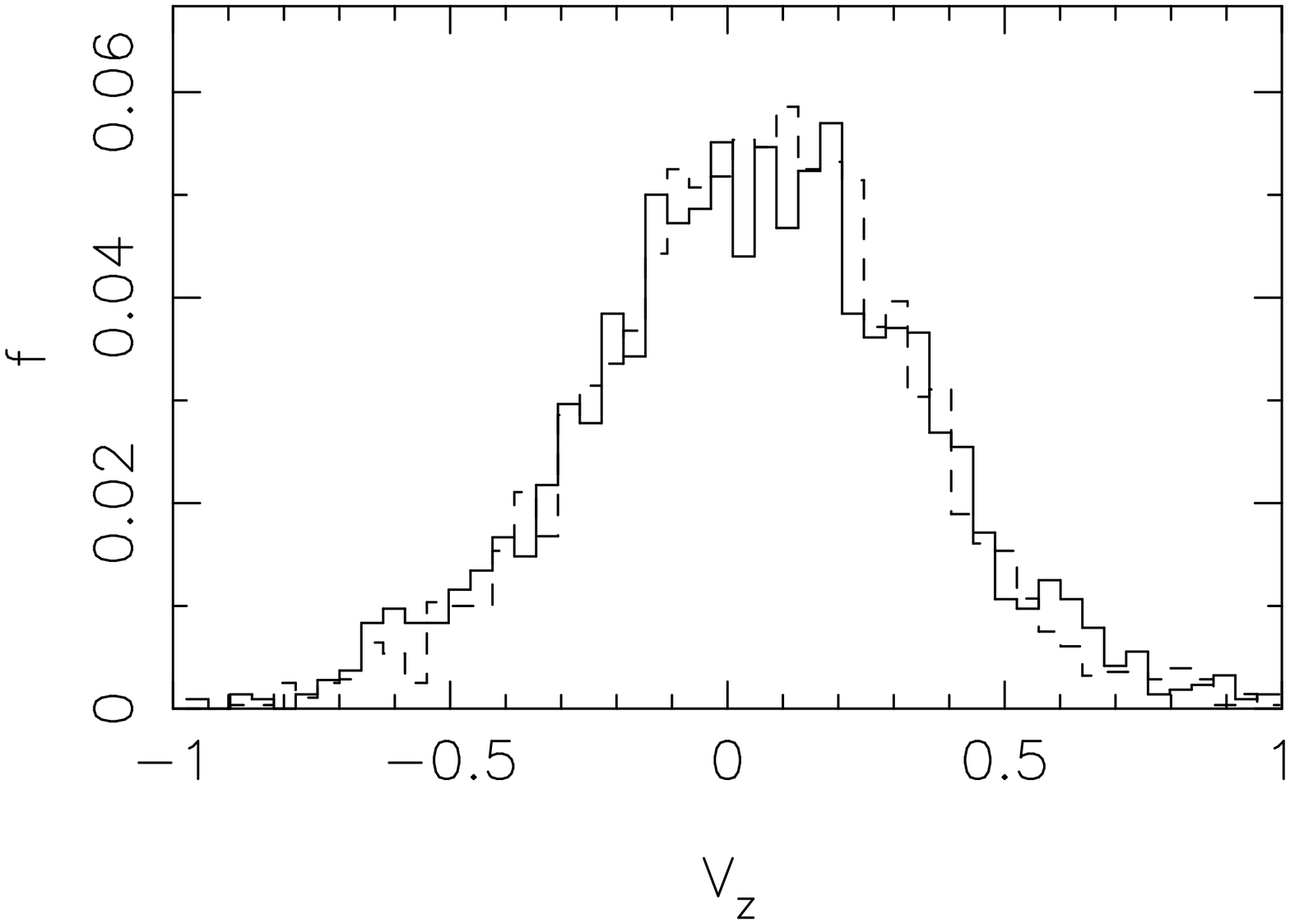}\\
    R2 
    \includegraphics[width=0.63in, angle=-90 ]{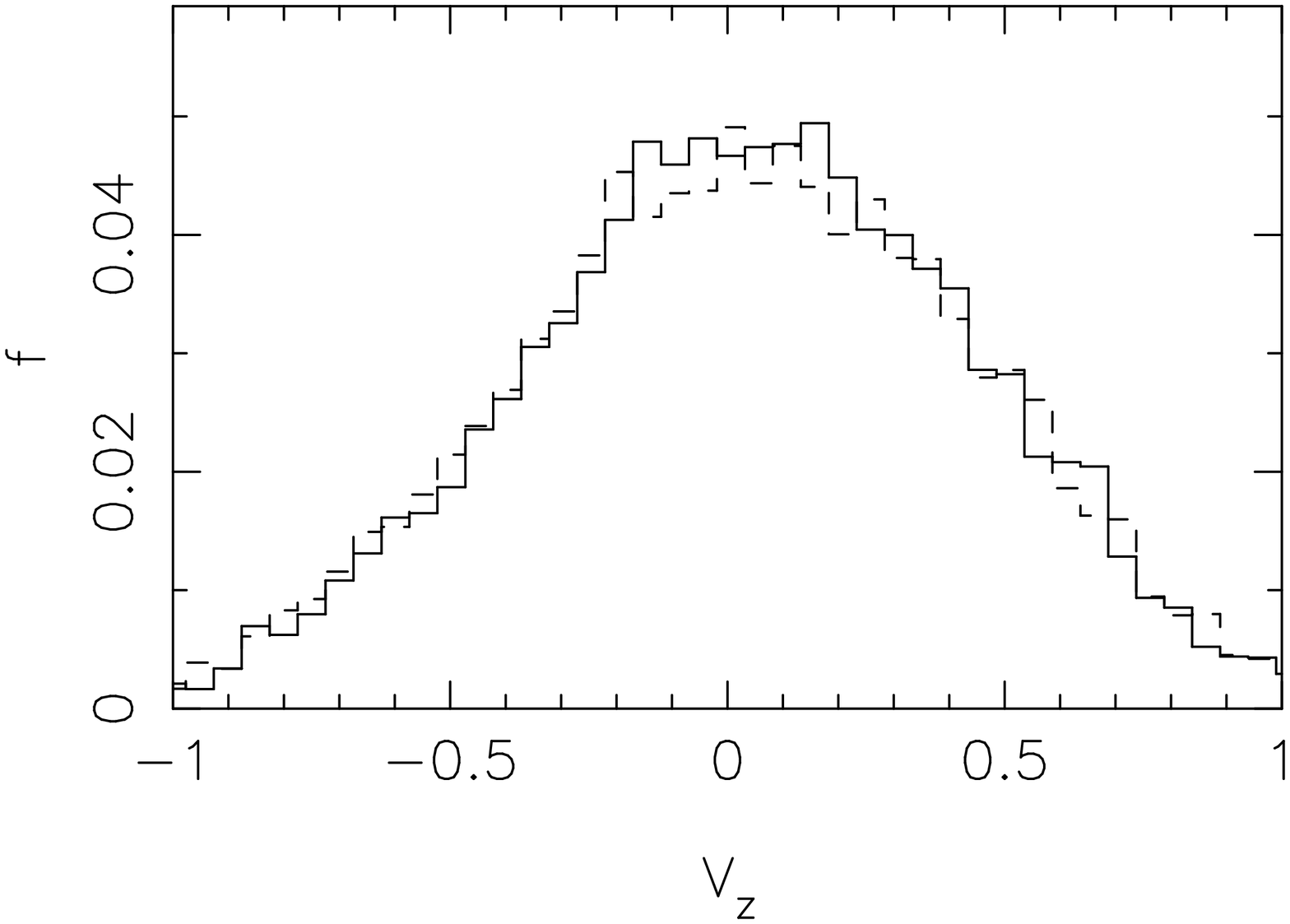}&
    \includegraphics[width=0.63in, angle=-90 ]{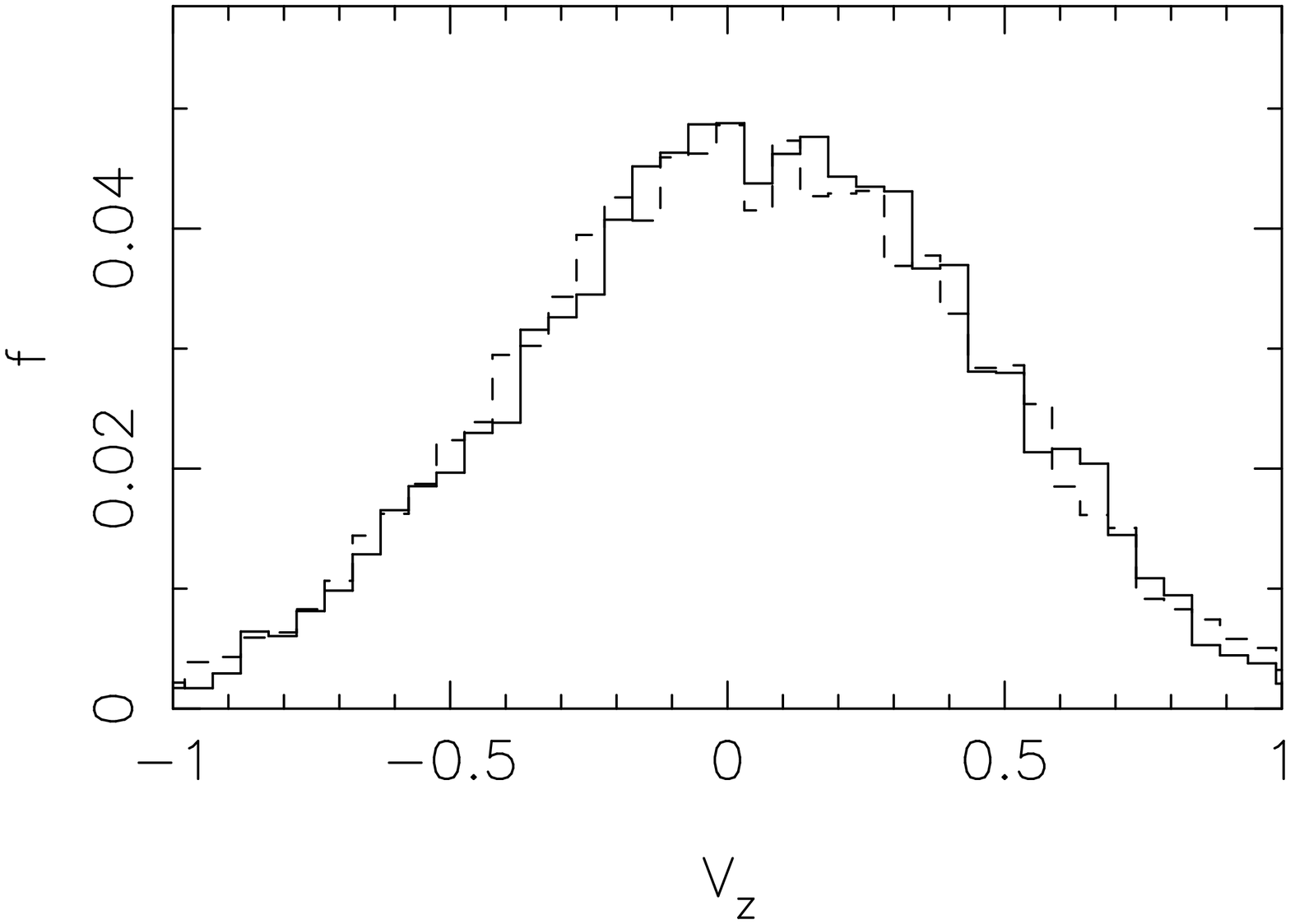}&
    \includegraphics[width=0.63in, angle=-90 ]{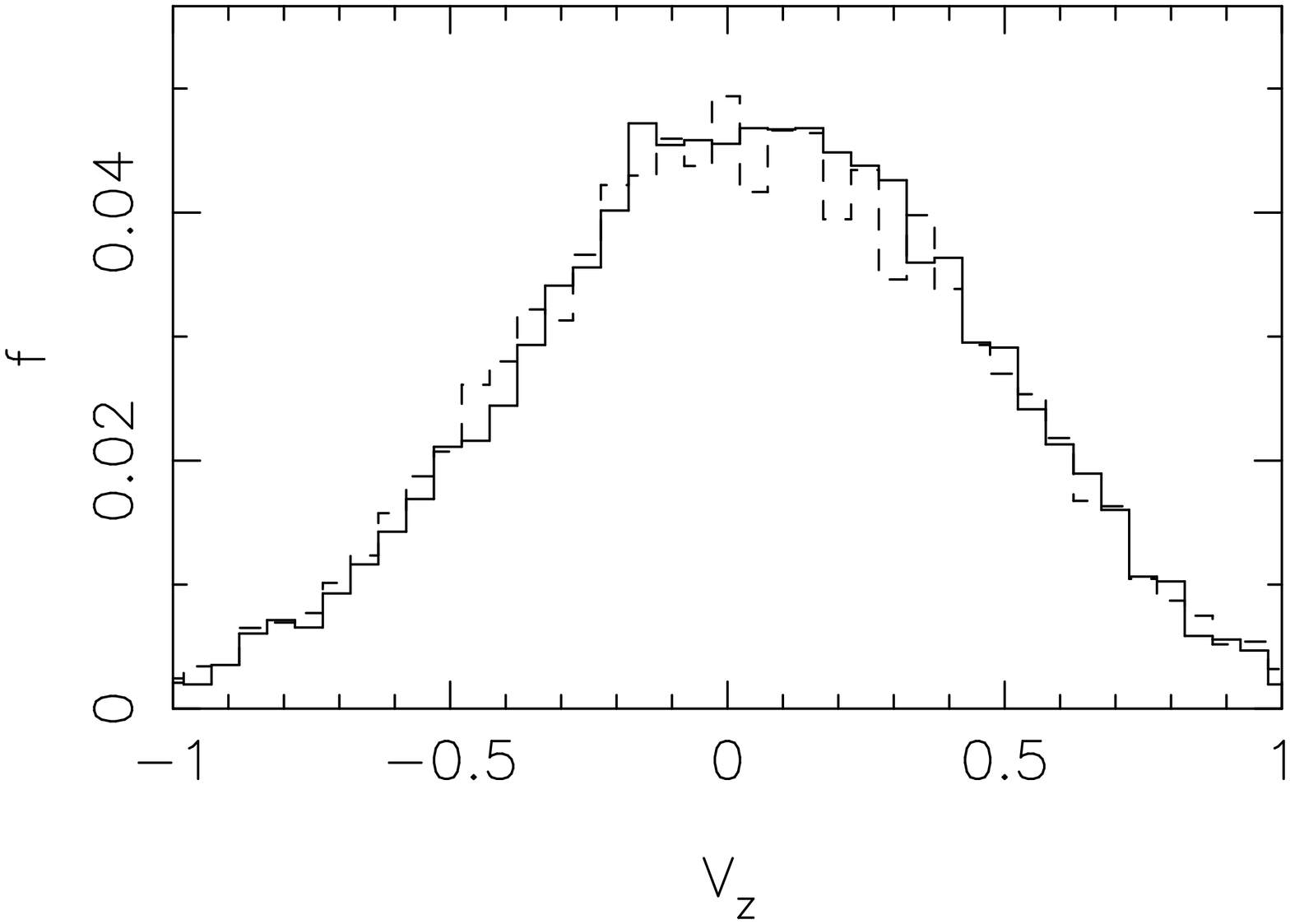}\\
    R6 
    \includegraphics[width=0.63in, angle=-90 ]{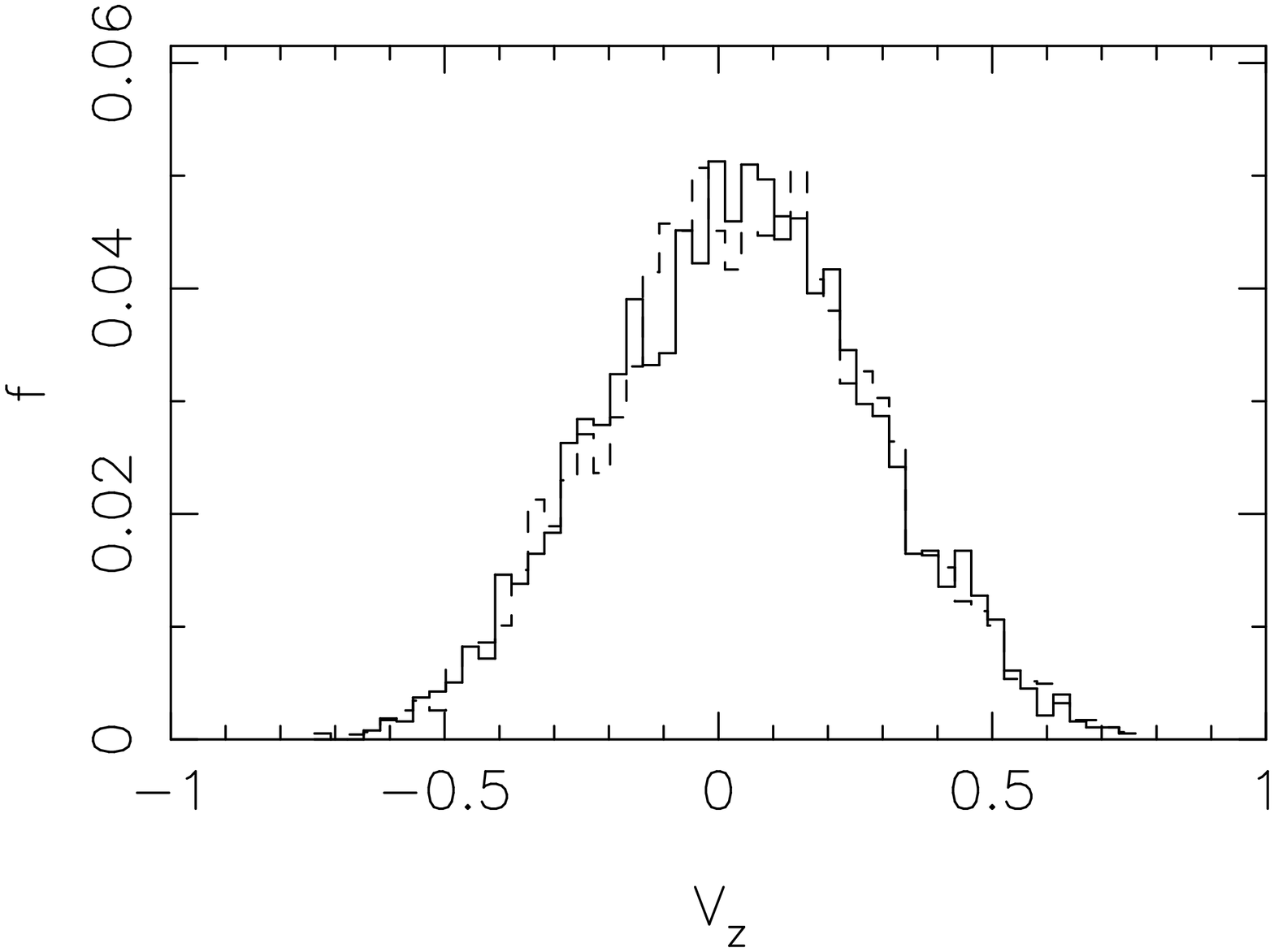}&
    \includegraphics[width=0.63in, angle=-90 ]{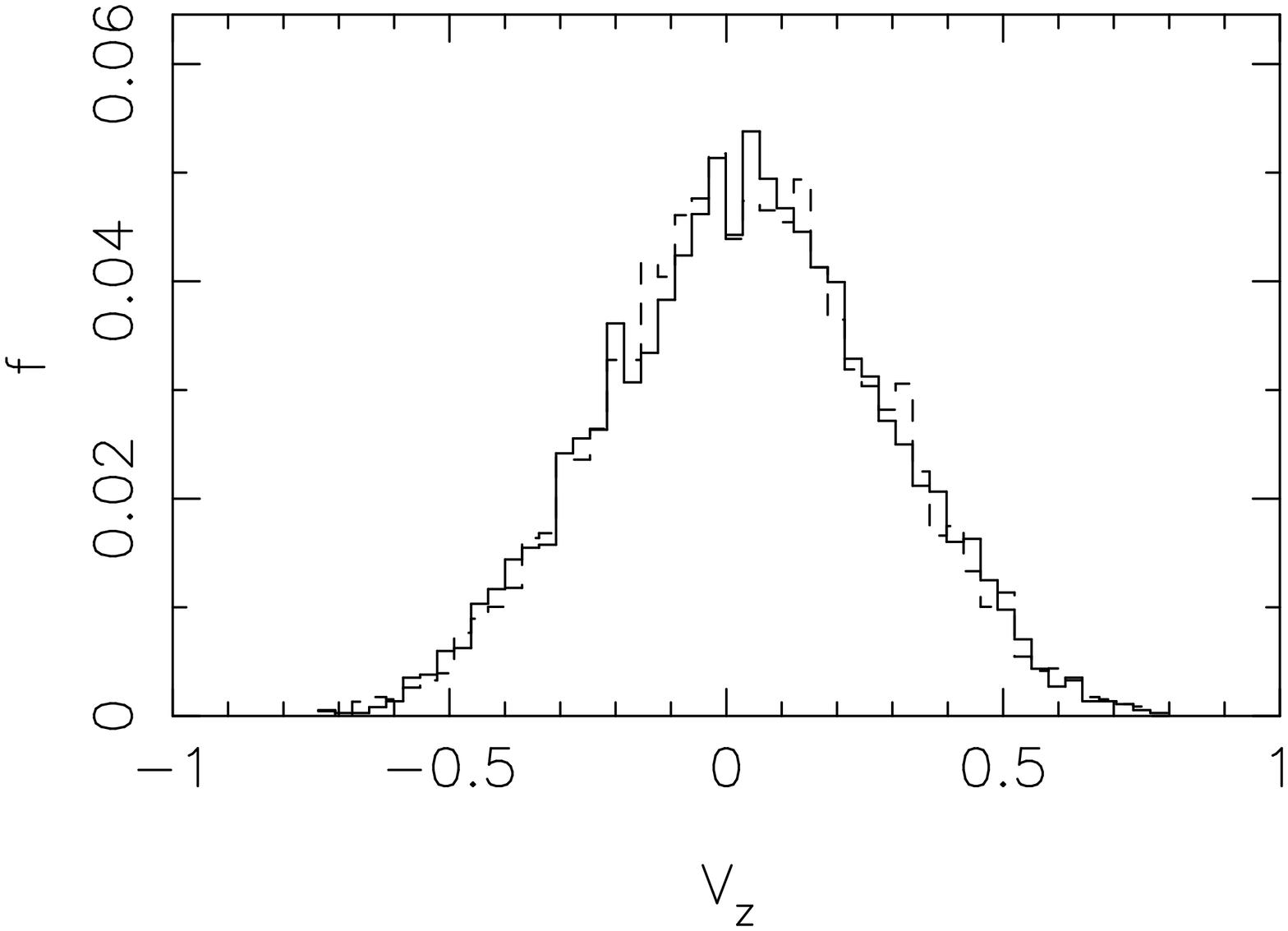}&
    \includegraphics[width=0.63in, angle=-90 ]{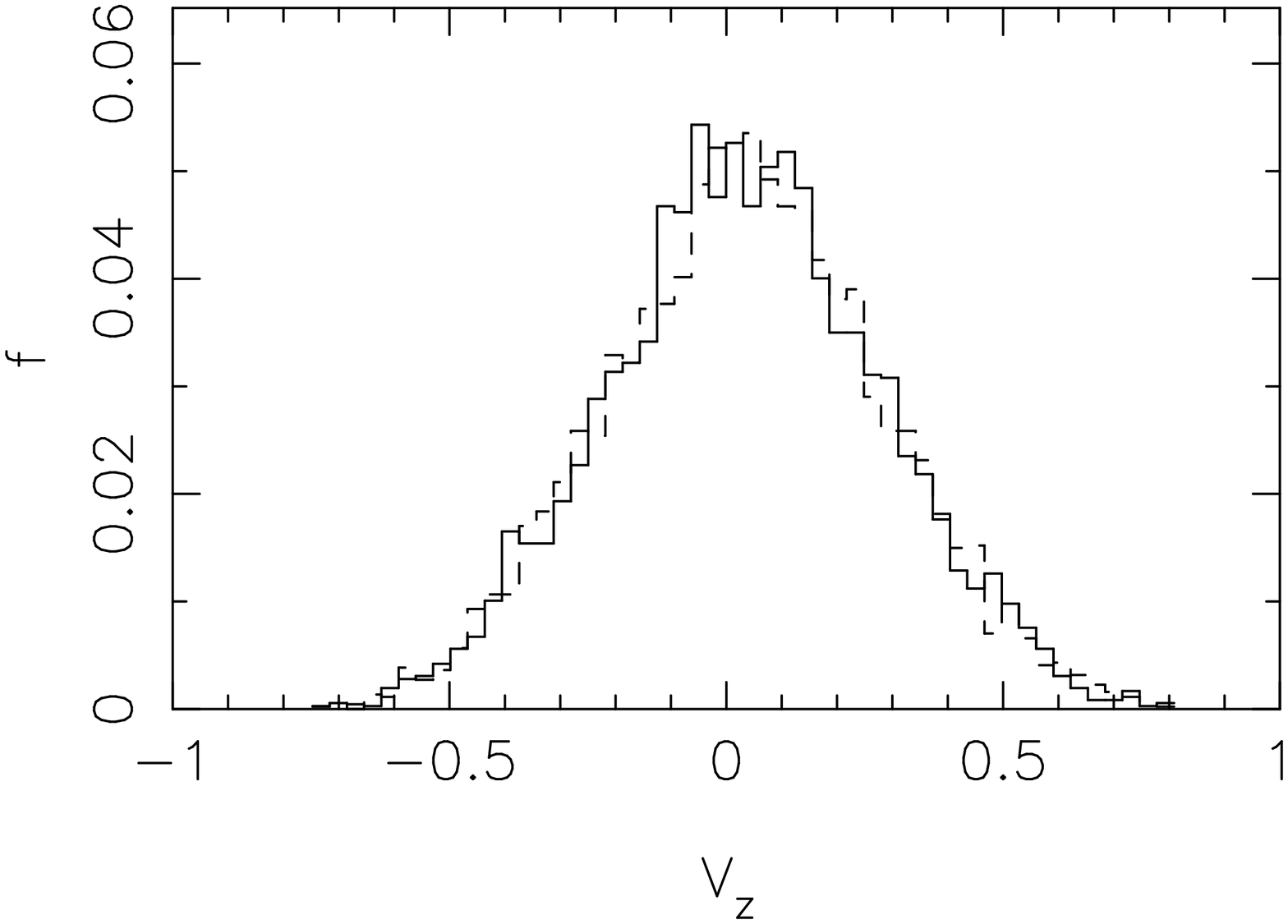}

\end{tabular}

\caption{Same as Fig. \ref{Avr} for vertical velocities, V$_z$. None
  of the models exhibit substantial near/far asymmetry.}
\label{Avz}
\end{figure}

An asymmetric bulge should show velocity distributions which change as
the viewing angle changes.  In Fig. \ref{Avr}, \ref{Avphi} and
\ref{Avz} we show the distributions of near and far side velocities
for bar angles $15\degr$, $25\degr$ and $35\degr$ at $(l,b) = (0\degr,
-6\degr)$.  For the X-shaped models, R1 and B3, the $V_{los}$
distributions are different and change considerably with bar angle;
the distributions are less separated and show less relative change in
the rest of the models.  This latitude therefore is a sensitive probe
of the bar angle in $V_{los}$.  Azimuthal (Fig. \ref{Avphi}) and
vertical velocities (Fig. \ref{Avz}) are mostly unaffected by the
change in bar angle.  The absence of a near/far side asymmetry shows
little dependence on bar angle.

\subsection{Conclusions from all models}

Models with X-shapes have a distinct signature in the difference
between near and far sides mean line-of-sight velocity only. This
signature is not present in models without X-shapes. Additionally, the
bar angle has an effect on these asymmetries.

\label{lastpage}

\end{document}